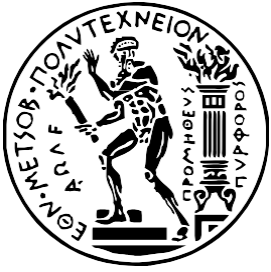

Εθνικό Μετσόβιο Πολυτεχνείο
Σχολή Ηλεκτρολόγων Μηχανικών
και Μηχανικών Υπολογιστών
Τομέας Τεχνολογίας Πληροφορικής και Υπολογιστών

# Κατασκευή ενός AI-agent βάσει LLM: Μια μεθοδολογία υψηλού επιπέδου για τη βελτίωση των LLMs με APIs

# Creating an LLM-based AI-agent: A high-level methodology towards enhancing LLMs with APIs

ΔΙΠΛΩΜΑΤΙΚΗ ΕΡΓΑΣΙΑ

**ΙΩΑΝΝΗΣ ΤΖΑΧΡΗΣΤΑΣ**

**Επιβλέπων :** Νικόλαος Παπασπύρου
Καθηγητής Ε.Μ.Π.

Αθήνα, Ιούνιος 2024

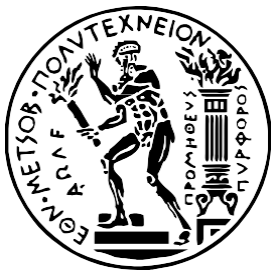

Εθνικό Μετσόβιο Πολυτεχνείο

Σχολή Ηλεκτρολόγων Μηχανικών
και Μηχανικών Υπολογιστών

Τομέας Τεχνολογίας Πληροφορικής και Υπολογιστών

# Κατασκευή ενός AI-agent βάσει LLM: Μια μεθοδολογία υψηλού επιπέδου για τη βελτίωση των LLMs με APIs

# Creating an LLM-based AI-agent: A high-level methodology towards enhancing LLMs with APIs

## ΔΙΠΛΩΜΑΤΙΚΗ ΕΡΓΑΣΙΑ

### ΙΩΑΝΝΗΣ ΤΖΑΧΡΗΣΤΑΣ

**Επιβλέπων :** Νικόλαος Παπασπύρου
Καθηγητής Ε.Μ.Π.

Εγκρίθηκε από την τριμελή εξεταστική επιτροπή τη 10η Ιουνίου 2024.

................................................  ................................................  ................................................
Νικόλαος Παπασπύρου          Κωνσταντίνος Σαγώνας         Δημήτριος Φωτάκης
Καθηγητής Ε.Μ.Π.                     Αν. Καθηγητής Ε.Μ.Π.              Καθηγητής Ε.Μ.Π.

Αθήνα, Ιούνιος 2024

..................................
**Ιωάννης Τζαχρήστας**
Διπλωματούχος Ηλεκτρολόγος Μηχανικός και Μηχανικός Υπολογιστών Ε.Μ.Π.

# Περίληψη

Τα Μεγάλα Γλωσσικά Μοντέλα (LLMs) έχουν επιφέρει ριζικές αλλαγές σε ποικίλους επιστημονικούς και τεχνολογικούς τομείς. Η χρησιμότητά τους συχνά περιορίζεται από την έλλειψη αλληλεπίδρασης με το εξωτερικό ψηφιακό περιβάλλον. Για να ξεπεραστεί αυτός ο περιορισμός και να ενταχθούν τα LLMs και η Τεχνητή Νοημοσύνη (AI) σε πραγματικές εφαρμογές, κατασκευάζονται καθημερινά εξειδικευμένοι πράκτορες Τεχνητής Νοημοσύνης (AI-agents). Στην παρούσα Διπλωματική Εργασία διατυπώνεται μια υψηλού επιπέδου αρχιτεκτονική αυτών των AI-agents και αναλύεται εκτενώς η διαδικασία εξόπλισης των LLMs με τη δυνατότητα αλληλεπίδρασης με Διεπαφές Προγραμματισμού Εφαρμογών (APIs). Βασιζόμενοι σε προηγούμενες έρευνες, παρουσιάζουμε μια μεθοδολογία 7 βημάτων που ξεκινά με την επιλογή κατάλληλων LLMs και την "αποσύνθεση" των εργασιών που είναι απαραίτητη για την επίλυση σύνθετων προβλημάτων. Αυτή η μεθοδολογία περιλαμβάνει τεχνικές για τη δημιουργία δεδομένων εκπαίδευσης (training data) για συγκεκριμένα APIs και ευρετικές για την επιλογή του κατάλληλου API από ένα πλήθος επιλογών. Αυτά τα βήματα τελικά οδηγούν στη δημιουργία κλήσεων των προγραμματιστικών διεπαφών (API calls) που είναι τόσο συντακτικά όσο και σημασιολογικά ευθυγραμμισμένες με την κατανόηση του LLM για μια δεδομένη εργασία. Επιπλέον, αναθεωρούμε υπάρχουσες πλατφόρμες και εργαλεία που διευκολύνουν αυτές τις διαδικασίες και τονίζουμε τα κενά στις τρέχουσες προσπάθειες. Σε αυτή την κατεύθυνση, προτείνουμε μια αρχιτεκτονική ενσωμάτωσης των LLMs σε φορητές συσκευές που στοχεύει στην αξιοποίηση της λειτουργικότητας των φορητών συσκευών χρησιμοποιώντας μικρά μοντέλα από την κοινότητα Hugging Face. Εξετάζουμε την αποτελεσματικότητα των εν λόγω προσεγγίσεων σε πραγματικές εφαρμογές διαφόρων τομέων, συμπεριλαμβανομένης της διαδικασίας δημιουργίας μιας μουσικής παρτιτούρας για πιάνο. Μέσω μιας εκτεταμένης ανάλυσης της βιβλιογραφίας και των διαθέσιμων τεχνολογιών, αυτή η διατριβή στοχεύει να αποτελέσει μπούσουλα για τους ερευνητές και τους μηχανικούς που φιλοδοξούν να διασυνδέσουν τα LLMs με εξωτερικά εργαλεία και εφαρμογές, διαμορφώνοντας έτσι τις κατάλληλες συνθήκες για πιο αυτόνομους, αξιόπιστους και εξειδικευμένους AI-agents.

## Λέξεις κλειδιά

Large Language Models (LLMs), Application Programming Interfaces (APIs), Artificial Intelligence (AI), Natural Language Processing (NLP), AI-agent Architecture, Classification, Semantic Vector Space, Semantic Alignment, Word Embedding, Human-Computer Interaction (HCI), Music Notation

# Abstract

Large Language Models (LLMs) have revolutionized various aspects of engineering and science. Their utility is often bottlenecked by the lack of interaction with the external digital environment. In order to overcome this limitation and achieve integration of LLMs and Artificial Intelligence (AI) into real-world applications, customized AI-agents are being constructed every day. Based on the technological trends and the existing techniques, we extract a high-level approach for constructing these AI-agents, focusing on their underlying architecture. This thesis serves as a comprehensive guide that elucidates a multi-faceted approach for empowering LLMs with the capability to leverage Application Programming Interfaces (APIs). We present a 7-step methodology that begins with the selection of suitable LLMs and the task decomposition that is necessary for complex problem-solving. This methodology includes techniques for generating training data for API interactions and heuristics for selecting the appropriate API among a plethora of options. These steps eventually lead to the generation of API calls that are both syntactically and semantically aligned with the LLM's understanding of a given task. Moreover, we review existing frameworks and tools that facilitate these processes and highlight the gaps in current attempts. In this direction, we propose an on-device architecture that aims to exploit the functionality of carry-on devices by using small models from the Hugging Face community. We examine the effectiveness of the aforementioned approaches on real-world applications of various domains, including the generation of a piano sheet. Through an extensive analysis of the literature and available technologies, this thesis aims to set a compass for researchers and practitioners to harness the full potential of LLMs augmented with external tool capabilities, thus paving the way for more autonomous, robust and context-aware AI-agents.

# Key words

# Ευχαριστίες

Ευχαριστώ το Θεό.
Ευχαριστώ τους γονείς μου.
Τη μητέρα μου, Δημητρούλα Ηγουμενίδη του Γεωργίου, μηχανικό και πιανίστα.
Τον πατέρα μου, Κωνσταντίνο Τζαχρήστα του Ιωάννη, Συνταγματάρχη Πεζικού.
Ευχαριστώ τα αδέρφια μου, Γεώργιο και Θεοφάνη.
Ευχαριστώ τον κ. Θεόδωρο Μπόλη, πρώην πρόεδρο της Ελληνικής Μαθηματικής Εταιρείας και συντοπίτη μου, που υπήρξε αφορμή να ασχοληθώ με τις Ολυμπιάδες Μαθηματικών.
Ευχαριστώ τον Liszt, τον Αττίκ.
Ευχαριστώ τον πατέρα Αλέξανδρο Κατερινόπουλο, τις κυρίες Στεργιανή και Νεραντζιά και συνολικά την ενορία μου στο Μόναχο, Salvatorkirche, που είναι το Κυριακάτικο πρωινό μου στη Γερμανία.
Ευχαριστώ τους φίλους μου.
Ευχαριστώ όλους τους ανθρώπους της ακαδημαϊκής κοινότητας του Εθνικού Μετσόβιου Πολυτεχνείου (ενδεικτικά αναφέρω: κ. Ανδρέα Μπουντουβή, κ. Κατερίνα Κριθινάκη, κ. Δημήτριο Φωτάκη, κ. Πέτρο Ποτίκα, κ. Βασίλειο Βεσκούκη, κ. Θανάση Λιανέα, κ. Ευστάθιο Ζάχο, κ. Σοφία Λαμπροπούλου, κ. Ηλία Γλύτση, κ. Ανάργυρο Φελλούρη, κ. Παρασκευή Τζούβελη, κ. Στέφανο Κόλλια, κ. Αριστείδη Παγουρτζή, κ. Κωστή Σαγώνα, κ. Δημήτριο Ασκούνη, κ. Αντώνιο Παπαβασιλείου, κ. Παναγιώτη Φράγκο, κ. Παναγιώτη Κωττή, κ. Αθανάσιο Παναγόπουλο, κ. Αριστοτέλη Χαραλαμπάκη, κ. Ειρήνη Κοιλανιώτη, κ. Θεοδώρα Σούλιου, κ. Νεκτάριο Κοζύρη) για τη διαρκή και έμπρακτη στήριξή τους και ιδιαίτερα τον επιβλέποντά μου, κ. Νικόλαο Σ. Παπασπύρου.
Ευχαριστώ το Huawei Munich Research Center και ιδιαίτερα τη supervisor μου Dr. Aifen Sui.
Η παρούσα Διπλωματική Εργασία εκπονήθηκε στο Ερευνητικό Κέντρο της Huawei στο Μόναχο (Riesstraße 25, 80992 München), στο οποίο εργάζομαι από τις 15 Ιουλίου 2023.
Ευχαριστώ τον καλλιτέχνη Ιωάννη Βλέτσα για το υπέροχο σκίτσο που μου χάρισε σε ένα πεντάλεπτο διάλειμμα φροντιστηρίου στη Β΄ Λυκείου.

## Αντί Προλόγου

Επισυνάπτω μουσικό απόσπασμα της τελετής βράβευσης του Huawei Munich Research Center (8 Μαρτίου 2024), όπου κλήθηκα να παίξω πιάνο. https://www.youtube.com/watch?v=13Zj6UqBi1E
Επισυνάπτω, επίσης, ένα κείμενο της μητέρας μου με τίτλο "Εκ βαθέων", που είναι η "διαισθητική απόδειξη" ότι η Τεχνητή Νοημοσύνη δεν θα αντικαταστήσει την ανθρώπινη ψυχή.
https://www.youtube.com/watch?v=m_0Xq2u75YY

Όπως είχε πει ο πατέρας μου: "Φοβάμαι τα πάντα, ελπίζω τα πάντα και είμαι ελεύθερος."
Όπως είχε πει ο αδερφός μου ο Γιώργος: "Η σκληρή δουλειά είναι το μεγαλύτερο ταλέντο."
Όπως είχε πει ο αδερφός μου ο Φάνης: "Αν ήμουν αριθμός, θα ήθελα να είμαι το "8", για να μπορώ να ξαπλώνω και να γίνομαι άπειρο."
Όπως είχε πει η μητέρα μου: "Η γρήγορη λύση είναι τέλεια λύση."
Όπως είχε πει ο Kung-Fu-Panda: "Δεν υπάρχει μυστικό συστατικό στη μυστικοσυστατικόσουπα."
Όπως είχε πει ο Pasteur: "Η τύχη ευνοεί τους προετοιμασμένους."
Καλή ανάγνωση...

<div style="text-align: right;">
Ιωάννης Τζαχρήστας,
Αθήνα, 10η Ιουνίου 2024
</div>

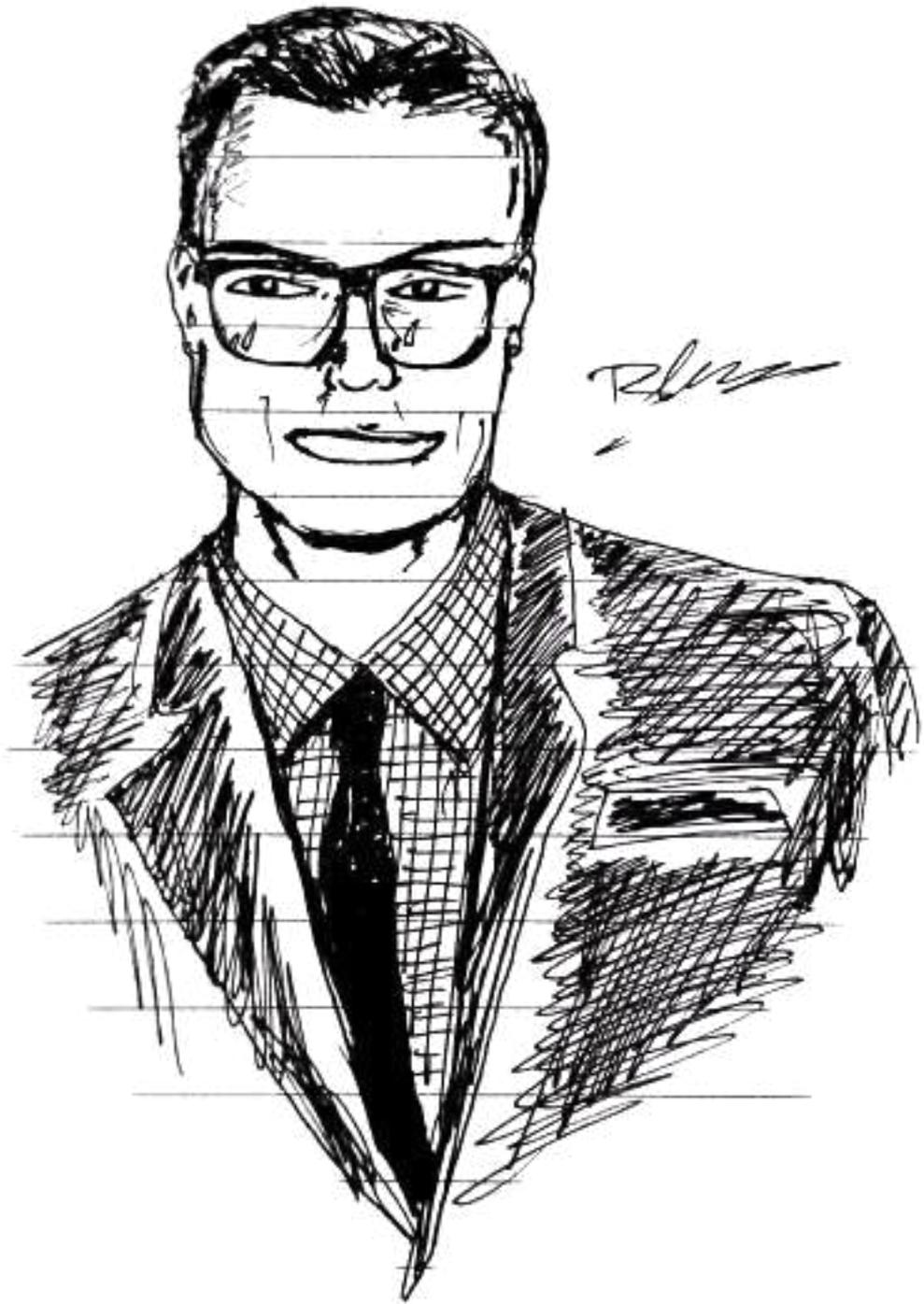

## «Εκ βαθέων»

Ο πόνος, τόσο σαν κόπος όσο και σαν θρήνος-λύπη-στεναγμός, είναι δυστυχώς η απαραίτητη οδός. Οδός προς την ανακούφιση και τη χαρά! Δυστυχώς η μέρα έρχεται μετά τη νύχτα, η χαρά μετά τον πόνο και η Ανάσταση μετά το θάνατο. Πώς να δεις λευκή γραμμή σε λευκό τοίχο;

Πονά η απώλεια πιο πολύ από την έλλειψη. Πονά η βεβαιότητα περισσότερο από την αμφιβολία. Πονά η αδιαφορία περισσότερο από την απόρριψη. Πονά η τύψη περισσότερο από την αδικία.

Πόνο θα βρεις παντού! Διάλεξε οποιοδήποτε μονοπάτι! Πέρνα χωρίς φόβο. Μετά τον πόνο υπάρχει η ανακούφιση και η χαρά!

Να πονέσεις όμως! Μη φοβηθείς! Ο φοβισμένος, ηττημένος. Ο πονεμένος, ένδοξος νικητής!

Ισορροπώ πάνω στη δοκό του παρόντος. Βλέπω γύρω μου μόνο το χάος του παρελθόντος και του μέλλοντος και μπροστά μου το τρέχον μονοπάτι παρόντος, ένα από τα άπειρα που θα μπορούσαν να ακολουθηθούν. Δεν μπορώ να θαυμάσω τον εαυτό μου από μακριά, από κάποια κερκίδα θεατή, δεν τολμώ καν να γυρίσω τα μάτια μου απρόσεκτα μακριά, πίσω και μπροστά... Ανύπαρκτα η αρχή και το τέλος της δοκού, αυτό που μόλις άφησα κι αυτό που σύντομα θα πατήσω... Μόνο ισορροπώ με ένστικτο και εμπιστοσύνη στο Θεό!

Τώρα τα ερωτήματα είναι πολλά. Γιατί γεννήθηκα; Γιατί γέννησα; Γιατί κάποια μέρα θα πεθάνω;

Δεν μπορώ να ζωγραφίσω τη δοκό μου. Όπως και δεν μπορώ να δω τον εαυτό μου όταν περπατώ ανάμεσα στα βότσαλα με τη λαχτάρα της θάλασσας στο σώμα και την ψυχή. Ούτε την ώρα που παραπατώ απρόσεχτη, ούτε την ώρα που γελάω επειδή αναπάντεχα κάποιο από τα παιδιά μου με βρέχει, ούτε όταν βυθίζομαι στην αγκαλιά του νερού και νιώθω τη δύναμη γύρω μου και τη δύναμη μέσα μου, την υδροστατική πίεση, την ατμοσφαιρική πίεση, τη δυναμική ενέργεια, τις δυνάμεις προώθησης, το κύμα, το ηλιοβασίλεμα, το φεγγάρι, τον άνεμο, το μικρό πετραδάκι που προκαλεί ομόκεντρους κύκλους πριν βυθιστεί σε άγνωστο πυθμένα... Δεν μπορώ τίποτε ουσιαστικό να δω, να φωτογραφίσω από τη δοκό ή από το πάτημα σε αυτή. Μπορώ μόνο να ζωγραφίσω το φανταστικό εαυτό μου ακροβάτη των αναμνήσεων και της ελπίδας, «αιθεροβάτη», όπως θα έλεγε κάποιος αγαπημένος μου ζωγράφος...

Όλα είναι ερωτηματικό, θαυμαστικό, αποσιωπητικά... Δεν υπάρχει κάτι σαν τελεία στη δοκό. Πιο πολύ καμαρώνει το ερωτηματικό...

Κι αν όλα τα απομνημονεύματα συναντηθούν σε μια μεγάλη δεξαμενή γνώσης κι αναλυθούν από όλους, μόνο ζωγραφιές ή προτάσεις με ερωτηματικά μπορούν να προκύψουν. Κι αν αφήναμε τον κόσμο μας να ερμηνευτεί από τα παιδιά, θα το βλέπαμε αυτό πεντακάθαρα. Είναι ωραίο να ζεις με την "essence", την εντύπωση, το απόσταγμα, την οσμή και την ώσμωση των γεγονότων... Αυτό είναι κάποτε πικρό σαν δηλητήριο...

...Κι άλλα ταξίδια για την Ιθάκη θα μπορούσα να περιγράψω, είτε ως διαδρομές φόβου κι αγωνίας (εξετάσεις, παραστάσεις, οικονομικά εγχειρήματα...), είτε ως διαδρομές επιθυμίας κι ελπίδας (σχέσεις, θέσεις, γνωριμίες, δώρα, ταξίδια...). Όλα φτιάχνουν ψίθυρους. Χίλιους ψίθυρους. Τριβελίζουν το μυαλό μέχρι κάποια στιγμή-ορόσημο, κάποια στιγμή τέλους ή συνειδητοποίησης... Κάθε τέλος... μια καινούρια αρχή!

Οι ψίθυροι αιωρούνται για λίγο γύρω από το ψυχικό μας περίγραμμα και ίσως φλερτάρουν με περιγράμματα άλλων ανθρώπων ή εποχών... Οι ψίθυροι μένουν στα αυτιά της ψυχής μας και μας συντονίζουν τα βήματα στη δοκό του παρόντος. Ψίθυροι αγωνίας και αγαλλίασης αναμειγνύονται με κάθε βήμα μας... Κι έπειτα, μας συνοδεύουν ακόμη και στα όνειρα. Ίσως και μετά το θάνατο;... Η συνείδηση είναι ένα σύνολο τέτοιων ψιθύρων κι ενώνεται μέσω της δοκού του παρόντος με τις συνειδήσεις των άλλων και με το Θεό!

Λόγια μυστικά φτιάχνουν χίλιους ψίθυρους. Οι ζωές μας είναι κι αυτές ψίθυροι στη βουή του σύμπαντος!

Ωραία...

Ουδέν κακόν αμιγές καλού;

Δύσκολη πράξη. Ωραία θεωρία, όμως. Η διάγνωση είναι αρχή της θεραπείας. Η θεωρία είναι αρχή της πράξης. Η αισιοδοξία είναι καθολική αρχή και κέντρο βάρους πάνω στη δοκό του παρόντος...

...Έτσι και αλλιώς, κάθε πτώση είναι αιώρηση... κάτι σαν πτήση! Ωραία!

Συγκινήθηκα κάθε φορά που ένιωσα την ανθρώπινη προσπάθεια για κάτι καλό. Η επιμέλεια που δεν έχει την ωφελιμιστική ανυπομονησία του χαρισμένου και του εύκολου με συγκλονίζει! Όταν συνδυάζεται με το χαρισμένο και το εύκολο μάλιστα, ακόμη περισσότερο. Αυτοπεριορισμός από επιλογή και όχι από τύχη ή φόβο. Πρόγραμμα, οργάνωση, υπομονή και σεβασμός είναι τα όπλα του πολέμου ενάντια στον κακό μας εαυτό. «Ανάσα έχω να δώσω σχήμα και μορφή στο γυαλί μου», λέει κάποιο τραγούδι. Μια πολιτεία του Πλάτωνα αλλά και του Αριστοτέλη και του Πυθαγόρα και του Νεύτωνα και του Αϊνστάιν και του Παστέρ, που είπε όταν κατηγορήθηκε ως τυχερός κι ευνοημένος στις ιδανικές θερμοκρασίες του Παρισιού: «Η τύχη στο χώρο των παρατηρήσεων ευνοεί τα προετοιμασμένα μυαλά». Προετοιμασμένα μυαλά θεωρώ αυτά που πιστεύουν κι ελπίζουν, αγαπούν και αναγνωρίζουν, επικοινωνούν και αγωνίζονται μέσα στο κοινωνικό σύνολο, με την πλήρη συναίσθηση ότι αποτελούν κομματάκι του! «Service to the society is the rent we pay to live on our planet» ήταν μια επιγραφή που συνάντησα σε νοσοκομείο της Βοστώνης! Δεν την ξεχνώ!

Κάθε παιδί που υπομονετικά περιμένει μια διαφωτιστική ανάδραση από το διαβήτη και το γνώμονά του, από ένα λεξικό, από μια ερώτηση σε κάποιον μεγαλύτερο και σοφότερο, κάθε δημιουργός που αφουγκράζεται το ακροατήριό του, κάθε μητέρα που αδιαφορεί για την προσωπική της εικόνα και τύχη και ακόμα για την απόσβεση των προσπαθειών της όταν μεγαλώνει παιδιά, κάθε μικρός «Αρχιμήδης» που αισθάνεται την ολιστική χαρά του «εύρηκα» που έγκειται στην εύρεση και όχι στο α΄ ενικό, δίνει ελπίδα στον περίεργο κόσμο μας.

Μάλλον άρχισα ήδη να γερνάω... Κι αυτό μου δίνει μεγάλη χαρά! Νιώθω σιγουριά. Περνάω στη σφαίρα του «μεγάλου» όχι προφανώς για να κάνω ό,τι θέλω, αλλά για να αισθάνομαι ευγνωμοσύνη για καθετί που μπορώ ακόμη να κάνω, να δίνω αξία στο καθετί και να αφαιρώ αναξιότητα από όλα. Χαίρομαι που γερνάω... Είδα τη χαρά στην καμπάνα του Εσπερινού και στο πρωινό τάισμα μιας οικογένειας από πάπιες. Περίφημοι γονείς αυτοί ξέρουν να ταΐζουν από στόμα σε στόμα. Χρυσή και ασημένια γύρω τους η λίμνη... η σωτήρια πρωινή όψη της λίμνης των Ιωαννίνων...

Τι είναι ήθος;

Ήθος είναι αυτό που σκέφτεσαι να ταυτίζεται με αυτό που λες κι αυτό που κάνεις. Ήθος είναι να αναγνωρίζεις τις αδυναμίες σου. Ήθος είναι να αποφεύγεις την εκδούλευση. Ήθος είναι η αλήθεια και κάποτε το «κατά συνθήκη αγάπης» ψεύδος. Ήθος είναι να εξομολογείσαι αντί να κατακρίνεις. Να μιμείσαι το παράδειγμα Αυτού που αγάπησε, ανέχτηκε, πόνεσε, υπέμεινε, συγχώρεσε. Ήθος είναι να μην αποφεύγεις τις συγκυρίες που πληγώνουν τον εγωισμό, αλλά να βελτιώνεις τον εγωισμό προς την κατεύθυνση του απρόσβλητου. Ήθος είναι να χαίρεσαι με την επιτυχία και τη χαρά του άλλου. Ήθος είναι να αναγνωρίζεις τα λάθη και τα πάθη σου. Να ελέγχεις με αυστηρότητα τον εαυτό σου και με επιείκεια τους άλλους. Να μη συνεχίζεις βασανιζόμενος από τύψεις, αλλά να εξομολογείσαι στο Θεό και να προσεύχεσαι. Να μην εθελοτυφλείς σε πράγματα και καταστάσεις και να μη «βαφτίζεις» τα πράγματα αλλιώς. Να έχεις παρρησία και ειλικρίνεια. Να προσπαθείς κατά το μέγιστο, αλλά να συμβιβάζεσαι ευτυχής με το ελάχιστο. Να περνάς με ταπεινότητα και όχι με ψευτοσεμνοτυφία το «μεθεόρτιο» χρόνο. Να οργανώνεις την επόμενη «εορτή» με δουλειά, χαρά, υπομονή, συνεργασία... Να μαθαίνεις συνέχεια και καθημερινά από όλους. Να χαίρεσαι όταν κάποιοι χαίρονται, ελπίζουν, ανακουφίζονται... Να μοιράζεσαι. Να ενημερώνεις και να εκφράζεσαι, γιατί η απόκρυψη της αλήθειας είναι ψεύδος... Να μη λυπάσαι που η ελιά σου δεν κάνει μήλα... Κι αν θέλεις μήλα, φύτεψε μηλιά. Κι αν δεν προκάνεις μήλα, να χαίρεσαι πάλι και να θυμάσαι ότι και η ελιά σού δόθηκε από κάπου (δανείζομαι εδώ και παραφράζω το όμορφο ποίημα ενός καθηγητή μου)... Να συγχωρείς, να παραβλέπεις, να αγαπάς. Και να αγαπάς με τη δύσκολη αγάπη. Εύκολο είναι, για παράδειγμα, να αγαπάς τον εαυτό σου και τα παιδιά σου. Δύσκολο είναι να αγαπάς τους μακρινότερους («πλησίον» παρά ταύτα) ανθρώπους. Δύσκολο ακόμη να αγαπάς τον «αποτυχημένο» εαυτό σου... Πρέπει να μετουσιώνεις την αδιαφορία και την καχυποψία σε θετική ουδετερότητα, το παράπονο σε προσευχή, το θυμό σε δύναμη δράσης, τις ανήθικες και πτωτικές σκέψεις και διαθέσεις σε έργα τέχνης, πολιτισμού και φιλανθρωπίας... Να σμιλεύεις την αγριάδα της φύσης σου... Να μην κολακεύεις και να μην κολακεύεσαι. Να σκέφτεσαι και το κοινό καλό πού και πού... Να είσαι κομμάτι του σύμπαντος και όχι σύμπαν του κομματιού σου. Και να αγαπάς τον εαυτό σου σαν κομμάτι του σύμπαντος: «Service to the society is the rent we pay to live on our planet»!...

- Μακάριοι οἱ πτωχοὶ τῷ πνεύματι, ὅτι αὐτῶν ἐστιν ἡ βασιλεία τῶν οὐρανῶν.
- Μακάριοι οἱ πενθοῦντες, ὅτι αὐτοὶ παρακληθήσονται.
- Μακάριοι οἱ πραεῖς, ὅτι αὐτοὶ κληρονομήσουσιν τὴν γῆν.
- Μακάριοι οἱ πεινῶντες καὶ διψῶντες τὴν δικαιοσύνην, ὅτι αὐτοὶ χορτασθήσονται.
- Μακάριοι οἱ ἐλεήμονες, ὅτι αὐτοὶ ἐλεηθήσονται.
- Μακάριοι οἱ καθαροὶ τῇ καρδίᾳ, ὅτι αὐτοὶ τὸν Θεὸν ὄψονται.
- Μακάριοι οἱ εἰρηνοποιοί, ὅτι αὐτοὶ υἱοὶ Θεοῦ κληθήσονται.
- Μακάριοι οἱ δεδιωγμένοι ἕνεκεν δικαιοσύνης, ὅτι αὐτῶν ἐστιν ἡ βασιλεία τῶν οὐρανῶν.
- Μακάριοί ἐστε ὅταν ὀνειδίσωσιν ὑμᾶς καὶ διώξωσιν καὶ εἴπωσιν πᾶν πονηρὸν καθ' ὑμῶν ψευδόμενοι ἕνεκεν ἐμοῦ.

Δεν ξέρω τι επιφυλάσσει το μέλλον, αλλά οι συμπτώσεις είναι υλικές ευκαιρίες του άυλου! Σημάδια και σύμβολα! Αφορμές και ορόσημα! Θα 'ταν αστείο να σκεφτούμε ότι ο Θεός ξεφυλλίζει το ημερολόγιό μας, παρακολουθεί τα ρολόγια μας, ρυθμίζει τις απρόοπτες συναντήσεις μας, τα ευτυχήματα και τα ατυχήματά μας. Ο χρονισμός τους, όμως, είναι η ευκαιρία που η καθημερινότητα δίνει σαν τροφή στο μυαλό και την ψυχή μας, για να δουν τη δόξα του Θεού και να ακούσουν τραγούδια υπέροχου φωτός!

Φεγγαράκι μου λαμπρό, χάρτινο φεγγαράκι μου κι αστέρι του βοριά, φως ιλαρό, φως εσπερινό... φέγγε μου να περπατώ, να μαθαίνω γράμματα, γράμματα σπουδάματα, του Θεού τα πράματα...

Ξαναπέρασα από το άγαλμα του Μελαδάκη. Πρόσεξα ότι ήταν Μιχαήλ του Ιωάννου, ότι ήταν οπλαρχηγός κι ότι έζησε από το 1889 έως το 1914. «Στου Μίνωα τη γη, ξένε, σαν πας και το τραγούδι ακούσεις του ριζίτη, πες τους στην Κορυτσά απόμεινα, φρουρός στην Ήπειρο, τιμή στην Κρήτη»! Το άγαλμα αυτό βρίσκεται στην πλατεία με το όνομα εκείνου που έγραψε το τελευταίο του ποίημα με το αληθινό αίμα της καρδιάς του, του Λορέντζου Μαβίλη! Δίπλα στην οικογένεια από πάπιες! Αυτές που ταΐζουν από στόμα σε στόμα! Και είναι σταθμός στον αγαπημένο μου περίπατο, αυτόν που μου δίνει κουράγιο κι ανταμοιβή για την κούραση της μέρας.

Κοιτάζω πίσω μου τη γραμμή του χρόνου. ...Εμένα πάλι μου θυμίζει πεντάγραμμο. Και ξέρω και ποιες νότες θα ήθελα να σβήσω. Δεν μπορώ να σβήσω. Μπορώ μόνο να γράφω νέες νότες στο πεντάγραμμο της ζωής που απομένει... ή ακόμη καλύτερα, στην παρασημαντική, χαρούμενους ψαλμούς ωραιότητας! Να σκέφτομαι την «ημέρα» (όπως λέει ο παπα-Θανάσης), τη «δοκό του παρόντος» (όπως λέω εγώ) και όταν αιωρηθώ στους «ασύμμετρους ζυγούς του μέλλοντος», να ακούω τη μουσική αυτή, που θα 'ναι γραμμένη με μελάνι από καλά «εφαλτήρια του παρελθόντος»...

...Για φαντάσου! Με τέτοιο μελάνι μόλις τελείωσα κάτι που δεν θυμάμαι να άρχισα...

# Περιεχόμενα

# Κατάλογος σχημάτων

**Σχήματα στο αγγλικό κείμενο**

# Κεφάλαιο 1

# Εισαγωγή

## 1.1 Γιατί AI-agents βασισμένοι σε LLMs;

Τα Μεγάλα Γλωσσικά Μοντέλα (LLMs) έχουν επιφέρει επανάσταση σε διάφορες πτυχές της μηχανικής και της επιστήμης και θεωρούνται ορόσημο στον τομέα της Τεχνητής Νοημοσύνης (AI) και της Επεξεργασίας Φυσικής Γλώσσας (NLP). Ξεκινώντας με απλούστερα μοντέλα ικανά να κατανοούν και να παράγουν κείμενο βάσει στατιστικών μεθόδων, αυτός ο τομέας ασχολείται με ολοένα και πιο εξελιγμένες αρχιτεκτονικές. Αυτές οι εξελίξεις έχουν οδηγήσει στην ανάπτυξη μοντέλων όπως τα GPT-3, GPT-4 [7], BERT [8] και LLaMA [9], τα οποία όχι μόνο κατανοούν λεπτομερώς την ανθρώπινη γλώσσα, αλλά παράγουν και συνεκτικό, ενσυνείδητο και εξειδικευμένο κείμενο. Τα πλεονεκτήματα αυτών των μοντέλων έγκεινται στη βαθιά κατανόηση των γλωσσικών παραδειγμάτων, την ικανότητά τους να παράγουν κείμενο που μοιάζει με ανθρώπινο και την εφαρμογή τους σε διάφορες εργασίες, από μετάφραση έως δημιουργία περιεχομένου.

Ωστόσο, παρά τις εντυπωσιακές τους δυνατότητες, τα LLMs αντιμετωπίζουν σημαντικούς περιορισμούς. Η γνώση τους είναι "παγωμένη" στο χρόνο εκπαίδευσης, καθιστώντας τα ανίκανα να προσπελάσουν ή να επεξεργαστούν πληροφορίες σε πραγματικό χρόνο ή να αλληλεπιδράσουν δυναμικά με το περιβάλλον. Επιπλέον, συχνά παράγουν απαντήσεις βάσει προτύπων στα δεδομένα εκπαίδευσής τους, γεγονός που μπορεί να οδηγήσει σε "ψευδαισθήσεις", ανακρίβειες ή αντιφάσεις με τα πραγματικά δεδομένα. Αρχικά, οι ερευνητές εστίασαν στην εκπαίδευση για προχωρημένες δεξιότητες επεξεργασίας γλώσσας, συχνά αγνοώντας τον κρίσιμο τομέα της χρήσης εργαλείων, οδηγώντας σε μοντέλα που δεν αναπαράγουν τον ανθρώπινο τρόπο μάθησης όταν δοκιμάζονται σε ανοιχτά περιβάλλοντα [10, 11, 12, 13]. Πρόσφατες μελέτες τείνουν να επιτυγχάνουν "πιο ανθρώπινα" αποτελέσματα [14, 7, 15, 16, 17, 18, 19, 9], καθώς οι ερευνητές ενισχύουν τα LLMs με δυνατότητες μνήμης, σχεδιασμού και συλλογιστικής. Μηχανισμοί που μπορούν να γεφυρώσουν το χάσμα μεταξύ στατικής γνώσης και δυναμικής, πραγματικής πληροφόρησης και επεξεργασίας είναι ιδιαίτερου ενδιαφέροντος. Η πιο δημοφιλής προσέγγιση προς αυτή την κατεύθυνση περιλαμβάνει την ενίσχυση των LLMs με τη δυνατότητα αλληλεπίδρασης με εξωτερικά εργαλεία και εφαρμογές. Η διασύνδεση των LLMs με Διεπαφές Προγραμματισμού Εφαρμογών (APIs) γεννά τον τομέα των AI-agents βασισμένων σε LLMs, έναν τομέα αυξανόμενης εξειδίκευσης και εκθετικής ανάπτυξης (Σχήμα 1.1). Αξίζει να σημειωθεί ότι (αισιόδοξοι) ερευνητές αναμένουν ότι η Γενική Τεχνητή Νοημοσύνη (AGI) θα επιτευχθεί μέσω αυτοκατευθυνόμενου σχεδιασμού και συμπεριφορών και οι αυτόνομοι AI-agents φαίνεται να είναι η μέθοδος για να φτάσουμε εκεί [19, 14].

## 1.2 LLMs & APIs

Τα APIs λειτουργούν ως γέφυρα που επεκτείνει τις δυνατότητες των LLMs πέρα από τους εγγενείς περιορισμούς τους, επιτρέποντας την αλληλεπίδραση με εξωτερικά συστήματα, βάσεις δεδομένων και υπηρεσίες, επιτρέποντας στα LLMs να αντλούν δεδομένα σε πραγματικό χρόνο, να εκτελούν υπολογισμούς και ακόμα και να εκτελούν συναλλαγές. Αυτή η ενσωμάτωση ανοίγει νέες δυνατότητες για τα LLMs, μετατρέποντάς τα από στατικά μοντέλα σε δυναμικούς AI-agents ικανούς για πιο σύνθετες και ενσυνείδητες λειτουργίες [20, 17, 5]. Σημεία αναφοράς όπως το API-bank έχουν δημιουργηθεί για να αξιολογήσουν τον αυξανόμενο αριθμό νέων μοντέλων. Η συνέργεια των LLMs και των APIs σηματοδοτεί μια κρίσιμη αλλαγή προς τη δημιουργία πιο ευφυών, προσαρμόσιμων και ευέλικτων συστημάτων AI. Η πορεία της ανάπτυξης των LLMs έχει χαρακτηριστεί από σημαντικές ανακαλύψεις και η εξέλιξή τους περιλαμβάνει καινοτομίες όπως η διασύνδεση με APIs, θέτοντας το υπόβαθρο για το επόμενο άλμα στις δυνατότητες της Τεχνητής Νοημοσύνης.

## 1.3 Σχετική Εργασία & Έρευνα

Οι τελευταίες εξελίξεις στα LLMs και την AI περιλαμβάνουν Αυτόνομους Μηχανικούς Λογισμικού AI που μπορούν να κατανοήσουν οδηγίες υψηλού επιπέδου από ανθρώπους, να τις διασπάσουν σε βήματα, να ερευνήσουν σχετικές πληροφορίες και να γράψουν κώδικα για να επιτύχουν έναν συγκεκριμένο στόχο, όπως το OpenDevin [21] και το Devika [22]. Επίσης, fora σχετικά με LLMs, όπως η κοινότητα Hugging Face, παρέχουν μια μεγάλη ποικιλία εξειδικευμένων προεκπαιδευμένων μοντέλων προσανατολισμένων σε συγκεκριμένες λειτουργίες [23]. Με την ταχεία ανάπτυξη αυτού του ερευνητικού τομέα, έχουν προκύψει πολλαπλές εκτεταμένες έρευνες, δίνοντας βαθιές γνώσεις σε ένα ευρύ φάσμα θεμάτων. Οι [24] και [25] παρέχουν μια συνολική εισαγωγή στο υπόβαθρο, σημαντικές ανακαλύψεις και τις κύριες τεχνολογίες. Από την άλλη πλευρά, η [26] εστιάζει ιδιαίτερα στις εφαρμογές των LLMs σε διάφορες συνήθεις διαδικασίες, καθώς και στα εμπόδια που σχετίζονται με την υιοθέτησή τους. Η ευθυγράμμιση των LLMs με την ανθρώπινη νοημοσύνη είναι ένας αυξανόμενος τομέας μελέτης που στοχεύει να ξεπεράσει ζητήματα όπως οι προκαταλήψεις και οι ψευδαισθήσεις [27]. Η συλλογιστική είναι ένα άλλο σημαντικό μέρος της νοημοσύνης που επηρεάζει τη λήψη αποφάσεων, την επίλυση προβλημάτων και τη γνωστική ικανότητα [28]. Σύμφωνα με το [29], τα Ενισχυμένα Γλωσσικά Μοντέλα (ALMs) μπορούν να βελτιώσουν τα γλωσσικά μοντέλα μέσω της συμπερίληψης δυνατοτήτων συλλογιστικής και εργαλείων. Η αξιολόγηση της απόδοσης των μοντέλων μεγάλης κλίμακας γίνεται πιο σημαντική καθώς η χρησιμότητά τους αυξάνεται και υπάρχουν διάφοροι δείκτες που χρησιμοποιούνται για το σκοπό αυτό [30]. Αυτή η εργασία εστιάζει στην ενίσχυση των δυνατοτήτων των LLMs μέσω της διασύνδεσης με APIs.

## 1.4 Δομή της Διατριβής

Η διατριβή είναι δομημένη ως εξής: Το Κεφάλαιο 1 καθορίζει το υπόβαθρο συζητώντας τη σημασία των AI-agents βασισμένων σε LLMs και την αλληλεπίδρασή τους με APIs. Το Κεφάλαιο 2 προτείνει μια μεθοδολογία 7 βημάτων που χρησιμοποιείται για την ενίσχυση των LLMs με APIs. Ξεκινά με την ''Επιλογή Μοντέλου,'' που αναλύει τα κριτήρια για την επιλογή κάποιου LLM που επιτρέπει τη διασύνδεση με APIs βάσει απόδοσης, διαλειτουργικότητας και προσαρμοστικότητας. Ακολουθεί η ''Ενίσχυση με Γνώση Εξωτερικών Εργαλείων,'' όπου συζητείται η εκπαίδευση των LLMs για να εκτελούν αποτελεσματικά εντολές σχετικές με API. Ο ''Σχεδιασμός Πολυσταδιακής Αρχιτεκτονικής'' περιγράφει μια δομημένη προσέγγιση για την ενσωμάτωση των APIs μέσω σταδίων όπως η αναγνώριση εργασιών και η εύρεση API. Ο ''Μηχανισμός Επιλογής API'' περιγράφει τα κριτήρια και τους μηχανισμούς για την επιλογή κατάλληλων APIs για συγκεκριμένες εργασίες. Η ''Δημιουργία Κλήσεων API'' συζητά τεχνικές για να διασφαλιστεί ότι οι κλήσεις API είναι συντακτικά σωστές και εννοιολογικά ευθυγραμμισμένες με τις προθέσεις των χρηστών. Η ''Διάσπαση Εργασιών'' καλύπτει στρατηγικές για τη διάσπαση σύνθετων εργασιών σε απλούστερες υπο-εργασίες και η ''Διαρκής Βελτίωση και Ανατροφοδότηση Χρηστών'' αφορά τη βελτίωση των αλληλεπιδράσεων με τα API μέσω συνεχόμενης ανατροφοδότησης από τους χρήστες. Το Κεφάλαιο 3 εμβαθύνει στην πρακτική εφαρμογή των μεθοδολογιών που συζητήθηκαν προηγουμένως, εισάγοντας μια καινοτόμα αρχιτεκτονική για την ενσωμάτωση των LLMs σε φορητές συσκευές. Αυτό το κεφάλαιο συζητά τις τεχνικές και τις επιχειρησιακές πτυχές της υλοποίησης της αρχιτεκτονικής αυτής, δίνοντας έμφαση στις απαιτήσεις αποθήκευσης, RAM και CPU που είναι απαραίτητες για την αποδοτική λειτουργία. Εξερευνά επίσης την ενίσχυση των AI-agents με τοπικές προσαρμοσμένες βάσεις δεδομένων, παρέχοντας μια λεπτομερή εξέταση των συνιστωσών και της λειτουργικής ροής της προτεινόμενης αρχιτεκτονικής.

Με την αντιμετώπιση των προκλήσεων μέσω στρατηγικών μετριασμού, αυτό το κεφάλαιο παρουσιάζει τα πλεονεκτήματα της ενσωμάτωσης των LLMs σε φορητές συσκευές, όπως οι βελτιωμένοι χρόνοι απόκρισης και η μειωμένη εξάρτηση από υπηρεσίες νέφους (cloud services). Στο Κεφάλαιο 4, εξετάζονται πρακτικές εφαρμογές και περιπτώσεις χρήσης για να καταδείξουν πώς αυτές οι αρχιτεκτονικές διευκολύνουν την ανάπτυξη των AI-agents στον πραγματικό κόσμο, ανοίγοντας το δρόμο για πιο ανθεκτικές και ενσυνείδητες εφαρμογές. Τέλος, το Κεφάλαιο 5, "Συμπέρασμα, Συζήτηση & Μελλοντική Έρευνα," συνοψίζει τα ευρήματα, συζητά τις επιπτώσεις και προτείνει μελλοντικές κατευθύνσεις έρευνας. Μαζί, αυτά τα κεφάλαια καλύπτουν ολοκληρωμένα το νέο θέμα της ενίσχυσης των LLMs με δυνατότητες αλληλεπίδρασης με APIs.

**Κεφάλαιο 2**

# Μεθοδολογία Ενσωμάτωσης Προγραμματιστικών Διεπαφών Εφαρμογών στα Μεγάλα Γλωσσικά Μοντέλα

## 2.1 ΒΗΜΑ 1: Επιλογή Μοντέλου

Η επιλογή Μεγάλων Γλωσσικών Μοντέλων (LLMs) που επιτρέπουν διασύνδεση με APIs περιλαμβάνει την αξιολόγηση πολλών κρίσιμων κριτηρίων για να διασφαλιστεί ότι τα μοντέλα δεν είναι μόνο επαρκή στην επεξεργασία φυσικής γλώσσας, αλλά και προσαρμοστικά και αποτελεσματικά στην αλληλεπίδραση με εξωτερικά APIs. Βασισμένοι στην τρέχουσα έρευνα [18, 4] και στα πρότυπα AI, οι ακόλουθες πτυχές πρέπει να ληφθούν υπόψιν:

**Απόδοση σε Γλωσσικές Εργασίες:**
Η ικανότητα ενός LLM να κατανοεί και να παράγει κείμενο που μοιάζει με ανθρώπινο είναι κρίσιμη. Το μοντέλο πρέπει να παρουσιάζει υψηλή απόδοση σε διάφορες γλωσσικές εργασίες όπως η κατανόηση, η μετάφραση, η περίληψη και η απάντηση σε ερωτήσεις.

**Διαλειτουργικότητα με APIs:**
Το επιλεγμένο LLM θα πρέπει να είναι ικανό να κατανοεί οδηγίες που περιλαμβάνουν εξωτερικά εργαλεία και να μετατρέπει αυτές τις οδηγίες σε κλήσεις API. Αυτό απαιτεί από το μοντέλο να μπορεί να εντοπίσει το σχετικό API και να μορφοποιήσει το αίτημα κατάλληλα.

**Προσαρμοστικότητα και Ικανότητα Μάθησης:**
Το LLM πρέπει να είναι προσαρμοστικό, ικανό να μαθαίνει από τις αλληλεπιδράσεις και να βελτιώνεται με την πάροδο του χρόνου. Η ικανότητα του μοντέλου να ενσωματώνει ανατροφοδότηση από τις αλληλεπιδράσεις API σε μελλοντικές απαντήσεις, βελτιώνοντας την απόδοσή του και την ακρίβειά του, είναι σημαντική.

**Γενίκευση σε Διάφορους Τομείς:**
Το μοντέλο πρέπει να δείχνει ισχυρές ικανότητες γενίκευσης, που σημαίνει ότι μπορεί να εφαρμόσει αποτελεσματικά τις γνώσεις του και τις δεξιότητες ενσωμάτωσης API σε ένα ευρύ φάσμα τομέων και εργασιών, όχι μόνο στα σενάρια στα οποία έχει προεκπαιδευτεί.

**Κλιμάκωση και Αποδοτικότητα:**
Δεδομένου του πιθανού υψηλού όγκου κλήσεων API και της πολυπλοκότητας της επεξεργασίας απαντήσεων, το επιλεγμένο LLM πρέπει να είναι κλιμακούμενο και αποδοτικό στις λειτουργίες του. Αυτό διασφαλίζει ότι η ενσωμάτωση παραμένει πρακτική και βιώσιμη, ακόμη και καθώς επεκτείνεται το εύρος των εργασιών.

**Ηθική:**
Το μοντέλο πρέπει να διαχειρίζεται τα δεδομένα ηθικά, ειδικά όταν αλληλεπιδρά με APIs που μπορεί να έχουν πρόσβαση σε ευαίσθητες ή προσωπικές πληροφορίες. Πρέπει επίσης να αξιολογείται για μεροληψίες, ώστε να μετριάζονται τυχόν αρνητικές επιπτώσεις που θα μπορούσαν να προκύψουν από τις απαντήσεις του.

**Επικαιρότητα:**
Το LLM πρέπει να έχει εκπαιδευτεί σε μια πρόσφατη συλλογή δεδομένων, διασφαλίζοντας ότι η βάση γνώσεών του είναι όσο το δυνατόν πιο ενημερωμένη. Αυτό είναι κρίσιμο για εργασίες που απαιτούν ενημερωμένες πληροφορίες, τις οποίες το μοντέλο μπορεί να χρειάζεται να ανακτήσει μέσω APIs.

**Απαιτήσεις Μνήμης και Αποθήκευσης:**
Το LLM θα πρέπει να έχει διαχειρίσιμες απαιτήσεις μνήμης και αποθήκευσης για να εξασφαλίζεται η συμβατότητα με την υπάρχουσα υποδομή χωρίς να απαιτούνται υπερβολικές αναβαθμίσεις. Το μοντέλο πρέπει να χρησιμοποιεί αποδοτικά τη διαθέσιμη μνήμη συστήματος (RAM), κάτι που είναι κρίσιμο όταν διαχειρίζεται μεγάλα σύνολα δεδομένων ή πολλαπλές ταυτόχρονες κλήσεις API. Αυτό επίσης επηρεάζει την ταχύτητα με την οποία το μοντέλο μπορεί να προσπελάσει και να επεξεργαστεί δεδομένα.

**Χρόνος Απόκρισης και Καθυστέρηση:**
Είναι σημαντικό το LLM να παρουσιάζει χαμηλή καθυστέρηση στην κατανόηση και επεξεργασία κλήσεων API. Οι γρήγοροι χρόνοι απόκρισης είναι απαραίτητοι για τη διατήρηση μιας ομαλής εμπειρίας χρήστη, ειδικά σε εφαρμογές που απαιτούν αλληλεπίδραση σε πραγματικό χρόνο ή γρήγορη επεξεργασία δεδομένων. Αυτό περιλαμβάνει τη βελτιστοποίηση της αρχιτεκτονικής του μοντέλου με σκοπό την ελαχιστοποίηση των καθυστερήσεων.

**Ευελιξία των Prompts και Περιορισμοί:**
Το LLM πρέπει να επιτρέπει στους χρήστες να κάνουν αιτήματα χρησιμοποιώντας τόσο συνοπτικές όσο και λεπτομερείς περιγραφές. Ωστόσο, είναι σημαντικό να καθοριστούν οι περιορισμοί σχετικά με το μήκος των prompts που μπορεί να επεξεργαστεί αποτελεσματικά το μοντέλο χωρίς πτώση στην απόδοση ή την ακρίβεια, καθώς αυτό επηρεάζει τη χρησιμότητά του σε διάφορες εφαρμογές.

**Ενσωμάτωση και Συμβατότητα με Υπάρχοντα Συστήματα:**
Το μοντέλο πρέπει να ενσωματώνεται εύκολα με τις υπάρχουσες αρχιτεκτονικές λογισμικού. Αυτό περιλαμβάνει τη συμβατότητα με διάφορες γλώσσες προγραμματισμού που χρησιμοποιούνται συνήθως για την αλληλεπίδραση με APIs.

**Οικονομική Αποδοτικότητα:**
Λαμβάνοντας υπόψιν τα λειτουργικά κόστη, συμπεριλαμβανομένων των υπολογιστικών πόρων και των πιθανών συνδρομητικών τελών, το επιλεγμένο LLM θα πρέπει να προσφέρει μια ισορροπία μεταξύ απόδοσης και κόστους. Αυτό το κριτήριο είναι κρίσιμο για να διασφαλιστεί ότι το μοντέλο παραμένει μια βιώσιμη λύση καθώς κλιμακώνεται η χρήση του.

**Υποστήριξη και Οικοσύστημα Κοινότητας:**
Μια ενεργή κοινότητα υποστήριξης και η διαθεσιμότητα ολοκληρωμένων οδηγιών χρήσης είναι ανεκτίμητα για την αντιμετώπιση προβλημάτων και τη βελτίωση του μοντέλου. Η πρόσβαση σε ένα ισχυρό οικοσύστημα, όπως αυτό που παρέχει η κοινότητα Hugging Face, μπορεί να διευκολύνει σημαντικά τη διαδικασία ενσωμάτωσης και να παρέχει πόρους για συνεχή βελτίωση και μάθηση. Τα παραπάνω κριτήρια παρέχουν μια ολιστική αξιολόγηση των πιθανών LLMs, λαμβάνοντας υπόψιν όχι μόνο τις "διανοητικές" τους ικανότητες αλλά και πρακτικές παραμέτρους που επηρεάζουν την ανάπτυξη και τη μακροπρόθεσμη χρηστικότητά τους σε σενάρια αλληλεπίδρασης με APIs. Στις έρευνες [18] και [5, 9] παρέχονται παραδείγματα αυτών των κριτηρίων παρουσιάζοντας μοντέλα ειδικά σχεδιασμένα για τη χρήση εργαλείων και APIs. Αυτά παρουσιάζουν ανώτερη απόδοση στην κατανόηση σύνθετων οδηγιών και στην εκτέλεση εργασιών που περιλαμβάνουν εξωτερικά εργαλεία, χρησιμεύοντας έτσι ως σημείο αναφοράς για τις ικανότητες που πρέπει να διαθέτει ένα LLM στο πλαίσιο της διασύνδεσης με APIs.

(Αναφορά στο Σχήμα 2.2.)

Μια πολύ συνηθισμένη προσέγγιση για την επιλογή ενός κατάλληλου LLM είναι η χρήση των πληροφοριών που είναι διαθέσιμες στον πίνακα κατάταξης της κοινότητας Hugging Face [23, 17]. Όπως φαίνεται στο Σχήμα 2.2, υπάρχει μια μεγάλη ποικιλία σημείων αναφοράς για εξειδικευμένες εργασίες, συμπεριλαμβανομένων της περίληψης, της ταξινόμησης, των ενσωματώσεων και της απάντησης σε ερωτήσεις. Η σελίδα περιλαμβάνει μια γραμμή αναζήτησης για την εύρεση συγκεκριμένων μοντέλων ή αδειών. Κάτω από αυτήν, υπάρχει μια ενότητα που επιτρέπει στους χρήστες να επιλέξουν ποια δεδομένα να εμφανίζονται, όπως η μέση απόδοση ή τα αποτελέσματα από συγκεκριμένα σημεία αναφοράς όπως το ARC, το Hellaswag, το MMLU και άλλα. Διαθέτει επίσης φίλτρα για την απόκρυψη ορισμένων τύπων μοντέλων, συμπεριλαμβανομένων ιδιωτικών, διαγραμμένων, συγχωνευμένων ή σημασμένων εγγραφών. Για περαιτέρω προσαρμογή, η πλατφόρμα προσφέρει φίλτρα ανά τύπο μοντέλου, όπως προεκπαιδευμένα ή συνεχώς προεκπαιδευόμενα μοντέλα, αυτά που είναι προσαρμοσμένα σε σύνολα δεδομένων συγκεκριμένων τομέων και άλλες κατηγορίες. Τα φίλτρα ακρίβειας επιτρέπουν την επιλογή βάσει της υπολογιστικής ακρίβειας των μοντέλων (π.χ. float16, bfloat16) και υπάρχουν επιλογές για φιλτράρισμα μοντέλων κατά μέγεθος, κατηγοριοποιημένα με βάση τον αριθμό των παραμέτρων. Αυτός ο πίνακας κατάταξης χρησιμεύει ως μια ολοκληρωμένη πηγή για τη σύγκριση και ανάλυση των δυνατοτήτων και επιδόσεων διάφορων μεγάλων γλωσσικών μοντέλων, εξυπηρετώντας ερευνητές και προγραμματιστές που ασχολούνται με Τεχνητή Νοημοσύνη. Μια χρήσιμη διαδικασία πριν από την επιλογή του επιθυμητού μοντέλου είναι η δημιουργία ενός εμπειρικού συνοπτικού πίνακα σύγκρισης. Αυτός ο πίνακας θα πρέπει να αποτελείται από ορισμένα υποψήφια μοντέλα που αξιολογούνται σε ορισμένες χρήσιμες λειτουργίες. Με αυτόν τον τρόπο, δημιουργείται μια γενική επισκόπηση των διαθέσιμων επιλογών. Για παράδειγμα, ένας τέτοιος πίνακας παρουσιάζεται στο Σχήμα 2.3.

Για να συμπληρωθεί αυτός ο πίνακας με ακριβή δεδομένα, θα πρέπει να ληφθούν υπόψιν τα ακόλουθα:

1. Εντοπισμός των εργασιών που είναι πιο σχετικές με το συγκεκριμένο έργο.
2. Συλλογή ή αναφορά εμπειρικών δεδομένων από τη βιβλιογραφία σχετικά με το πώς αποδίδει κάθε μοντέλο σε αυτές τις εργασίες.
3. Ενημέρωση του πίνακα ανάλογα.

(Αναφορά στο Σχήμα 2.3.)

## 2.2 ΒΗΜΑ 2: Ενίσχυση με Εξωτερική Γνώση Εργαλείων

Αυτό το στάδιο περιγράφει τη διαδικασία εκπαίδευσης των LLMs ώστε να καθίσταται εφικτή η αλληλεπίδραση με εξωτερικά APIs. Η εκπαίδευση περιλαμβάνει την κατανόηση της δομής και της λειτουργίας των APIs, καθώς και την ικανότητα του μοντέλου να εκτελεί κλήσεις API με συνέπεια και ακρίβεια. Η διαδικασία αυτή βοηθά τα μοντέλα να "μάθουν" πώς να αντιδρούν και να αλληλεπιδρούν με το εξωτερικό περιβάλλον μέσω των APIs, επιτρέποντάς τους να εκτελούν πιο σύνθετες εργασίες. Η ενίσχυση των LLMs με εξωτερική γνώση εργαλείων είναι το πιο κρίσιμο βήμα προς τη δημιουργία πιο ευέλικτων και πρακτικών συστημάτων AI. Αυτή η διαδικασία περιλαμβάνει την εκπαίδευση ή την προσαρμογή των LLMs για να ερμηνεύουν και να εκτελούν εντολές που περιλαμβάνουν αλληλεπίδραση με εξωτερικά APIs. Αν υπάρχουν δεδομένα πραγματικών χρηστών για το API, τότε αυτά χρησιμοποιούνται για την προσαρμογή ενός LLM. Αυτό, όμως, δεν ισχύει πάντα. Τα APIs που δεν είναι ευρέως χρησιμοποιούμενα ή που έχουν κυκλοφορήσει πρόσφατα δεν έχουν αρκετά δεδομένα πραγματικών χρηστών που να σχετίζονται με αυτά. Αυτό το πρόβλημα λύνεται με τη μεθοδολογία που προτείνεται στο [5]. Πιο συγκεκριμένα, ένα ισχυρό LLM (όπως το ChatGPT4 [7]) χρησιμοποιείται για τη δημιουργία καλής ποιότητας δεδομένων για τα συγκεκριμένα APIs, ακολουθώντας μια προσεκτική διαδικασία πολλαπλών σταδίων. Αυτά τα δεδομένα καλής ποιότητας χρησιμοποιούνται στη συνέχεια για την εκπαίδευση ενός δημοσίως διαθέσιμου μοντέλου. Το τελικό μοντέλο είναι πολύ καλά εκπαιδευμένο στο πλαίσιο του API. Η παραπάνω μέθοδος (όπως φαίνεται και στο Σχήμα 2.4) θυμίζει το κοινό ρητό της επιστήμης δεδομένων:

*"Η ποιότητα του μοντέλου εξαρτάται από την ποιότητα των δεδομένων εκπαίδευσης."*
(Αναφορά στο Σχήμα 2.4.)

Οι πιο παραδοσιακές μέθοδοι περιλαμβάνουν:

**Εκπαίδευση με Τεκμηρίωση και Παραδείγματα:**
Μία από τις κύριες μεθόδους για την ενσωμάτωση της γνώσης των APIs στα LLMs περιλαμβάνει τη χρήση τεκμηρίωσης (οδηγιών χρήσης) των APIs και παραδειγμάτων κλήσεων API ως μέρος των δεδομένων εκπαίδευσης. Αυτό επιτρέπει στο μοντέλο να μάθει τη δομή, τη σύνταξη και τα πρότυπα χρήσης διαφόρων APIs. Κατανοώντας πώς διατυπώνονται τα αιτήματα στα APIs και ποιες απαντήσεις παράγουν, τα LLMs μπορούν να προβλέψουν καλύτερα τις απαραίτητες ενέργειες που πρέπει να εκτελεστούν όταν ανατίθεται μια λειτουργία που απαιτεί αλληλεπίδραση με API.

**Προσαρμογή σε Σενάρια Συγκεκριμένων Εργασιών:**
Πέρα από τη βασική εκπαίδευση, η προσαρμογή των LLMs για σενάρια συγκεκριμένων εργασιών που περιλαμβάνουν πρόσβαση σε API μπορεί να βελτιώσει δραστικά την ικανότητά τους να επικοινωνούν με εξωτερικά εργαλεία. Αυτό απαιτεί την κατασκευή συνόλων δεδομένων που μιμούνται πραγματικές δράσεις που περιλαμβάνουν κλήσεις API και στη συνέχεια τη χρήση τους για την προσαρμογή του μοντέλου. Αυτή η εξατομικευμένη προσέγγιση διδάσκει στο μοντέλο τις λεπτομέρειες του πότε και πώς να αξιοποιεί τα APIs για την επίτευξη συγκεκριμένων στόχων.

**Περιβάλλοντα Προσομοίωσης Αλληλεπίδρασης με APIs:**

Μια άλλη στρατηγική είναι η χρήση περιβαλλόντων προσομοίωσης όπου τα LLMs μπορούν να αλληλεπιδρούν με ψευδο-APIs. Αυτά τα APIs προσομοίωσης μπορούν να παρέχουν δομημένες απαντήσεις στα αιτήματα του μοντέλου, επιτρέποντάς του να εξασκείται στις αλληλεπιδράσεις με APIs σε ελεγχόμενο περιβάλλον. Αυτή η μέθοδος βοηθά στην ενίσχυση της κατανόησης του μοντέλου χωρίς την ανάγκη για πραγματικές κλήσεις API, οι οποίες μπορεί να έχουν περιορισμούς ρυθμού κλήσεων ή αυξημένο κόστος.

**Ενσωμάτωση Σημασιολογικής Κατανόησης των APIs:**

Ένας από τους πιο σημαντικούς παράγοντες της αλληλεπίδρασης LLM-API είναι η σημασιολογική σχέση μεταξύ φυσικών γλωσσικών οδηγιών και λειτουργικότητας των APIs. Αυτό περιλαμβάνει τη χαρτογράφηση των προθέσεων του χρήστη σε συγκεκριμένες ενέργειες API, μια διαδικασία που μπορεί να διευκολυνθεί μέσω τεχνικών εποπτευόμενης μάθησης όπου το μοντέλο εκπαιδεύεται σε ζεύγη φυσικών γλωσσικών οδηγιών και των αντίστοιχων κλήσεων API.

**Αξιοποίηση της Μεταφοράς Μάθησης:**

Τεχνικές μεταφοράς μάθησης μπορούν να χρησιμοποιηθούν για τη μεταφορά γνώσης από προεκπαιδευμένα μοντέλα που είναι ήδη ικανά στις αλληλεπιδράσεις API σε νέα μοντέλα. Αυτή η προσέγγιση μπορεί να μειώσει την ποσότητα των δεδομένων που απαιτούνται για την εκπαίδευση και να επιταχύνει τη διαδικασία προσαρμογής, αξιοποιώντας την προϋπάρχουσα γνώση που είναι ενσωματωμένη στα μοντέλα.

**Συνεχής Μάθηση και Προσαρμογή:**

Τέλος, η δυνατότητα των LLMs να μαθαίνουν συνεχώς από τις αλληλεπιδράσεις τους μπορεί να ενισχύσει περαιτέρω τη γνώση τους για τα APIs. Αναλύοντας την επιτυχία και την αποτυχία των κλήσεων API και ενσωματώνοντας αυτή την ανατροφοδότηση στους επόμενους κύκλους εκπαίδευσης, τα LLMs μπορούν να προσαρμοστούν και να βελτιώσουν τις δυνατότητες αλληλεπίδρασης με τα APIs.

## 2.3 ΒΗΜΑ 3: Σχεδιασμός Πολυσταδιακής Αρχιτεκτονικής

Στο τρίτο βήμα, εξετάζεται η δημιουργία ενός πολυσταδιακού σχεδίου για την ενσωμάτωση των APIs στα LLMs. Αυτή η διαδικασία περιλαμβάνει τον σχεδιασμό πλάνου δράσης, την επιλογή του κατάλληλου API, τη δημιουργία κλήσεων προς το API και την επεξεργασία των αποκρίσεων του API. Ο σχεδιασμός αυτός εξασφαλίζει την αποδοτικότητα της διαδικασίας και τη συμφωνία με τις προθέσεις και επιθυμίες του χρήστη.

Εμπνευσμένη από τις διαδικασίες που συζητήθηκαν στα [3] και [18], αυτή η μεθοδολογία εξασφαλίζει ότι τα LLMs μπορούν να κατανοούν, να επιλέγουν και να χρησιμοποιούν αποτελεσματικά τα APIs για την εκτέλεση πραγματικών εργασιών, επεκτείνοντας σημαντικά τις δυνατότητές τους πέρα από τη γεννήτρια κειμένου ή την κατανόηση. Εδώ, περιγράφουμε τα κρίσιμα στάδια αυτής της διαδικασίας και τη σημασία κάθε βήματος.

(Αναφορά στο Σχήμα 2.5.)

**1. Αναγνώριση και Κατανόηση Εργασίας**

Το αρχικό στάδιο περιλαμβάνει την ακριβή αναγνώριση της πρόθεσης του χρήστη και την κατανόηση της συγκεκριμένης εργασίας. Αυτό το βήμα είναι κρίσιμο, καθώς θέτει τα θεμέλια για την επιλογή του κατάλληλου API για την εκτέλεση της εργασίας. Περιλαμβάνει τεχνικές κατανόησης φυσικής γλώσσας για την ανάλυση του αιτήματος του χρήστη και την εξαγωγή σχετικών πληροφοριών, όπως η επιθυμητή ενέργεια και οποιεσδήποτε συγκεκριμένες παράμετροι απαιτούνται για την εργασία.

**2. Επιλογή API**

Αφού η εργασία έχει προσδιοριστεί σωστά, το επόμενο βήμα είναι να επιλεγεί το καταλληλότερο API

για την ολοκλήρωσή της. Αυτό περιλαμβάνει την αντιστοίχιση των χαρακτηριστικών που παρέχονται από τα διαθέσιμα APIs με τις απαιτήσεις της εργασίας. Τα κριτήρια απόφασης περιλαμβάνουν την ικανότητα του API να ολοκληρώσει την εργασία, την αξιοπιστία του, την ταχύτητα απόκρισης και πιθανώς το κόστος. Αυτό το στάδιο μπορεί να χρησιμοποιήσει μια συλλογή όπου τα APIs είναι οργανωμένα ανάλογα με τη λειτουργικότητα και τους τομείς τους (π.χ. το "Rapid's API Hub" [31] όπως χρησιμοποιείται στα [5] και [3]).
(Αναφορά στο Σχήμα 2.6.)

**3. Δημιουργία Κλήσης API**
Με το κατάλληλο API επιλεγμένο, η διαδικασία επικεντρώνεται στη δημιουργία της κλήσης API. Αυτό το βήμα απαιτεί τη μετατροπή των παραμέτρων και των απαιτήσεων της εργασίας σε μια μορφή που να ταιριάζει με τις προδιαγραφές του API. Περιλαμβάνει τη δημιουργία του URL του endpoint, τη μέθοδο (GET, POST, κ.λπ.) και τυχόν απαιτούμενες κεφαλίδες ή παραμέτρους σώματος. Αυτό το βήμα είναι κρίσιμο, καθώς επηρεάζει άμεσα την επιτυχία της αλληλεπίδρασης με το API.

**4. Επεξεργασία Απάντησης API**
Μετά την εκτέλεση της κλήσης API, το LLM πρέπει να επεξεργαστεί την απάντηση που έλαβε. Αυτό το στάδιο περιλαμβάνει την ανάλυση της απάντησης του API, την εξαγωγή των σχετικών δεδομένων και τη μετατροπή τους σε μια μορφή που μπορεί να κατανοηθεί και να χρησιμοποιηθεί εύκολα από τον χρήστη. Ανάλογα με την πολυπλοκότητα της απάντησης του API, αυτό μπορεί να περιλαμβάνει φιλτράρισμα, ταξινόμηση ή συσσώρευση δεδομένων.

**5. Ενσωμάτωση Ανατροφοδότησης και Επανάληψη**
Το τελικό στάδιο περιλαμβάνει την ενσωμάτωση της ανατροφοδότησης από την αλληλεπίδραση με το API στις μελλοντικές λειτουργίες. Αυτό μπορεί να περιλαμβάνει την προσαρμογή της διαδικασίας επιλογής API με βάση την απόδοση ή την ενημέρωση της μεθόδου για τη δημιουργία κλήσεων API με βάση τα ποσοστά επιτυχίας. Μηχανισμοί συνεχούς μάθησης μπορούν να εφαρμοστούν για τη βελτίωση κάθε σταδίου της διαδικασίας βάσει δεδομένων αλληλεπίδρασης πραγματικού κόσμου.

Εμπνευσμένη από το [3], αυτή η δομημένη προσέγγιση προς την ενσωμάτωση των API είναι πρωταρχικής σημασίας για τη δημιουργία AI-agents που μπορούν να χρησιμοποιούν αποτελεσματικά εξωτερικά APIs. Εξασφαλίζει μια συστηματική διαδικασία από την κατανόηση των αιτημάτων του χρήστη μέχρι την εκτέλεση των εργασιών μέσω APIs, ενισχύοντας έτσι τη λειτουργικότητα και την εφαρμοσιμότητα των LLMs σε πραγματικά σενάρια. Ένα τέτοιο σενάριο που έχει πρακτικό ενδιαφέρον είναι η "κράτηση διακοπών". Παρακάτω εξηγούμε πώς εφαρμόζεται κάθε βήμα στην εργασία της κράτησης διακοπών μέσω ενός API χρησιμοποιώντας ένα LLM.

Αν ένας χρήστης εισαγάγει "Θέλω να κλείσω διακοπές σε παραλία στην Ελλάδα για την πρώτη εβδομάδα του Ιουλίου", το LLM διακρίνει την εργασία (κράτηση διακοπών), τον τύπο (παραλία), την τοποθεσία (Ελλάδα) και τον χρόνο (πρώτη εβδομάδα του Ιουλίου). Για την κράτηση διακοπών, το LLM επιλέγει APIs από έναν ταξιδιωτικό τομέα για πτήσεις και ξενοδοχεία, εξασφαλίζοντας ότι καλύπτουν ελληνικούς προορισμούς και προσφέρουν επιλογές για παραθαλάσσια καταλύματα. Για να βρει παραθαλάσσια ξενοδοχεία στην Ελλάδα, το LLM δημιουργεί ένα αίτημα GET στο API κράτησης ξενοδοχείων, περιλαμβάνοντας παραμέτρους για την τοποθεσία ("Ελλάδα"), τις ημερομηνίες ("πρώτη εβδομάδα του Ιουλίου") και φίλτρα για "παραθαλάσσια ακίνητα". Στη συνέχεια, το LLM επεξεργάζεται τη λίστα των παραθαλάσσιων ξενοδοχείων, διαλέγοντας μερικές επιλογές βάσει αξιολογήσεων, διαθεσιμότητας και τιμής. Τέλος, παρουσιάζει αυτές τις επιλογές στον χρήστη με συνοπτικό και ενημερωτικό τρόπο. Είναι σημαντικό να σημειωθεί ότι με βάση την επιλογή του χρήστη, το σύστημα μαθαίνει προτιμήσεις, όπως προτεραιότητα σε επιλογές χαμηλού κόστους ή συγκεκριμένες ανέσεις, βελτιώνοντας τις μελλοντικές εκτελέσεις εργασιών. Αυτή η σύντομη ανάλυση κάθε βήματος, που παρουσιάζεται στο παράδειγμα "κράτηση διακοπών", υπογραμμίζει την πολυπλοκότητα και τη δυνατότητα εξοπλισμού των LLMs με εξωτερικά APIs για την εκτέλεση πραγματικών εργασιών, παρουσιάζοντας μια σημαντική πρόοδο στις δυνατότητες των AI-agents.
(Αναφορά στο Σχήμα 2.7.)

## 2.4 ΒΗΜΑ 4: Μηχανισμός Επιλογής API

Το τέταρτο βήμα αφορά τη δημιουργία ενός μηχανισμού για την επιλογή του κατάλληλου API βάσει συγκεκριμένων κριτηρίων όπως η συνάφεια με την εργασία, η απόδοση, η αξιοπιστία και το κόστος. Αυτή η διαδικασία εξασφαλίζει ότι το επιλεγμένο API είναι το πιο κατάλληλο για την εκάστοτε εργασία, βελτιστοποιώντας την αποδοτικότητα και την επιτυχία των αλληλεπιδράσεων.

Τα ακόλουθα είναι τα κύρια κριτήρια και οι μηχανισμοί που πρέπει να ληφθούν υπόψιν:

### 2.4.1 Κριτήρια για την Επιλογή API

**1. Συνάφεια με την Εργασία:** Οι στόχοι της εργασίας πρέπει να υποστηρίζονται άμεσα από το API.
**2. Κάλυψη και Εύρος:** Το API πρέπει να καλύπτει όλες τις απαραίτητες υπηρεσίες και γεωγραφικές περιοχές.
**3. Απόδοση και Αξιοπιστία:** Προτιμώνται APIs με υψηλή αξιοπιστία, χαμηλή καθυστέρηση και γρήγορους χρόνους απόκρισης, για να προσφέρουν μια ομαλή εμπειρία χρήστη.
**4. Ποιότητα και Ακρίβεια Δεδομένων:** Το API πρέπει να παρέχει αξιόπιστες, ενημερωμένες και ακριβείς πληροφορίες, κάτι που είναι απαραίτητο για εργασίες που απαιτούν δεδομένα σε πραγματικό χρόνο ή εξειδικευμένες λεπτομέρειες.
**5. Κόστος και Όρια Ρυθμού Κλήσεων:** Τα κόστη του API, καθώς και τυχόν όρια ρυθμού κλήσεων, πρέπει να ευθυγραμμίζονται με τον προϋπολογισμό του έργου και τον προβλεπόμενο όγκο χρήσης.
**6. Τεκμηρίωση και Υποστήριξη:** Καλά τεκμηριωμένα APIs με σαφείς οδηγίες χρήσης και ενεργά φόρουμ υποστήριξης είναι απαραίτητα για την αποτελεσματική ενσωμάτωση και την αντιμετώπιση προβλημάτων.
**7. Ασφάλεια και Απόρρητο:** Το API πρέπει να ακολουθεί αυστηρά πρωτόκολλα ασφάλειας και απορρήτου, ιδιαίτερα όταν χειρίζεται ευαίσθητα δεδομένα χρηστών.

Ας εξετάσουμε το παράδειγμα της "Κράτησης Διακοπών" βάσει των παραπάνω παραμέτρων:
Όταν ένας χρήστης ζητά βοήθεια για την κράτηση διακοπών, το LLM θα αναγνωρίσει πρώτα τις απαιτήσεις της εργασίας, όπως ο προορισμός, οι ημερομηνίες και οι προτιμήσεις. Στη συνέχεια, θα αναζητήσει τον κατάλογο με τα APIs, φιλτράροντας τα ταξιδιωτικά APIs που καλύπτουν τον συγκεκριμένο προορισμό και τις ημερομηνίες. Ο μηχανισμός επιλογής μπορεί να δώσει προτεραιότητα στα APIs βάσει των προτιμήσεων του χρήστη για επιλογές χαμηλού κόστους, υψηλές αξιολογήσεις και θετικά σχόλια από προηγούμενες ενσωματώσεις. Αφού δημιουργηθεί μια σύντομη λίστα κατάλληλων APIs, το LLM επιλέγει το API που ταιριάζει καλύτερα στα κριτήρια. Μέσω αυτών των κριτηρίων και μηχανισμών, η διαδικασία επιλογής API γίνεται μια σύνθετη, πολυδιάστατη διαδικασία λήψης αποφάσεων, διασφαλίζοντας ότι τα LLMs μπορούν να αξιοποιήσουν αποτελεσματικά εξωτερικά APIs για να επεκτείνουν τις δυνατότητές τους και να εκτελέσουν ένα ευρύ φάσμα εργασιών με μεγαλύτερη ακρίβεια, αποδοτικότητα και ικανοποίηση χρήστη.

Το **"Κριτήριο Συνάφειας με την Εργασία"** είναι θεμελιώδες στον μηχανισμό επιλογής API για τη διασύνδεση Μεγάλων Γλωσσικών Μοντέλων (LLMs) με εξωτερικά εργαλεία και APIs. Αυτό το κριτήριο διασφαλίζει ότι το επιλεγμένο API ευθυγραμμίζεται ουσιαστικά με τους συγκεκριμένους στόχους της εργασίας, διασφαλίζοντας ότι το LLM μπορεί να αξιοποιήσει αποτελεσματικά το API για να ικανοποιήσει το αίτημα του χρήστη. Εδώ, εξετάζουμε την αξιολόγηση της συνάφειας ενός API με μια δεδομένη εργασία, χρησιμοποιώντας το παράδειγμα της κράτησης ταξιδιών για να εξηγήσουμε την έννοια λεπτομερώς.
**Λειτουργίες Συγκεκριμένες για την Εργασία:**
Ο πρωταρχικός παράγοντας στον καθορισμό της συνάφειας ενός API είναι το σύνολο των λειτουργιών του. Για μια εργασία όπως η κράτηση ταξιδιών, το API πρέπει να προσφέρει συγκεκριμένες δυνατότητες που σχετίζονται με την αναζήτηση πτήσεων, την κράτηση καταλυμάτων και ίσως επιπλέον υπηρεσίες όπως ενοικίαση αυτοκινήτων ή κρατήσεις δραστηριοτήτων. Οι βασικές λειτουργίες του API πρέπει να ευθυγραμμίζονται με τις απαιτήσεις της εργασίας, για να διασφαλιστεί ότι μπορεί να εξυπηρετήσει τον επιδιωκόμενο σκοπό.

**Γεωγραφική Κάλυψη:**

Για εργασίες με γεωγραφικό στοιχείο, όπως η κράτηση ταξιδιών, η γεωγραφική κάλυψη του API είναι κρίσιμη. Πρέπει να υποστηρίζει ερωτήματα και συναλλαγές για τους συγκεκριμένους προορισμούς που εμπλέκονται στην εργασία. Ένα API που προσφέρει εκτενείς καταχωρίσεις ξενοδοχείων στην Ευρώπη αλλά περιορισμένες επιλογές στην Ασία δεν θα ήταν σχετικό για τον προγραμματισμό ενός ταξιδιού στο Τόκιο, για παράδειγμα.

**Πολυπλοκότητα Ενσωμάτωσης και Συμβατότητα:**

Η χρησιμότητα ενός API επηρεάζεται επίσης από το πόσο εύκολο είναι να ενσωματωθεί στο υπάρχον πλαίσιο, ιδιαίτερα από το πόσο καλά λειτουργεί με τις μορφές δεδομένων και την αρχιτεκτονική του LLM. Αν και ένα API μπορεί τεχνικά να υποστηρίξει τους στόχους της εργασίας, μπορεί να μην είναι εξίσου κατάλληλο αν χρειάζεται σημαντική προσαρμογή ή εργασία μετατροπής.

Αξιολόγηση Συνάφειας

**Ανασκόπηση Τεκμηρίωσης:**

Μια λεπτομερής εξέταση της τεκμηρίωσης του API μπορεί να αποκαλύψει τις δυνατότητές του, τα μοντέλα δεδομένων και τις υποστηριζόμενες λειτουργίες, παρέχοντας πληροφορίες για το αν πληροί τις ανάγκες της εργασίας.

**Ανάλυση Περιπτώσεων Χρήσης και Παραδειγμάτων:**

Πολλά APIs παρέχουν παραδείγματα ή περιπτώσεις χρήσης που απεικονίζουν τυπικές εφαρμογές. Η ανασκόπηση αυτών μπορεί να βοηθήσει στην αξιολόγηση του πόσο καλά οι δυνατότητες του API ικανοποιούν τις απαιτήσεις της εργασίας.

**Δοκιμή και Έλεγχος:**

Σε ορισμένες περιπτώσεις, η διεξαγωγή μιας δοκιμαστικής ενσωμάτωσης ή ενός δοκιμαστικού ερωτήματος μπορεί να προσφέρει άμεση αξιολόγηση της συνάφειας του API, επιτρέποντας μια εμπειρική εκτίμηση της καταλληλότητάς του για την εργασία.

Στο παράδειγμα "Κράτηση Διακοπών", ας υποθέσουμε ότι ένας χρήστης ζητά βοήθεια για την κράτηση διακοπών στην Ελλάδα, συμπεριλαμβανομένων των πτήσεων, των καταλυμάτων και των δραστηριοτήτων. Το LLM, στο οποίο έχει ανατεθεί να εκπληρώσει αυτό το αίτημα, πρέπει να επιλέξει APIs που υποστηρίζουν άμεσα αυτούς τους στόχους:

• **API Κράτησης Πτήσεων:**

Πρέπει να προσφέρει μια ολοκληρωμένη βάση δεδομένων πτήσεων προς και από την Ελλάδα, επιτρέποντας αναζητήσεις βάσει ημερομηνιών, προτιμήσεων (π.χ. χωρίς στάσεις) και τιμών.

• **API Κράτησης Ξενοδοχείων:**

Πρέπει να παρέχει πρόσβαση σε μια ευρεία γκάμα καταλυμάτων στην Ελλάδα, με δυνατότητες που επιτρέπουν στους χρήστες να φιλτράρουν κατά τοποθεσία, αξιολόγηση, τιμή και ανέσεις.

• **API Δραστηριοτήτων:**

Πρέπει να καλύπτει ψυχαγωγικές δραστηριότητες που είναι διαθέσιμες στην Ελλάδα, όπως περιηγήσεις, μουσεία και άλλες εμπειρίες, με δυνατότητες κράτησης.

Κάθε επιλεγμένο API πρέπει όχι μόνο να προσφέρει τις σωστές λειτουργίες, αλλά και να εξασφαλίζει ενημερωμένη και ολοκληρωμένη κάλυψη επιλογών στην Ελλάδα, για να θεωρείται σχετικό για την εργασία. Αξιολογώντας προσεκτικά τη συνάφεια ενός API με τους στόχους της εργασίας, μπορεί κανείς να βελτιώσει σημαντικά την ικανοποίηση του χρήστη.

**Λεπτομέρειες Υλοποίησης - Μετρική Ομοιότητας Συνημιτόνου:**

Σε αυτό το σημείο, δίνουμε μια σύντομη επισκόπηση του πώς εφαρμόζεται η έννοια της ομοιότητας συνημιτόνου στη διαδικασία επιλογής API βάσει συνάφειας, με τη χρήση γενικών αρχών στη μηχανική μάθηση και την επεξεργασία φυσικής γλώσσας (NLP). Η ομοιότητα συνημιτόνου είναι μια μετρική που χρησιμοποιείται για τη μέτρηση της ομοιότητας δύο διανυσμάτων σε έναν πολυδιάστατο χώρο, συχνά χρησιμοποιούμενη στο πλαίσιο της ανάλυσης κειμένου. Υπολογίζει το συνημίτονο της γωνίας μεταξύ δύο διανυσμάτων που, σε αυτή την περίπτωση, μπορεί να αντιπροσωπεύουν την περιγραφή της εργασίας ενός χρήστη και την περιγραφή της λειτουργικότητας ενός API. Η τιμή της μετρικής ομοιότητας συνημιτόνου κυμαίνεται από -1 έως 1, όπου το 1 υποδηλώνει ταυτόσημη κατεύθυνση (ή τέλεια ομοιότητα), το 0 υποδηλώνει ορθογωνιότητα (καμία ομοιότητα) και το -1 υποδηλώνει αντίθετη κατεύθυνση.

**Εφαρμογή στην Επιλογή API βάσει Συνάφειας:**

Κατά την επιλογή APIs με βάση τη συνάφεια με μια δεδομένη εργασία, οι περιγραφές τόσο της εργασίας όσο και των APIs μπορούν να μετατραπούν σε διανύσματα χρησιμοποιώντας τεχνικές όπως το TF-IDF (Term Frequency-Inverse Document Frequency) ή ενσωματώσεις που δημιουργούνται από μοντέλα γλώσσας. Μόλις μετατραπούν σε διανύσματα, μπορεί να υπολογιστεί η συγγένεια cosine μεταξύ του διανύσματος της εργασίας και κάθε διανύσματος API για να προσδιοριστεί πόσο στενά ευθυγραμμίζεται η λειτουργικότητα του κάθε API με τις απαιτήσεις της εργασίας.

**1. Διανυσματοποίηση:**

Μετατροπή των περιγραφών κειμένου της εργασίας και των APIs σε διανύσματα. Αυτό μπορεί να περιλαμβάνει:

• Εξαγωγή χαρακτηριστικών χρησιμοποιώντας το TF-IDF, δίνοντας έμφαση στις λέξεις που είναι συχνές σε ένα έγγραφο αλλά όχι σε όλα τα έγγραφα.

• Χρησιμοποίηση προεκπαιδευμένων ενσωματώσεων λέξεων από μοντέλα όπως το Word2Vec ή το BERT [8] για την αναπαράσταση λέξεων ή φράσεων σε διανυσματικό χώρο.

**2. Υπολογισμός Ομοιότητας Συνημιτόνου:**

Για κάθε API, υπολογισμός της συγγένειας cosine μεταξύ του διανύσματός του και του διανύσματος της εργασίας χρησιμοποιώντας τον τύπο:

$$cosine\ similarity = \frac{\mathbf{A} \cdot \mathbf{B}}{\|\mathbf{A}\| \|\mathbf{B}\|}$$

όπου **A** και **B** είναι οι διανυσματικές αναπαραστάσεις του API και της εργασίας, αντίστοιχα και $\|\mathbf{A}\|$ και $\|\mathbf{B}\|$ είναι τα μέτρα τους.

**3. Κατάταξη και Επιλογή:**

Κατάταξη των APIs με βάση την ομοιότητα συνημιτόνου με το διάνυσμα της εργασίας. Το API ή τα APIs με την υψηλότερη βαθμολογία θεωρούνται τα πιο σχετικά με την εργασία και επιλέγονται για ενσωμάτωση.

**Εξειδικευμένες Λεπτομέρειες**

• **Μείωση Διαστάσεων:** Τα υψηλής διάστασης διανύσματα μπορεί να απαιτούν τεχνικές μείωσης διαστάσεων (π.χ. PCA) για τη βελτίωση της υπολογιστικής αποδοτικότητας χωρίς σημαντική παραχώρηση στην ικανότητα σύλληψης της σημασιολογικής ομοιότητας.

• **Ενσωματώσεις Συμφραζομένων:** Για πολύπλοκες εργασίες, οι ενσωματώσεις συμφραζομένων από μοντέλα όπως το BERT [8] ή το GPT [7] μπορεί να προτιμώνται έναντι απλούστερων μεθόδων διανυσματοποίησης, καθώς αποτυπώνουν καλύτερα τις γλωσσικές οντότητες.

• **Οριοθέτηση:** Η εφαρμογή ενός ορίου για την ομοιότητα συνημιτόνου μπορεί να βοηθήσει στο φιλτράρισμα των APIs που βρίσκονται κάτω από ένα συγκεκριμένο επίπεδο συνάφειας, διασφαλίζοντας ότι μόνο τα πιο σχετικά APIs λαμβάνονται υπόψιν για επιλογή.

### 2.4.2 Μηχανισμοί Επιλογής API

**1. Κατάλογος με APIs και Μεταδεδομένα:**

Η διατήρηση μιας συλλογής APIs και μεταδεδομένων σχετικά με τη λειτουργικότητά τους, τα καλυπτόμενα θέματα, τους δείκτες απόδοσης και τους όρους χρήσης, επιταχύνει τη διαδικασία επιλογής.

• Κατηγοριοποίηση: Τα APIs οργανώνονται σε κατηγορίες βάσει του τομέα εφαρμογής τους (π.χ. ταξίδια, οικονομικά, υγειονομική περίθαλψη).

• Σχολιασμός Μεταδεδομένων: Κάθε API συνοδεύεται από μεταδεδομένα που περιλαμβάνουν τις δυνατότητές του, τους δείκτες απόδοσης, τα κόστη χρήσης και τους περιορισμούς.

## 2. Αυτοματοποιημένοι Αλγόριθμοι Αντιστοίχισης:

Η διαδικασία επιλογής μπορεί να αυτοματοποιηθεί χρησιμοποιώντας αλγόριθμους για την αντιστοίχιση των απαιτήσεων της εργασίας με τα μεταδεδομένα των APIs. Αυτοί οι αλγόριθμοι είναι σε θέση να κατατάσσουν τα APIs σύμφωνα με το πόσο καλά ικανοποιούν τις απαιτούμενες προδιαγραφές.

• Επεξεργασία Φυσικής Γλώσσας (NLP): Ανάλυση της περιγραφής της εργασίας για την εξαγωγή των βασικών απαιτήσεων.

• Βαθμολόγηση και Κατάταξη: Οι αλγόριθμοι αξιολογούν τα APIs με βάση τα κριτήρια αντιστοίχισης και τα κατατάσσουν ανάλογα. Τεχνικές όπως η ομοιότητα συνημιτόνου μπορούν να χρησιμοποιηθούν εδώ.

## 3. Προτιμήσεις Χρηστών και Ιστορικά Δεδομένα:

Η διαδικασία επιλογής μπορεί να βελτιωθεί περιλαμβάνοντας προτιμήσεις χρηστών και δεδομένα από προηγούμενες επιλογές, δίνοντας προτεραιότητα στα APIs που πληρούν τα καθορισμένα κριτήρια των χρηστών ή έχουν δείξει αποτελεσματικότητα στο παρελθόν.

• Προφίλ Προτιμήσεων: Οι χρήστες μπορούν να καθορίσουν προτιμήσεις που χρησιμοποιούνται για το φιλτράρισμα ή την προτεραιοποίηση των επιλογών API.

• Ανάλυση Συμφραζομένων: Το σύστημα λαμβάνει υπόψιν το πλαίσιο του αιτήματος (π.χ. γεωγραφική τοποθεσία, ώρα της ημέρας) για την επιλογή των καταλληλότερων APIs.

## 4. Δυναμική Αξιολόγηση:

Οι τεχνικές δυναμικής αξιολόγησης που δοκιμάζουν τα APIs σε πραγματικό χρόνο για να αξιολογήσουν την τρέχουσα απόδοση και την ποιότητα των δεδομένων τους μπορεί να είναι χρήσιμες για εργασίες όπου πολλά APIs μπορεί να είναι κατάλληλα.

• Δοκιμή API: Αποστολή δειγμάτων αιτημάτων στα APIs για την εκτίμηση της απόδοσής τους και της συνάφειας των απαντήσεών τους.

• Προσαρμοστική Επιλογή: Το σύστημα προσαρμόζει δυναμικά τις επιλογές API με βάση τα αποτελέσματα αυτών των δοκιμών.

## 5. Βρόχοι Ανατροφοδότησης:

Η συνεχής βελτίωση διασφαλίζεται μέσω της εγκατάστασης βρόχων ανατροφοδότησης όπου η αποτελεσματικότητα και η επιτυχία των προσπαθειών ενσωμάτωσης API παρακολουθούνται και χρησιμοποιούνται για την ενημέρωση του καταλόγου των APIs και των αλγορίθμων επιλογής.

• Ενσωμάτωση Ανατροφοδότησης Χρηστών: Οι αξιολογήσεις και τα σχόλια των χρηστών για την απόδοση των APIs χρησιμοποιούνται για τη ρύθμιση των μελλοντικών επιλογών.

• Ανάλυση Απόδοσης: Ανάλυση του ποσοστού επιτυχίας και της αποδοτικότητας των ενσωματώσεων των APIs για την ενημέρωση των προσαρμογών στη διαδικασία επιλογής.

## 6. Φίλτρα Ασφάλειας και Συμμόρφωσης:

Η εφαρμογή προτύπων ασφάλειας και απαιτήσεων συμμόρφωσης αποτελεί φιλτράρισμα στη διαδικασία επιλογής API και διασφαλίζει ότι τα επιλεγμένα APIs πληρούν τα οργανωτικά ή ηθικά πρότυπα.

• Έλεγχοι Συμμόρφωσης: Επαλήθευση ότι τα APIs συμμορφώνονται με τους σχετικούς κανονισμούς προστασίας δεδομένων και απορρήτου.

• Αξιολόγηση Ασφάλειας: Αξιολόγηση των μέτρων ασφαλείας και των τρωτών σημείων των APIs.

Μια ενδιαφέρουσα μέθοδος επιλογής API σε 2 επίπεδα χρησιμοποιείται σε πρόσφατες προσεγγίσεις όπως το [18] και το [5]. Στο πρώτο βήμα της μεθόδου επιλέγεται ένας τομέας που περιέχει πολλά APIs (π.χ. από τις 49 κατηγορίες της συλλογής του [31] όπως φαίνεται στο Σχήμα 2.9). Στη συνέχεια, επιλέγεται το πιο κατάλληλο API για την εργασία με βάση μια προκαθορισμένη μετρική "Συνάρτηση Απώλειας" (όπως εξηγείται λεπτομερώς στο [18]). Με λίγα λόγια, οι κλήσεις API φιλτράρονται δυναμικά με βάση τη συμβολή τους στην ακρίβεια πρόβλεψης, ποσοτικοποιημένη από μια συνάρτηση απώλειας $L_i(z)$. Αυτή η συνάρτηση ορίζεται ως $L_i(z) = -\sum_{j=i}^{n} w_{j-i} \log p_M(x_j|z, x_{1:j-1})$, όπου $w_{j-i}$ είναι βάρη που μειώνονται με την απόσταση από τη θέση $i$ για να δοθεί έμφαση στις πιο άμεσες συμβολές και $\log p_M(x_j|z, x_{1:j-1})$ υποδηλώνει το λογάριθμο της πιθανότητας παρατήρησης του στοιχείου $x_j$ δεδομένου του συμφραζομένου $z$ και όλων των προηγούμενων στοιχείων. Οι κλήσεις API δημιουργούνται σε διάφορες θέσεις της ακολουθίας και η χρησιμότητά τους αξιο-λογείται με βάση τη βελτίωση που παρέχουν στην απόδοση πρόβλεψης του μοντέλου, δηλαδή τη μείωση της υπολογιζόμενης απώλειας. Μόνο οι κλήσεις API που οδηγούν σε σημαντική μείωση της απώλειας διατηρούνται, επιτρέποντας στο μοντέλο να χρησιμοποιεί εξωτερικές πηγές πληροφοριών αποτελεσματικά και να βελτιώνει τη συνολική διαδικασία λήψης αποφάσεων. Αυτή η τεχνική θυμίζει αλγορίθμους "περικοπής του χώρου αναζήτησης".

Αυτοί οι μηχανισμοί, όταν συνδυάζονται, σχηματίζουν ένα ολοκληρωμένο πλαίσιο για την επιλογή API, διασφαλίζοντας ότι τα επιλεγμένα APIs δεν είναι μόνο τεχνικά ικανά να εκπληρώσουν την εργασία αλλά και ευθυγραμμίζονται με τις προτιμήσεις των χρηστών, τις προσδοκίες απόδοσης και τις απαιτήσεις ασφάλειας.

(Αναφορά στα Σχήματα 2.9 και 2.10.)

## 2.5 ΒΗΜΑ 5: Γεννήτρια Κλήσεων API - Ανάληψη Δράσης

Το πέμπτο βήμα περιγράφει τη διαδικασία δημιουργίας και εκτέλεσης κλήσεων API. Σε αυτό το στάδιο, τα LLMs χρησιμοποιούν τη γνώση που έχουν αποκτήσει για να διαμορφώσουν και να εκτελέσουν κλήσεις προς τα APIs, αποσκοπώντας στην επίλυση συγκεκριμένων εργασιών. Η ακρίβεια και η συνέπεια στη διαμόρφωση των κλήσεων είναι κρίσιμη, καθώς επηρεάζει άμεσα την ποιότητα των αποτελεσμάτων της αλληλεπίδρασης με τα APIs.

Είναι σημαντικό να κατανοήσουμε τη φύση της εργασίας πριν προχωρήσουμε στην εκτέλεση. Υπάρχουν "απλές εργασίες" που απαιτούν μόνο κειμενική έξοδο (π.χ. "Γράψε ένα σύντομο παραμυθένιο ποίημα που να περιλαμβάνει έναν δράκο και μια πριγκίπισσα..."), υπάρχουν εργασίες που απαιτούν επικοινωνία με εξωτερικά εργαλεία (π.χ. "Θα βρέξει αύριο;") και τέλος, υπάρχουν εργασίες που απαιτούν προχωρημένη συλλογιστική και - κατά κάποιο τρόπο - μια "πρώιμη" μορφή της Γενικής Τεχνητής Νοημοσύνης (AGI) (π.χ. "Δεδομένης της εικόνας του ψυγείου μου, μπορείς να προτείνεις μια γρήγορη συνταγή που δεν απαιτεί τη χρήση του φούρνου μου;" ή "Δεδομένης μιας παρτιτούρας, μπορείς να προτείνεις έναν λογικό δακτυλισμό ώστε να μπορεί ένας πιανίστας να παίξει το κομμάτι;"). Παρακάτω αναλύουμε αυτές τις 3 κατηγορίες εργασιών λεπτομερώς.

(Αναφορά στα Σχήματα 2.11 και 2.12.)

### 2.5.1 Κειμενική έξοδος

Τα πρόσφατα LLMs έχουν καταφέρει αποκρίσεις που μοιάζουν με ανθρώπινο κείμενο. Μοντέλα όπως το ChatGPT και το PanGu είναι εξαιρετικά στην παραγωγή κειμένου που είναι συναφές και συνεκτικό. Αυτή η επανάσταση έχει ανοίξει ένα ευρύ φάσμα εφαρμογών όπου η προηγμένη, ευέλικτη επεξεργασία γλώσσας είναι απαραίτητη, όπως η εξυπηρέτηση πελατών και η δημιουργία περιεχομένου. Ένα αξιοσημείωτο χαρακτηριστικό των LLMs είναι η ικανότητά τους να παράγουν καθοδηγούμενες αποκρίσεις. Αυτό σημαίνει ότι οι χρήστες μπορούν να προσαρμόσουν τις αποκρίσεις για να ανταποκριθούν σε συγκεκριμένες ανάγκες και καταστάσεις, κάτι που είναι κρίσιμο για ιδιαίτερα εξειδικευμένες εργασίες όπως η δημιουργική γραφή ή η εξατομικευμένη επικοινωνία. Η ευελιξία και η χρησιμότητα των μοντέλων αποδεικνύεται από τον τρόπο ενσωμάτωσής τους σε πολλές βιομηχανίες, γεγονός που αντιπροσωπεύει μια σημαντική πρόοδο στην κατανόηση και παραγωγή ανθρώπινης γλώσσας από μηχανές. Καθώς αυτές οι τεχνολογίες εξελίσσονται, έχουν τη δυνατότητα να μεταμορφώσουν πλήρως τις εφαρμογές της AI βελτιώνοντας την αποτελεσματικότητα και την διαισθητικότητα των ψηφιακών πλατφορμών και γεφυρώνοντας το χάσμα μεταξύ ανθρώπινης και μηχανικής επικοινωνίας. Για να υπογραμμίσουμε αυτή την τεχνολογική πρόοδο, παρέχουμε ένα μικρό παράδειγμα παραγωγής κειμένου στο Σχήμα 2.12.

(Αναφορά στο Σχήμα 2.13.)

### 2.5.2 Χρήση εργαλείων

Τα εργαλεία παρέχουν μια πολύτιμη επέκταση των δυνατοτήτων των LLMs, επιτρέποντας πιο αποτελεσματική και ενημερωμένη ολοκλήρωση των εργασιών. Για παράδειγμα, όταν ένας χρήστης ρωτά για τον καιρό, το LLM θα πρέπει να μπορεί να συνδεθεί με ένα "Εργαλείο Πρόβλεψης Καιρού" για να παρέχει μια λογική απάντηση, όπως φαίνεται στο Σχήμα 2.13. Τα LLMs έχουν τεράστιο όγκο γνώσεων, αλλά δεν έχουν τη δυνατότητα να έχουν πρόσβαση σε δεδομένα πραγματικού κόσμου σε απευθείας σύνδεση. Αυτός ο περιορισμός μπορεί συχνά να οδηγήσει σε παρανοήσεις των ερωτημάτων των χρηστών ή σε ψευδείς γνώσεις. Για να μετριαστούν αυτοί οι περιορισμοί, κατασκευάζουμε AI-agents που μπορούν να επικοινωνούν με το εξωτερικό περιβάλλον μέσω εργαλείων και APIs. Ένα άλλο πλεονέκτημα της ικανότητας χρήσης εργαλείων από τους AI-agents είναι εμφανές σε τομείς υψηλής σημασίας (π.χ. υγειονομική περίθαλψη, χρηματοοικονομικά). Σε τέτοιες εφαρμογές η διαφάνεια είναι κρίσιμη, αλλά η διαδικασία λήψης αποφάσεων των LLMs είναι συχνά σαν ένα "μαύρο κουτί". Ο εξοπλισμός των LLMs με εργαλεία οδηγεί σε αυξημένη ερμηνευσιμότητα και ανθεκτικότητα και παρέχει ένα πιο αξιόπιστο πλαίσιο για τη λήψη αποφάσεων, αντανακλώντας μια δομημένη προσέγγιση σε πολύπλοκα προβλήματα. Τα LLMs πρέπει να έχουν μια πλήρη κατανόηση των εφαρμογών των εργαλείων ώστε να τα χρησιμοποιούν αποτελεσματικά. Τα LLMs μπορούν να μιμηθούν τις διαδικασίες μάθησης των ανθρώπων (π.χ. αναθεώρηση εγχειριδίων ή παρακολούθηση άλλων χρηστών) και να βελτιώσουν τις δεξιότητες χρήσης εργαλείων τους μέσω τεχνικών όπως η εκμάθηση με λίγα παραδείγματα. Η μετάβαση από τα LLMs σε AI-agents χρήσης εργαλείων υποδηλώνει μια αλλαγή προς όλο και πιο έξυπνα και αυτόνομα συστήματα AI.

### 2.5.3 Ενσωματωμένη διάδραση

Στην εποχή των LLMs και των τεχνολογιών AI, είναι λογικό να φανταζόμαστε το μέλλον της Γενικής Τεχνητής Νοημοσύνης (AGI). Η "Υπόθεση της Ενσωμάτωσης", που υποστηρίζει ότι η νοημοσύνη αναπτύσσεται μέσω συνεχιζόμενης περιβαλλοντικής αλληλεπίδρασης, αποτελεί σημαντική πηγή έμπνευσης για την Ενσωματωμένη διάδραση, η οποία δίνει έμφαση στη σημαντική μετάβαση από τον ψηφιακό στον φυσικό κόσμο. Αυτή η ακριβής υπόθεση θυμίζει τη διάσημη φιλοσοφική σχολή του Πλάτωνα, του Αρχαίου Έλληνα Φιλοσόφου. Ο Πλάτωνας ανέπτυξε τη θεωρία ότι μέσω της "Ανάμνησης" (που είναι μια εξερεύνηση της μνήμης του ατόμου) ένα άτομο μπορεί να αποκτήσει γνώση σημαντικών αληθειών (ή αλλιώς "Γνώση"). Οι παραδοσιακοί αλγόριθμοι μηχανικής μάθησης (ML) περιορίζονται στο επιχειρησιακό τους πεδίο σε ψηφιακές εξόδους με βάση δομημένες πληροφορίες. Η πρόσφατη έρευνα εκμεταλλεύεται το ψηφιακό περιβάλλον των AI-agents, επιτρέποντάς τους να αναπτύξουν ανθρώπινη συμπεριφορά μάθησης μέσω παρατήρησης, χειρισμού, πλοήγησης και άλλων "ενσωματωμένων ενεργειών". (Αναφορά στο Σχήμα 2.14.)

Η μακρά προεκπαίδευση των LLMs είναι προϋπόθεση για αυξημένες ικανότητες γενίκευσης και συλλογιστικής και επιτρέπουν την αλληλεπίδραση των AI-agents με οντότητες του πραγματικού κόσμου. Για να ενημερωθούν η μνήμη και οι εσωτερικές καταστάσεις των AI-agents, συλλέγονται πληροφορίες από αισθητήρες. Έπειτα από περίπλοκες διεργασίες, παρέχεται μια εκτελεστική διαδρομή για τις ενέργειες του AI-agent. Ο συνδυασμός των "ενσωματωμένων ενεργειών" με τις δυνατότητες των LLMs οδηγεί σε ευέλικτα, ανεξάρτητα και αποτελεσματικά συστήματα AI που πρωτοπορούν σε ό,τι αφορά την αλληλεπίδραση Ανθρώπου-Υπολογιστή (HCI). (Αναφορά στο Σχήμα 2.15.)

Μια ενδιαφέρουσα προσέγγιση για την αξιολόγηση της απόδοσης των ενισχυμένων με εργαλεία LLMs παρουσιάζεται στο [3]. Σε αυτό το άρθρο, 53 συνήθη APIs (επιλεγμένα με ποικιλία, συμπεριλαμβανομένων μηχανών αναζήτησης, αριθμομηχανών, ερωτήσεων ημερολογίου, ελέγχου έξυπνου σπιτιού, διαχείρισης προγράμματος, διαχείρισης δεδομένων υγείας, ροής εργασιών πιστοποίησης λογαριασμού κ.λπ.) μελετώνται πειραματικά στο πλαίσιο της διασύνδεσης των LLMs με APIs. Τα σύνολα δεδομένων περιλαμβάνουν 264 σχολιασμένους διαλόγους που περιλαμβάνουν 568 κλήσεις API. Η επιλογή του κατάλληλου API γίνεται με παρόμοιο τρόπο όπως στο "ΒΗΜΑ 4" της παρούσας διατριβής. (Αναφορά στο Σχήμα 2.16.) Η διαδικασία λήψης αποφάσεων αξιολογείται με βάση ορισμένες μετρικές και τα πιο σημαντικά αντικείμενα της απόφασης είναι "Εάν χρειάζεται κλήση API", "Εάν το επιλεγμένο API είναι αρκετά συναφές" και "Εάν η παραγόμενη κλήση API είναι αρκετά καλή για τη συγκεκριμένη εργασία". Είναι σημαντικό να σημειωθεί ότι ορισμένες εργασίες όπως "Προγραμμάτισε ένα ταξίδι στην Ελλάδα" μπορεί να περιλαμβάνουν πολλές κλήσεις API και πολλαπλά APIs που πρέπει να συντονιστούν. Αυτό το πρόβλημα λύνεται στο επόμενο βήμα.

## 2.6 ΒΗΜΑ 6: Διάσπαση Εργασιών

Στο έκτο στάδιο, το σύστημα υποβάλλεται σε μια σειρά από επαληθεύσεις και δοκιμές για να διασφαλίσει ότι οι κλήσεις προς τα APIs λειτουργούν σύμφωνα με τις προδιαγραφές και ότι τα αποτελέσματα των διαδικασιών είναι αξιόπιστα και συνεπή. Αυτή η φάση περιλαμβάνει τη δοκιμή του συστήματος σε ποικίλες συνθήκες για να αναλυθεί η αντοχή και η αξιοπιστία του. Τα δεδομένα από αυτές τις δοκιμές χρησιμοποιούνται για να προσαρμοστούν και να βελτιωθούν οι μηχανισμοί κλήσης και απόκρισης του μοντέλου.

Όπως αναφέρθηκε στην προηγούμενη παράγραφο, ορισμένες εργασίες είναι υψηλής πολυπλοκότητας και απαιτούν πολλαπλές κλήσεις API και πολλαπλά APIs. Για τέτοιες εργασίες, δημιουργούμε μια σειρά υποεργασιών και παρέχουμε έναν προγραμματισμό τους. Αυτό είναι που θα έκανε και ένας άνθρωπος. Ο προγραμματισμός είναι μια βασική τακτική που χρησιμοποιούν οι άνθρωποι για να ξεπεράσουν δύσκολες καταστάσεις. Ο προγραμματισμός βοηθά στην οργάνωση της σκέψης, στη θέσπιση στόχων και στον καθορισμό του σχεδίου δράσης για την επίτευξη μεγαλύτερων στόχων. Οι AI-agents πρέπει να μπορούν να προγραμματίζουν με οργανωμένο τρόπο όπως οι άνθρωποι. Οι AI-agents χρησιμοποιούν συλλογιστική για να διασπάσουν δύσκολες εργασίες σε μικρότερες, πιο διαχειρίσιμες εργασίες και στη συνέχεια να δημιουργούν στρατηγικές ειδικά για κάθε μία από αυτές. Στη φάση διαμόρφωσης του σχεδίου, οι AI-agents συνήθως διασπούν μια κύρια εργασία σε πολλές μικρότερες υποεργασίες και έχουν προταθεί διαφορετικές στρατηγικές για αυτό το στάδιο. Μια προσέγγιση προτείνει ότι τα προβλήματα μπορούν να λυθούν ακολουθώντας ένα σχέδιο που δημιουργείται στην αρχή. Στη συνέχεια, το σχέδιο εκτελείται βήμα προς βήμα. Μια άλλη προσέγγιση είναι η χρήση προσαρμοστικών στρατηγικών για την αντιμετώπιση σύνθετων εργασιών πιο ρευστά, προγραμματίζοντας και αντιμετωπίζοντας τις υποεργασίες ξεχωριστά. Επιπλέον, ορισμένες παλαιότερες εργασίες τονίζουν τη σημασία του ιεραρχικού προγραμματισμού και της δομής των υποεργασιών σε δενδροειδή μορφή. Η ανάλυση εξαρτήσεων συμπληρώνει την ιεραρχική διάσπαση εντοπίζοντας τις εξαρτήσεις μεταξύ των διάφορων υποεργασιών. Εξασφαλίζει ότι οι εργασίες εκτελούνται με λογική σειρά, αποτρέποντας λάθη λόγω ανεκπλήρωτων προαπαιτήσεων. Αυτό είναι κρίσιμο σε εργασίες όπου το αποτέλεσμα μιας κλήσης API επηρεάζει τις επόμενες κλήσεις, όπως η κράτηση πτήσεων πριν από τα ξενοδοχεία σε μια εφαρμογή προγραμματισμού ταξιδιών. Οι στρατηγικές σειριακής και παράλληλης επεξεργασίας καθορίζουν τους πιο αποδοτικούς τρόπους εκτέλεσης των διασπασμένων εργασιών. Η σειριακή επεξεργασία είναι απαραίτητη για εξαρτώμενες εργασίες, διασφαλίζοντας ότι κάθε βήμα ολοκληρώνεται με τη σωστή σειρά, ενώ η παράλληλη επεξεργασία επιταχύνει την ολοκλήρωση της συνολικής εργασίας χειριζόμενη ανεξάρτητες εργασίες ταυτόχρονα. Παρόλο που οι AI-agents που βασίζονται σε LLM παρουσιάζουν ένα ευρύ φάσμα γενικών γνώσεων, περιστασιακά αντιμετωπίζουν προβλήματα όταν εκτελούν εργασίες που απαιτούν εξειδικευμένη γνώση. Για αυτόν τον λόγο, είναι συχνά καλή ιδέα να ενσωματώνονται εξειδικευμένοι προγραμματιστές υποεργασιών από ορισμένους τομείς. Μετά τον καθορισμό ενός σχεδίου, είναι κρίσιμο να εξεταστούν και να αξιολογηθούν τα οφέλη του. Οι AI-agents που βασίζονται σε LLM βελτιώνουν και αναβαθμίζουν τις στρατηγικές και τις τεχνικές προγραμματισμού τους χρησιμοποιώντας εσωτερικούς μηχανισμούς ανατροφοδότησης, συχνά μαθαίνοντας από προηγούμενα μοντέλα και εργασίες. Οι AI-agents αλληλεπιδρούν ενεργά με τους ανθρώπους για να ευθυγραμμιστούν καλύτερα με τις ανθρώπινες προτιμήσεις. Αυτό επιτρέπει στους ανθρώπους να διασαφηνίζουν παρανοήσεις και να ενσωματώνουν αυτή την εξατομικευμένη ανατροφοδότηση στη διαδικασία προγραμματισμού τους. Επιπλέον, οι χρήστες μπορούν να χρησιμοποιήσουν δεδομένα από πραγματικά ή εικονικά περιβάλλοντα, όπως ενδείξεις από ολοκληρωμένες εργασίες ή παρατηρήσεις μετά τη δράση, για να βοηθήσουν στην επεξεργασία και βελτίωση των σχεδίων τους. Παρακάτω αναφέρουμε ορισμένες βασικές ερευνητικές προσεγγίσεις προς τον προγραμματισμό και τη διάσπαση εργασιών: (Αναφορά στο Σχήμα 2.17.)

### 2.6.1 Συλλογιστική Μονοπάτιου

Αυτό περιλαμβάνει τη διάσπαση εργασιών σε διαδοχικά βήματα.
**Αλυσίδα Σκέψης (CoT)**: Ενσωματώνει βήματα συλλογιστικής για να καθοδηγεί τα LLMs βήμα προς βήμα.

**Αλυσίδα Σκέψης Μηδενικής Εκκίνησης**: Χρησιμοποιεί προτάσεις-εκκινήσεις όπως "σκέψου βήμα προς βήμα" για να εμπνεύσει τα LLMs να δημιουργούν διαδικασίες συλλογιστικής ανεξάρτητα, χωρίς προεπιλεγμένα παραδείγματα.

**Επαναπροτροπή**: Ελέγχει αν κάθε βήμα πληροί τα προαπαιτούμενα πριν συνεχίσει, αναδημιουργώντας σχέδια εάν χρειαστεί.

**ReWOO**: Παρέχει το σχέδιο πρώτα και στη συνέχεια ενσωματώνει παρατηρήσεις.

**HuggingGPT**: Υποστόχοι λύνονται αναδρομικά μέσω πολλαπλών ερωτημάτων προς τα LLMs.

### 2.6.2 Πολυδιαδρομική Συλλογιστική

Δομεί βήματα συλλογιστικής σε δενδροειδή δομή, επιτρέποντας πολλαπλά επόμενα μονοπάτια σε κάθε σημείο απόφασης.

**Αυτοσυνεπής CoT (CoT-SC)**: Χρησιμοποιεί CoT για να δημιουργήσει πολλαπλά μονοπάτια συλλογιστικής, επιλέγοντας την πιο συχνή απάντηση ως την τελική έξοδο.

**Δέντρο Σκέψεων (ToT)**: Σχέδια δημιουργούνται χρησιμοποιώντας δενδροειδή δομή με κόμβους που αντιπροσωπεύουν ενδιάμεσα βήματα και ολοκληρώνονται μέσω στρατηγικών BFS ή DFS.

**RecMind**: Εφαρμόζει μηχανισμό αυτοέμπνευσης που επαναχρησιμοποιεί απορριφθείσες πληροφορίες σχεδιασμού για τη δημιουργία νέων βημάτων.

**Γράφος Σκέψεων (GoT) και Αλγόριθμος Σκέψεων (AoT)**: Επεκτείνουν το ToT εισάγοντας πιο σύνθετες δομές και αλγοριθμικά παραδείγματα στις προτροπές.

**RAP**: Χρησιμοποιεί την Αναζήτηση Δέντρου Monte Carlo (MCTS) για να προσομοιώσει και να επιλέξει το καλύτερο σχέδιο από πολλαπλές επαναλήψεις.

### 2.6.3 Εξωτερικός Προγραμματιστής

Συνδυάζει τις δυνατότητες προγραμματισμού των LLMs με εξωτερικά εργαλεία για εξειδικευμένες εργασίες.

**LLM+P και LLM-DP**: Μετατρέπουν τις περιγραφές εργασιών σε Γλώσσα Ορισμού Τομέα Προγραμματισμού (PDDL) και χρησιμοποιούν εξωτερικούς προγραμματιστές για την εκτέλεση σχεδίων.

**CO-LLM**: Αντιμετωπίζει τους περιορισμούς των LLMs στην εκτέλεση ελέγχου χαμηλού επιπέδου ενσωματώνοντας εξωτερικό προγραμματιστή βασισμένο σε ευρετικές για την εκτέλεση ενεργειών. (Αναφορά στο Σχήμα 2.18.)

Σε αυτό το σημείο, δίνουμε κάποια ιδιαίτερη προσοχή στη μέθοδο που χρησιμοποιείται στο [5]. Η **Αναζήτηση Δέντρου Απόφασης με Βάση την Ανάλυση Βάθους (DFSDT)** λειτουργεί ως οδηγός για ισχυρά LLMs και παρέχει δυνατότητες συλλογιστικής χρησιμοποιώντας τεχνικές προτροπής. Φαίνεται να είναι μια ανώτερη τεχνική σε σύγκριση με τη CoT ([32]) ή τη REACT ([33]), επειδή ξεπερνά περιορισμούς όπως η διάδοση σφαλμάτων (μια λανθασμένη ενέργεια μπορεί να προκαλέσει περαιτέρω σφάλματα και να παγιδεύσει το μοντέλο σε έναν εσφαλμένο βρόχο) και η περιορισμένη εξερεύνηση (η CoT και η ReACT εξετάζουν μόνο μια πιθανή κατεύθυνση, οδηγώντας σε περιορισμένη εξερεύνηση του συνολικού χώρου δράσης) και βελτιώνει την αποδοτικότητα και αποτελεσματικότητα της αναζήτησης προς την Εύρεση Διαδρομής Λύσης. Επιτρέπει στα LLMs να αξιολογούν πολλαπλές ακολουθίες συλλογιστικής και να επεκτείνουν τον χώρο αναζήτησης. Μια άλλη τεχνική για συλλογιστική που προτείνεται από το Πανεπιστήμιο του Princeton και την Google DeepMind είναι το "Δέντρο Σκέψεων" ([34]). Κάθε στρατηγική αντιμετωπίζει διαφορετικές πτυχές και πολυπλοκότητες του προγραμματισμού, από απλά διαδοχικά βήματα έως σύνθετες, αναδρομικές και πολλαπλών διαδρομών συλλογιστικές διαδικασίες, βελτιώνοντας την ευελιξία και την αποτελεσματικότητα των LLMs στην επίλυση προβλημάτων.

## 2.7 ΒΗΜΑ 7: Επαναληπτική Βελτίωση και Ανατροφοδότηση Χρήστη

Το τελευταίο στάδιο της μεθοδολογίας είναι η ενσωμάτωση του συστήματος σε ένα περιβάλλον πραγματικής λειτουργίας όπου το μοντέλο και τα APIs θα λειτουργούν συνδυαστικά για την εκτέλεση πραγματικών εργασιών. Αυτό περιλαμβάνει την τελική αξιολόγηση της απόδοσης του μοντέλου ως προς την ακρίβεια, την ταχύτητα και τη συνολική αποτελεσματικότητα στην παραγωγή των επιθυμητών αποτελεσμάτων. Επίσης, αξιολογείται η ικανότητα του συστήματος να προσαρμόζεται σε αλλαγές και να διαχειρίζεται πιθανά σφάλματα ή απρόβλεπτες συνθήκες, εξασφαλίζοντας την ομαλή και απρόσκοπτη λειτουργία του.

### 2.7.1 Γιατί χρειαζόμαστε Ανατροφοδότηση;

Ο μακροπρόθεσμος προγραμματισμός για την ολοκλήρωση σύνθετων εργασιών είναι απαραίτητος σε πολλές καταστάσεις του πραγματικού κόσμου. Αυτή είναι μια πολύ δύσκολη εργασία, καθώς απαιτεί να ληφθούν υπόψη πολλές περιβαλλοντικές παράμετροι και να εφαρμοστεί προηγμένη συλλογιστική. Λόγω αυτού, η δημιουργία του αρχικού σχεδίου και η προσκόλληση σε αυτό μπορεί συνήθως να οδηγήσει σε αποτυχία ([4],[1]). Ας εμπνευστούμε από τον ανθρώπινο τρόπο δημιουργίας ενός σχεδίου. Οι άνθρωποι μερικές φορές δημιουργούν και τροποποιούν τα σχέδιά τους επανειλημμένα σε απόκριση εξωτερικών ερεθισμάτων. Πρόσφατες έρευνες προτείνουν τη δημιουργία εξωτερικών μονάδων προγραμματισμού που επιτρέπουν στον AI-agent να λαμβάνει ανατροφοδότηση, επιτρέποντας την ανάπτυξη ανθρώπινων δεξιοτήτων.

Η Επαναληπτική Βελτίωση και Ανατροφοδότηση Χρήστη είναι κρίσιμη για το συνεχή κύκλο μάθησης στην Τεχνητή Νοημοσύνη και τη Μηχανική Μάθηση, περιλαμβάνοντας τακτικές ενημερώσεις και βελτιώσεις βάσει εισροών χρηστών και μετρικών απόδοσης. Αυτό εξασφαλίζει ότι τα συστήματα ανταποκρίνονται και προσαρμόζονται στις εξελισσόμενες ανάγκες και περιβάλλοντα των χρηστών. Η ανατροφοδότηση των χρηστών μπορεί να καθοδηγήσει τον AI-agent προς τις προτιμήσεις και τις ανάγκες του χρήστη. Είναι επίσης απαραίτητη για την εξάλειψη των ψευδαισθήσεων. Το πιο απλό παράδειγμα που δείχνει την ανάγκη για ανατροφοδότηση χρήστη είναι η ερώτηση χρήστη "Θα βρέξει αύριο;", στην οποία ο χρήστης έπρεπε να καθορίσει την τοποθεσία του για να παράσχει ο AI-agent χρήσιμες πληροφορίες (Σχήμα 2.13).

### 2.7.2 Συνεχής Μάθηση και Προσαρμογή

Στα συστήματα AI και μηχανικής μάθησης, η συνεχής μάθηση και προσαρμογή είναι κρίσιμη για διάφορους λόγους. Είναι απαραίτητο να προσαρμόζονται στις ανάγκες των χρηστών, επειδή οι άμεσες εισροές κρατούν τα συστήματα ενημερωμένα και ευθυγραμμισμένα με τις προσδοκίες των χρηστών. Χρησιμοποιώντας την παρακολούθηση των μετρικών απόδοσης για τον εντοπισμό προβληματικών σημείων μπορεί να βελτιωθεί η ακρίβεια του συστήματος, οδηγώντας σε βελτιωμένη ακρίβεια και αξιοπιστία. Οι επαναληπτικές ενημερώσεις προωθούν την εξατομίκευση επιτρέποντας στα συστήματα να προσαρμόζονται στις μοναδικές προτιμήσεις και συμπεριφορές των χρηστών. Τέλος, η διατήρηση της τεχνολογικής συνέχειας εξασφαλίζει ότι τα συστήματα μπορούν εύκολα να αξιοποιήσουν τα πιο πρόσφατα εργαλεία και APIs και να παραμένουν ενημερωμένα.
(Αναφορά στο Σχήμα 2.19.)

### 2.7.3 Άλλοι Μηχανισμοί Συλλογής Ανατροφοδότησης

Η απόκτηση χρήσιμων ανατροφοδοτήσεων από τους χρήστες απαιτεί τη χρήση τεχνολογιών όπως τα αναλυτικά στοιχεία και οι φόρμες ανατροφοδότησης. Οι AI-agents καθοδηγούνται αποτελεσματικά χρησιμοποιώντας τους Βασικούς Δείκτες Απόδοσης (KPIs) για την παρακολούθηση της απόδοσης. Επίσης, η δοκιμή A/B χρησιμοποιείται για να λαμβάνονται αποφάσεις βάσει δεδομένων και να βελτιώνεται η εμπειρία των χρηστών κατανοώντας τις προτιμήσεις και τις συμπεριφορές τους. Η υιοθέτηση μιας σχεδιαστικής φιλοσοφίας επικεντρωμένης στον χρήστη εξασφαλίζει ότι τα συστήματα προσαρμόζονται για να ικανοποιούν τις ανάγκες του, οδηγώντας στη δημιουργία διεπαφών που βελτιώνουν την απόδοση του συστήματος και την ικανοποίηση των χρηστών.

### 2.7.4 Ανατροφοδότηση Περιβάλλοντος

Η ανατροφοδότηση περιβάλλοντος είναι ένα ουσιαστικό συστατικό της διαδικασίας λήψης αποφάσεων που βασίζεται σε AI-agents, επηρεάζοντας τον προγραμματισμό και τις ενέργειές τους. Μπορεί να αποκτηθεί από τον εικονικό ή πραγματικό κόσμο μέσω σημάτων ολοκλήρωσης εργασιών και αξιολογήσεων μετά την ενέργεια. Η τρέχουσα έρευνα περιλαμβάνει αυτή την τεχνική. Για παράδειγμα, η υψηλού επιπέδου σκέψη και προγραμματισμός του ReAct εκμεταλλεύεται τριπλέτες σκέψης-δράσης-παρατήρησης, ενώ ο Voyager ακολουθεί μια προσέγγιση "δοκιμής και λάθους" που περιλαμβάνει ενδιάμεση πρόοδο και αυτοεπιβεβαίωση ανατροφοδότησης.

### 2.7.5 Ανατροφοδότηση Μοντέλου

Εκτός από τη χρήση ανατροφοδότησης από τους χρήστες και το περιβάλλον, πρόσφατες έρευνες προτείνουν την εσωτερική ανατροφοδότηση από τους ίδιους τους AI-agents. Ο AI-agent μπορεί να δημιουργήσει έξοδο, να λάβει ανατροφοδότηση και στη συνέχεια να βελτιώσει την έξοδο βάσει της ανατροφοδότησης που έλαβε. Αυτή η διαδικασία τερματίζεται όταν επιτευχθεί το επιθυμητό αποτέλεσμα.

### 2.7.6 Προκλήσεις Ενσωμάτωσης Ανατροφοδότησης

Ορισμένες σημαντικές παρατηρήσεις κατά την ενίσχυση των AI-agents με ανατροφοδότηση είναι:
**Ισορροπία Ανατροφοδότησης και Στόχων**:
Είναι κρίσιμο να ευθυγραμμιστεί η ανατροφοδότηση των χρηστών με τους στόχους του συνολικού συστήματος, ενώ διασφαλίζεται η τεχνική εφικτότητα και η λειτουργικότητα του συστήματος.
**Ιδιωτικότητα Δεδομένων**:
Η ηθική διαχείριση και η συμμόρφωση με τους κανονισμούς είναι απαραίτητες όταν γίνεται διαχείριση της ανατροφοδότησης και των δεδομένων των χρηστών.
**Ποιότητα Ανατροφοδότησης**:
Η αντικειμενική και αντιπροσωπευτική ανατροφοδότηση είναι δύσκολο να αποκτηθεί, αλλά εξέχουσας σημασίας για τη συνεχή βελτίωση του συστήματος.

**Κεφάλαιο 3**

# Παρουσίαση μιας Αρχιτεκτονικής Ενσωμάτωσης των Μεγάλων Γλωσσικών Μοντέλων σε φορητές συσκευές

Η παραπάνω μεθοδολογία των 7 βημάτων μπορεί να χρησιμοποιηθεί για την κατασκευή AI-agents που απαιτούν ένα σχετικά "μεγάλο" μοντέλο που είναι προσβάσιμο μέσω υπηρεσιών νέφους. Τι σημαίνει "μεγάλο"; Παρακάτω συζητάμε τα "σημαντικά μεγέθη" στη μελέτη μας. Για να είμαστε πιο συγκεκριμένοι, δίνουμε μια σύντομη επισκόπηση των απαιτήσεων αποθήκευσης, RAM και CPU των τοπικών LLMs ([35], [36], [37]).

## 3.1 Απαιτήσεις Αποθήκευσης και RAM των LLMs

### 3.1.1 Απαιτήσεις Αποθήκευσης

Ένα μοντέλο με 2,7 δισεκατομμύρια παραμέτρους θα απαιτούσε κανονικά μια χωρητικότητα αποθήκευσης που μπορεί να υποστηρίξει τις παραμέτρους και το λειτουργικό του φορτίο για να αναπτυχθεί τοπικά, π.χ. σε μια κινητή συσκευή. Ανάλογα με τον τύπο δεδομένων των παραμέτρων (τιμές κινητής υποδιαστολής ή ακέραιοι μετά από ποσοτικοποίηση), ένα μοντέλο με 2,7 δισεκατομμύρια παραμέτρους μπορεί να απαιτεί ακόμη αρκετά gigabytes χώρου αποθήκευσης μετά την ποσοτικοποίηση και βελτιστοποίηση για χρήση σε κινητά.

### 3.1.2 Απαιτήσεις RAM

Κατά τη διάρκεια της έγκρισης, η ανάγκη για RAM σχετίζεται στενά με τα λειτουργικά χαρακτηριστικά του μοντέλου. Απαιτούνται ουσιαστικές μέθοδοι μείωσης και βελτιστοποίησης του μοντέλου, όπως η ποσοτικοποίηση, για να λειτουργήσει ένα μοντέλο σωστά σε κινητές συσκευές με μόλις 4GB RAM. Μειώνοντας την ακρίβεια των παραμέτρων του μοντέλου μέσω της ποσοτικοποίησης, απαιτείται λιγότερη RAM για τη φόρτωση και λειτουργία του μοντέλου.

### 3.1.3 Χρήση CPU

Όταν αναπτύσσεται ένα LLM τοπικά, οι δυνατότητες της CPU πρέπει να λαμβάνονται προσεκτικά υπόψιν. Η χρήση CPU για τη λειτουργία ενός LLM μπορεί να είναι υψηλή, ιδιαίτερα όταν δεν υπάρχουν επιταχυντές υλικού όπως GPUs:
**Φόρτος CPU**:
Κάθε βήμα λειτουργίας απαιτεί υπολογιστικές πράξεις υψηλού κόστους, όπως πολλαπλασιασμούς πινάκων και διανυσμάτων και άλλες λειτουργίες τενσόρων. Ο αριθμός των παραμέτρων, η αποδοτικότητα της αρχιτεκτονικής και η πολυπλοκότητα του μοντέλου θα επηρεάσουν τον φόρτο της CPU.
**Τεχνικές Βελτιστοποίησης**:
Η αποκοπή μοντέλου, η ποσοτικοποίηση και η χρήση αποσταγμένων μοντέλων μπορούν να μειώσουν τον φόρτο της CPU κάνοντας τις υπολογιστικές πράξεις που απαιτούνται για τη λειτουργία απλούστερες. Αυτές οι μέθοδοι βοηθούν στην προσαρμογή του μοντέλου για να ταιριάζει στους περιορισμούς λιγότερο ισχυρών συσκευών, όπως tablets ή smartphones.

### 3.1.4 Στρατηγικές Διαχείρισης Μνήμης

**Μνήμη Flash για Αποθήκευση**:
Μεγαλύτερα μοντέλα μπορεί να υπερβαίνουν τη χωρητικότητα DRAM των τοπικών συσκευών. Μια λύση που χρησιμοποιούν μερικοί μηχανικοί είναι η αποθήκευση των παραμέτρων του μοντέλου σε μνήμη flash και η δυναμική φόρτωσή τους στην DRAM κατά τη διάρκεια της λειτουργίας. Αυτή η στρατηγική περιλαμβάνει μια "ανταλλαγή" μεταξύ χωρητικότητας αποθήκευσης και ταχύτητας, καθώς η μνήμη flash είναι πιο αργή από την DRAM.

**Δυναμική Φόρτωση**:
Οι τεχνικές δυναμικής φόρτωσης περιλαμβάνουν τη φόρτωση μόνο των απαραίτητων παραμέτρων στη γρήγορη μνήμη ανάλογα με την ανάγκη. Αυτή η προσέγγιση μπορεί να μειώσει σημαντικά την ποσότητα της απαιτούμενης RAM ανά πάσα στιγμή, επιτρέποντας την εκτέλεση πολύ μεγαλύτερων μοντέλων σε συσκευές με περιορισμένη RAM.

**Βελτιώσεις Αποδοτικότητας**:
Στρατηγικές όπως η χρήση φόρτωσης ευαισθητοποιημένης σε αραιότητα, όπου φορτώνονται μόνο οι μη μηδενικές παράμετροι, μπορούν να βελτιστοποιήσουν τόσο τη χρήση CPU όσο και μνήμης. Αυτές οι μέθοδοι εκμεταλλεύονται τις καλές ιδιότητες των νευρωνικών δικτύων, όπου πολλές παράμετροι μπορεί να έχουν αμελητέες τιμές που δεν επηρεάζουν σημαντικά τα αποτελέσματα.

## 3.2 Ενίσχυση των AI-agents με Τοπικές Προσαρμοσμένες Βάσεις Δεδομένων

Στο εξελισσόμενο τοπίο της Τεχνητής Νοημοσύνης (AI), η εξάρτηση από μοντέλα βασισμένα σε υπηρεσίες νέφους παρουσιάζει σημαντικές προκλήσεις όσον αφορά την καθυστέρηση, την ιδιωτικότητα και τη συνδεσιμότητα. Τα LLMs "σχεδόν χωρούν" σε συσκευές στις μέρες μας, αλλά η ανάγκη για συνδεσιμότητα σε υπηρεσίες νέφους είναι ένα από τα εμπόδια των δυνατοτήτων τους. Η προσέγγισή μας στοχεύει στην ενίσχυση των κινητών συσκευών με δυνατότητες τύπου LLM, χρησιμοποιώντας την έννοια της "τοπικά προ-αποθηκευμένης μακροεντολής/συνάρτησης". Σε αυτό το κεφάλαιο παρουσιάζουμε αυτή τη νέα αρχιτεκτονική εντός συσκευής που στοχεύει στην ενσωμάτωση AI-agents βασισμένων σε LLMs στις συσκευές. Χρησιμοποιούνται τοπικοί υπολογιστικοί πόροι και προσαρμοσμένες βάσεις δεδομένων για αυτή την ενσωμάτωση. Αυτή η προσέγγιση όχι μόνο στοχεύει στη μείωση της εξάρτησης από τις υπηρεσίες νέφους, αλλά επίσης διασφαλίζει την αποδοτική, ασφαλή και σύγχρονη επεξεργασία εργασιών απευθείας στις συσκευές των χρηστών.

### 3.2.1 Επισκόπηση Αρχιτεκτονικής Εντός Συσκευής

Ο στόχος είναι να δημιουργηθεί ένας AI-agent βασισμένος σε LLM που μπορεί να κατανοήσει την πρόθεση του χρήστη και να εκτελέσει ενέργειες βάσει αυτής. Η είσοδος του χρήστη είναι πολυμορφική (αλλά για απλότητα ας υποθέσουμε ότι είναι μόνο κειμενικής μορφής προς το παρόν). Μπορούμε να επιτύχουμε το παραπάνω χρησιμοποιώντας μακροεντολές/συναρτήσεις που αντιστοιχούν σε μια σειρά κλήσεων API. Κάθε φορά που ένας χρήστης χρειάζεται να χρησιμοποιήσει τον AI-agent για μια εργασία που δεν έχει εκτελεστεί ποτέ πριν, παρέχεται μια "Διεπαφή Εκπαίδευσης" στον χρήστη. Στη συνέχεια, ο χρήστης περιγράφει λεκτικά ποιον στόχο (ποια πρόθεση) εκπληρώνει η νέα εργασία. Σε αυτό το στάδιο, ο χρήστης και ο AI-agent "επικοινωνούν" μέσω συνεχούς ανατροφοδότησης για να παρέχουν μια λεπτομερή σειρά ενεργειών (ουσιαστικά μια σειρά κλήσεων API) που πρέπει να γίνουν για να εκπληρωθεί η συγκεκριμένη εργασία. Σε αυτό το σημείο, προστίθεται μια νέα εγγραφή στη Προσαρμοσμένη Βάση Δεδομένων. Αυτή η εγγραφή αποτελείται από 3 κομμάτια πληροφοριών: τον Τίτλο, την Περιγραφή και τη Σειρά Κλήσεων API που αντιστοιχούν στη νέα λειτουργία (Σχήμα 3.3). Ουσιαστικά, ο AI-agent μπορεί τώρα να "θυμάται" ότι για αυτή τη συγκεκριμένη εργασία, εκτελεί τις ακριβείς κλήσεις API που υπάρχουν στη Βάση Δεδομένων. Με αυτό τον τρόπο, ο AI-agent χρειάζεται μόνο να εκτελέσει μια εργασία ταξινόμησης αντί για μια παραγωγική εργασία. Ως αποτέλεσμα, οι εργασίες που ζητούνται πολύ συχνά από τον χρήστη θα εκτελούνται από τον AI-agent πολύ γρήγορα! Τέλος, ένα από τα πιο σημαντικά μέρη αυτής της προσέγγισης είναι η αντιστοίχιση της κειμενικής εισόδου του χρήστη σε μια σειρά από μακροεντολές/συναρτήσεις (προκαθορισμένες εργασίες). Αυτό γίνεται μέσω ενός σημασιολογικού διανυσματικού χώρου ή ενσωματώσεων λέξεων.
(Αναφορά στο Σχήμα 3.1.)

**Ένα απλό παράδειγμα χρήσης**:
Ας υποθέσουμε ότι ο χρήστης θέλει να παραγγείλει είδη παντοπωλείου στη διεύθυνση του σπιτιού του και η είσοδος του χρήστη είναι:
"Παρακαλώ, παράγγειλε 6 από τα φθηνότερα κεσεδάκια γιαούρτι και 0,5 κιλά οποιουδήποτε τυριού από το κοντινότερο μάρκετ στο σπίτι μου".
Αυτή η εργασία θα μπορούσε να αντιστοιχεί στη λειτουργία: *"ORDER_FROM_NEAR-BY_MARKET ( X, Y )"*,
όπου για παράδειγμα το Χ αντιστοιχεί σε *"6 από τα φθηνότερα κεσεδάκια γιαούρτι και 0,5 κιλά οποιουδήποτε τυριού"* και το Υ αντιστοιχεί σε *"το σπίτι μου"*. Χρησιμοποιώντας τεχνικές σημασιολογικής ομοιότητας, ο AI-agent θα μπορούσε να καταλάβει ότι λειτουργίες όπως "PLAY_POPULAR_MUSIC" ή "SEND_EMAIL" δεν είναι σχετικές με τη συγκεκριμένη εργασία και μετά από σημασιολογική αναζήτηση, θα ενεργοποιηθεί η κατάλληλη λειτουργία που αναφέρθηκε παραπάνω.
(Αναφορά στο Σχήμα 3.2.)

### 3.2.2 Συστατικά

**Τοπικά Mini LLMs**:
Μικρότερες, αποδοτικές εκδόσεις LLMs βελτιστοποιημένες για λειτουργία σε περιορισμένους υπολογιστικούς πόρους χωρίς επιπτώσεις στην απόδοση.
**Προσαρμοσμένη Βάση Δεδομένων**:
Μια τοπικά αποθηκευμένη βάση δεδομένων που περιέχει προκαθορισμένες μακροεντολές/συναρτήσεις και δεδομένα εκπαίδευσης, επιτρέποντας στον AI-agent να εκτελεί εργασίες ακόμη και όταν είναι εκτός σύνδεσης.
(Αναφορά στο Σχήμα 3.3.)
**Διεπαφή Εκπαίδευσης και Ανάπτυξης**:
Μια φιλική προς τον χρήστη διεπαφή που επιτρέπει στους χρήστες να εκπαιδεύουν το μοντέλο με νέες εργασίες και να αναπτύσσουν αυτές τις εργασίες ως τοπικές λειτουργίες για άμεση εκτέλεση.
(Αναφορά στο Σχήμα 3.4.)

### 3.2.3 Λειτουργική Ροή

**Επεξεργασία Εισόδου**:
Ο AI-agent επεξεργάζεται τις εισόδους των χρηστών τοπικά, χρησιμοποιώντας ενσωματωμένα εργαλεία NLP για να κατανοήσει και να κατηγοριοποιήσει αιτήματα.
**Ταίριασμα και Εκτέλεση Εργασίας**:
Αξιοποιώντας την τοπική βάση δεδομένων, ο AI-agent ταιριάζει τα αιτήματα των χρηστών με τις σχετικές εργασίες και εκτελεί τις συσχετισμένες ενέργειες χρησιμοποιώντας τις προκαθορισμένες μακροεντολές/συναρτήσεις.

### 3.2.4 Περιπτώσεις Χρήσης και Πρακτικές Εφαρμογές

**Λειτουργικότητα Εκτός Σύνδεσης**:
Υπάρχουν πολλά σενάρια όπου ο AI-agent εντός συσκευής μπορεί να λειτουργεί πλήρως εκτός σύνδεσης, βελτιώνοντας την προσβασιμότητα και την εμπειρία χρήστη.
**Επεξεργασία Δεδομένων σε Πραγματικό Χρόνο**:
Σε σενάρια αντιμετώπισης έκτακτης ανάγκης ή περιβάλλοντα λήψης αποφάσεων σε πραγματικό χρόνο η άμεση επεξεργασία δεδομένων είναι κρίσιμη.
(Αναφορά στο Σχήμα 3.5.)

### 3.2.5 Προκλήσεις και Στρατηγικές Αντιμετώπισης

**Περιορισμοί Υλικού**:
Οι αρχιτεκτονικές εντός συσκευής συχνά αντιμετωπίζουν προκλήσεις που σχετίζονται με περιορισμένη υπολογιστική ισχύ, χωρητικότητα και ενεργειακή αποδοτικότητα. Αυτοί οι περιορισμοί απαιτούν τη χρήση βελτιστοποιημένων μοντέλων AI μέσω τεχνικών όπως η "αποκοπή μοντέλου", η οποία μειώνει την πολυπλοκότητα και το μέγεθος των μοντέλων και η "απόσταξη γνώσης", η οποία μεταφέρει γνώση από μεγάλα μοντέλα σε μικρότερα, πιο αποδοτικά μοντέλα. Αυτές οι στρατηγικές επιτρέπουν στα μοντέλα να λειτουργούν αποτελεσματικά εντός των περιορισμών των πόρων των κινητών και ενσωματωμένων συσκευών.

**Προβλήματα Ασφάλειας**:
Ο κίνδυνος παραβιάσεων δεδομένων και μη εξουσιοδοτημένης πρόσβασης είναι κεντρικός στην ασφάλεια της αρχιτεκτονικής εντός συσκευής και αυτός ο κίνδυνος αυξάνεται όταν τα ευαίσθητα δεδομένα γίνονται αντικείμενο επεξεργασίας τοπικά. Οι ασφαλείς μηχανισμοί πρόσβασης και οι ισχυροί αλγόριθμοι κρυπτογράφησης είναι απαραίτητοι για να μειωθούν αυτοί οι κίνδυνοι. Επιπλέον, η εφαρμογή ολοκληρωμένων μέτρων ασφαλείας όπως η ασφαλής εκκίνηση και ο έλεγχος ταυτότητας πολλών παραγόντων μπορεί να βοηθήσει στην προστασία των δεδομένων από επιθέσεις στον κυβερνοχώρο.

**Ενίσχυση με Υπάρχοντα Πλαίσια**:
Η ενίσχυση της τοπικής επεξεργασίας με τα υπάρχοντα πλαίσια LLM επιτρέπει έναν υβριδικό λειτουργικό τρόπο, συνδυάζοντας τα πλεονεκτήματα της τοπικής επεξεργασίας με τις δυνατότητες που παρέχουν οι υπηρεσίες νέφους. Αυτή η ενσωμάτωση βελτιώνει την αξιοπιστία και την ανταπόκριση του συστήματος, ειδικά σε περιβάλλοντα με διακεκομμένη συνδεσιμότητα. Επιπλέον, επιτρέπει στη συσκευή να εκτελεί κρίσιμες λειτουργίες εκτός σύνδεσης ενώ εξακολουθεί να συγχρονίζεται με τις υπηρεσίες νέφους για ενημερώσεις και προηγμένη επεξεργασία όταν είναι συνδεδεμένη.

**Σύγκριση με Μοντέλα Βασισμένα σε υπηρεσίες νέφους**:
Αυτή η προσέγγιση παρέχει μειωμένη καθυστέρηση και αυξημένη ιδιωτικότητα. Από την άλλη πλευρά, η ανάγκη για περιοδικές ενημερώσεις και η πολυπλοκότητα της αρχικής εγκατάστασης περιορίζουν τη χρησιμότητά της.

### 3.2.6 Πλεονεκτήματα αυτής της προσέγγισης

**Διαθεσιμότητα Εκτός Σύνδεσης**:
Η βασική λειτουργικότητα του AI-agent, συμπεριλαμβανομένης της εκτέλεσης προκαθορισμένων μακροεντολών/συναρτήσεων, δεν απαιτεί συνεχή σύνδεση στο διαδίκτυο με LLMs. Αυτό διασφαλίζει ότι ο AI-agent παραμένει λειτουργικός ακόμη και σε σενάρια εκτός σύνδεσης, εφόσον οι εργασίες δεν απαιτούν δεδομένα σε πραγματικό χρόνο.

**Μειωμένη Καθυστέρηση:**
Αυτή η αρχιτεκτονική παρέχει ελάχιστη καθυστέρηση όταν προκύπτει αίτημα χρήστη. Αυτό οφείλεται στο ότι η σημασιολογική αναζήτηση για την κατάλληλη προκαθορισμένη εργασία θα είναι πολύ ταχύτερη από την πρόσβαση σε ένα LLM ως υπηρεσία νέφους και τη δημιουργία της εργασίας "επί τόπου".

**Προσαρμογή**:
Οι χρήστες μπορούν να προσαρμόσουν τον AI-agent στις συγκεκριμένες ανάγκες τους δημιουργώντας ή τροποποιώντας μακροεντολές/συναρτήσεις. Αυτό το επίπεδο προσαρμογής σημαίνει ότι ο AI-agent μπορεί να εξελιχθεί για να χειριστεί ένα ευρύ φάσμα εργασιών όπως υπαγορεύονται από τις μεταβαλλόμενες απαιτήσεις του χρήστη.

**Επεκτάσεις που εξελίσσονται από την κοινότητα**:
Μια πλατφόρμα ή forum για την κοινή χρήση, συζήτηση και βελτίωση μακροεντολών/συναρτήσεων θα μπορούσε να ενισχύσει σημαντικά τη χρησιμότητα του AI-agent. Χρήστες χωρίς μεγάλη προγραμ-ματιστική εμπειρία θα μπορούσαν να αξιοποιήσουν τη συλλογική γνώση της κοινότητας για να βρουν ή να προσαρμόσουν μακροεντολές που ικανοποιούν τις ανάγκες τους, προωθώντας ένα συνεργατικό οικοσύστημα.

**Κλιμάκωση**:

Η αρχιτεκτονική έχει σχεδιαστεί για να φιλοξενεί μια επεκτεινόμενη βιβλιοθήκη εργασιών και λειτουργιών, κλιμακούμενη από προσωπικές χρήσεις έως - πιθανώς - εφαρμογές επιπέδου επιχείρησης. Η διεπαφή εκπαίδευσης και ο μηχανισμός αντιστοίχισης προθέσεων είναι βασικά συστατικά που διευκολύνουν αυτή την κλιμάκωση.

### 3.2.7 Παραδείγματα Εφαρμογών

**Προσωπικοί Βοηθοί:**
Προσαρμογή του AI-agent για την εκτέλεση προσωπικών εργασιών, όπως η διαχείριση του ημερολογίου, η αποστολή μηνυμάτων ή η παραγγελία προϊόντων. Αυτές οι λειτουργίες μπορούν να προσαρμοστούν ώστε να ικανοποιούν τις μοναδικές ανάγκες κάθε χρήστη, βελτιώνοντας την απόδοση και την εμπειρία χρήστη.

**Βοήθεια σε Εργασιακά Περιβάλλοντα:**
Ενσωμάτωση του AI-agent σε εργασιακά περιβάλλοντα για την αυτοματοποίηση επαναλαμβανόμενων εργασιών, όπως η δημιουργία αναφορών, η διαχείριση δεδομένων ή η αλληλεπίδραση με συστήματα CRM. Αυτό θα μπορούσε να βελτιώσει την αποδοτικότητα και να μειώσει τον φόρτο εργασίας των υπαλλήλων.

**Υποστήριξη σε Εφαρμογές Υγειονομικής Φροντίδας:**
Η χρήση AI-agents για την παροχή υποστήριξης σε ιατρικές εφαρμογές, όπως η παρακολούθηση των συνταγών φαρμάκων, η διαχείριση ραντεβού και η παροχή πληροφοριών για την υγεία. Η τοπική επεξεργασία θα μπορούσε να διασφαλίσει την ιδιωτικότητα των ασθενών και την ασφαλή διαχείριση των δεδομένων υγείας.

**Έξυπνα Σπίτια:**
Ενσωμάτωση των AI-agents σε συστήματα έξυπνων σπιτιών για την αυτοματοποίηση καθημερινών δραστηριοτήτων, όπως η διαχείριση φωτισμού, η παρακολούθηση ασφαλείας και η διαχείριση ενέργειας. Αυτό μπορεί να βελτιώσει την άνεση και την ασφάλεια των κατοίκων.

**Εκπαίδευση:**
Χρήση του AI-agent για την υποστήριξη εκπαιδευτικών εφαρμογών, όπως η παροχή μαθημάτων, η διαχείριση της προόδου των μαθητών και η παροχή εξατομικευμένης βοήθειας στη μάθηση. Αυτό θα μπορούσε να ενισχύσει την εμπειρία μάθησης, αφού ο AI-agent θα ήταν προσαρμορμένος στις ανάγκες κάθε μαθητή.

## 3.3 Συμπέρασμα

Η εισαγωγή της νέας αρχιτεκτονικής εντός συσκευής προσφέρει μια πρωτοποριακή προσέγγιση για την ενσωμάτωση των δυνατοτήτων των LLMs σε κινητές και ενσωματωμένες συσκευές. Χρησιμοποιώντας τοπικούς πόρους και προσαρμοσμένες βάσεις δεδομένων, αυτή η προσέγγιση μπορεί να μειώσει την εξάρτηση από τις υπηρεσίες νέφους, να βελτιώσει την απόδοση και την ασφάλεια και να προσφέρει λειτουργικότητα εκτός σύνδεσης. Ενώ υπάρχουν προκλήσεις που πρέπει να αντιμετωπιστούν, οι στρατηγικές βελτιστοποίησης και οι μηχανισμοί ασφαλείας μπορούν να βοηθήσουν στην υπέρβασή τους, καθιστώντας αυτή την αρχιτεκτονική μια πρακτική και αποδοτική λύση για την ανάπτυξη AI-agents σε πραγματικές εφαρμογές.

# Κεφάλαιο 4

# Πειραματικό Μέρος

Αυτό το κεφάλαιο αποτελεί το Πειραματικό Μέρος αυτής της εργασίας. Εστιάζουμε σε δύο πτυχές. Η πρώτη είναι η επίδειξη του πώς μοιάζει το "ΒΗΜΑ 5" της μεθοδολογίας 7 βημάτων σε ένα σενάριο σύνθεσης μουσικής. Το δεύτερο μέρος εξετάζει την αρχιτεκτονική εντός συσκευής που προτάθηκε στο προηγούμενο κεφάλαιο και αποτελεί "proof of concept" για αυτή την προσέγγιση.

## 4.1 Το παράδειγμα της παρτιτούρας για πιάνο

Ας υποθέσουμε ότι ένας χρήστης χρειάζεται την παρτιτούρα για πιάνο ενός μουσικού κομματιού ή έναν λογικό δακτυλισμό (ανάθεση δακτύλων στις νότες της παρτιτούρας) για μια υπάρχουσα παρτιτούρα. Αυτή η εργασία απαιτεί έναν εξειδικευμένο AI-agent και θεωρείται πολύ περίπλοκη. Παρακάτω παρουσιάζουμε τις προσπάθειες του ChatGPT 4 σε αυτές τις σύνθετες εργασίες και παρέχουμε τα αποτελέσματα της ρύθμισης αυτού του μοντέλου με δεδομένα πραγματικού κόσμου μουσικών φύλλων και τη σύνδεσή του με ένα API για δημιουργία παρτιτούρας πιάνου (π.χ. MuseScore [6]).

### 4.1.1 Αρχική προσπάθεια

Ζητήσαμε από το ChatGPT 4 να παράσχει την παρτιτούρα ενός απλού μουσικού κομματιού (Σχήμα 4.2). Η μελωδία είναι απολύτως σωστή! Αυτό είναι μια μικρή απόδειξη των ενσωματωμένων δυνατοτήτων των σύγχρονων LLMs. Ας προχωρήσουμε σε πιο απαιτητικές εργασίες, όπως η παροχή δακτυλισμού και η μεταγραφή σε μουσικό πεντάγραμμο.

### 4.1.2 Προσθήκη δακτυλισμού

Η παροχή μιας λογικής τοποθέτησης δακτύλων για μια παρτιτούρα πιάνου μπορεί να είναι μια απαιτητική διαδικασία, καθώς υπάρχουν πολλοί κανόνες που πρέπει να ληφθούν υπόψιν (Σχήμα 4.4). Ζητήσαμε από το ChatGPT 4 να παράσχει δακτυλισμό για αυτό το κομμάτι (Σχήμα 4.3). Η μελωδία είναι απολύτως σωστή και η τοποθέτηση δακτύλων είναι λογική (μπορεί να εκτελεστεί από έναν πιανίστα) αλλά όχι βέλτιστη (υπάρχει πιο αποδοτική τοποθέτηση δακτύλων). Για το επόμενο βήμα, δοκιμάσαμε την εκμάθηση με λίγα παραδείγματα στο GPT 4 βασισμένη σε παρτιτούρες πιάνου από το MuseScore και ζητήσαμε μια καλύτερη τοποθέτηση δακτύλων. Το αποτέλεσμα ήταν μια τοποθέτηση δακτύλων που φαίνεται βέλτιστη για έναν πιανίστα (Σχήμα 4.5).

### 4.1.3 Παροχή της σημειογραφίας σε μουσικό πεντάγραμμο

Το τελευταίο βήμα αυτής της εργασίας είναι η σύνδεση του LLM με το API του MuseScore για την απόκτηση της επιθυμητής παρτιτούρας πιάνου. Τα αποτελέσματα αυτής της ενσωμάτωσης φαίνονται στο Σχήμα 4.6.
Το παραπάνω πείραμα είναι απλώς μια σύντομη επίδειξη των απεριόριστων δυνατοτήτων των AI-agents.

### 4.1.4 Προκλήσεις & Μελλοντική Εργασία

Είναι σημαντικό να σημειωθεί ότι το τρέχον παράδειγμα του "Φεγγαράκι μου λαμπρό" είναι το απλούστερο παράδειγμα παρτιτούρας πιάνου. Προχωρημένα κομμάτια στο πιάνο απαιτούν δακτυλισμό για συγχορδίες (όταν παίζονται πολλές νότες ταυτόχρονα). Για αυτό τον λόγο, πολλαπλές μελωδίες όπως αυτή που παρέχεται στο Σχήμα 4.6 συνδυάζονται σε κλασικά κομμάτια. Επίσης, ο χρόνος μεταξύ των διαδοχικών νοτών πρέπει να ληφθεί υπόψιν. Υπάρχουν περιπτώσεις στις οποίες είναι δύσκολο ακόμη και για προχωρημένους πιανίστες να παράσχουν καλή τοποθέτηση δακτύλων ή ακόμα και να επιλέξουν το κατάλληλο χέρι με το οποίο πρέπει να παιχτεί μια νότα. Αυτή η ερευνητική περιοχή είναι πολύ νέα και αναμένεται ότι οι μηχανικοί συνεργαζόμενοι με μουσικούς θα ξεπεράσουν τους προαναφερθέντες περιορισμούς. Η λεπτομερής ρύθμιση των LLMs στα κατάλληλα APIs μουσικής σημειογραφίας και η ενίσχυσή τους με συλλογές δεδομένων από μουσικές παρτιτούρες είναι απαραίτητες για την επίτευξη της επιθυμητής αυτοματοποιημένης σύνθεσης τέτοιων.

## 4.2 Αξιοποίηση της Αρχιτεκτονικής Τοπικής Προσαρμοσμένης Βάσης Δεδομένων για την Εκπλήρωση Αιτημάτων Χρηστών

Αυτή η ενότητα αποτελεί "proof of concept" για τη νέα μεθοδολογία εντός συσκευής που αναλύθηκε στο Κεφάλαιο 3. Αρχικά, χρησιμοποιούμε το ChatGPT 4 για τη δημιουργία εισόδων χρήστη που προσομοιάζουν πραγματικά δεδομένα. Στη συνέχεια, αντιστοιχίζουμε την πρόθεση του χρήστη στις προκαθορισμένες μακροεντολές/συναρτήσεις της προσαρμοσμένης βάσης δεδομένων του Σχήματος 3.3. Για τον σκοπό αυτό, αρχικά χρησιμοποιούμε το μοντέλο BERT για την ταξινόμηση. Στη συνέχεια, εκμεταλλευόμαστε μικρά παραγωγικά μοντέλα. Συγκεκριμένα, χρησιμοποιούμε την έκδοση των 8 δισεκατομμυρίων παραμέτρων του Llama 3 από τη Meta και την έκδοση των 2,7 δισεκατομμυρίων παραμέτρων του Phi-2 από τη Microsoft. Το πλεονέκτημα των παραγωγικών μοντέλων είναι ότι μπορούν πολύ εύκολα να συνδυάσουν δύο ή περισσότερες προκαθορισμένες μακροεντολές/συναρτήσεις ή να παράξουν απλές και "προφανείς" κλήσεις API για την εκτέλεση της εργασίας.

### 4.2.1 Δημιουργία φυσικών εισόδων χρήστη που προσομοιάζουν πραγματικά δεδομένα

Αξιοποιούμε τις δυνατότητες του ChatGPT 4 για να δημιουργήσουμε 100 παραδείγματα εισόδων χρήστη που μοιάζουν με πραγματικές περιπτώσεις χρήσης (Σχήμα 4.9).

### 4.2.2 Αξιοποίηση της Τοπικής Προσαρμοσμένης Βάσης Δεδομένων

Για τις προαναφερθείσες εισόδους χρήστη, μιμούμαστε τη λειτουργικότητα ενός τοπικού AI-agent που χρησιμοποιεί τη βάση δεδομένων του Σχήματος 3.3. Όπως προτείνει η μεθοδολογία, τα χαρακτηριστικά ενδιαφέροντος δεν περιλαμβάνουν τη λίστα των κλήσεων API. Εστιάζουμε στις στήλες "Use Case", "Scenario Description" και "Task Macro/Function" (Σχήμα 4.7).
Αρχικά χρησιμοποιούμε το BERT για την ταξινόμηση. Στη συνέχεια δοκιμάζουμε την ίδια εργασία χρησιμοποιώντας την έκδοση των 8 δισεκατομμυρίων παραμέτρων του Llama 3 από τη Meta και την έκδοση των 2,7 δισεκατομμυρίων παραμέτρων του Phi-2 από τη Microsoft. Τα αποτελέσματα των πειραμάτων εμφανίζονται στο Σχήμα 4.8.
Παρατηρούμε ότι και στις τρεις περιπτώσεις η ταξινόμηση είναι επιτυχής. Ωστόσο, υπάρχει ένα πλεονέκτημα στη χρήση ενός παραγωγικού μοντέλου. Ενώ το BERT ήταν επιτυχές στην εργασία ταξινόμησης, το Llama 3 ήταν σε θέση να χειριστεί τα αιτήματα των χρηστών με μεγάλη ευκολία και επίσης να παράσχει πρόσθετες χρήσιμες εξηγήσεις κατά τη διάρκεια της εκτέλεσης, ακόμη και σε απαιτητικές εργασίες. Το Phi-2 κατάφερε επίσης να ολοκληρώσει την εργασία ταξινόμησης και παρείχε επιπλέον πληροφορίες/εξηγήσεις σε απλές εργασίες. Το γεγονός ότι το μέγεθος του Phi-2 μπορεί εύκολα να χωρέσει σε ένα μέσο smartphone του 2024 είναι μια ένδειξη των απεριόριστων δυνατοτήτων των LLMs και της τεχνολογίας AI. Παρόλο που αυτό το πείραμα δεν αποτελεί απόδειξη της αποτελεσματικότητας της προαναφερθείσας μεθόδου, υποδηλώνει έντονα ότι η ενσωμάτωση των LLMs σε φορητές συσκευές είναι μια έγκυρη, φουτουριστική και χρήσιμη ιδέα.

# Κεφάλαιο 5

# Συμπεράσματα, Συζήτηση & Μελλοντική Έρευνα

## 5.1 Συμπεράσματα

Σε αυτή τη Διπλωματική Εργασία παρουσιάστηκε μια ολοκληρωμένη μεθοδολογία για την ενίσχυση των Μεγάλων Μοντέλων Γλώσσας (LLMs) μέσω της ενσωμάτωσης Διεπαφών Προγραμματισμού Εφαρμογών (APIs), αντιμετωπίζοντας ένα κρίσιμο εμπόδιο στην εφαρμογή τους: την έλλειψη δυναμικής αλληλεπίδρασης με το εξωτερικό ψηφιακό περιβάλλον. Μέσω μιας μεθοδολογία 7 βημάτων, παρέχεται μια δομημένη προσέγγιση για την επιλογή και χρήση των APIs από LLMs, επεκτείνοντας έτσι τη χρησιμότητα και την αποτελεσματικότητά τους σε πραγματικές εφαρμογές. Ένα σημαντικό αποτέλεσμα αυτής της έρευνας είναι η πρόταση μιας αρχιτεκτονικής εντός συσκευής που έχει σχεδιαστεί για να χρησιμοποιεί μικρά μοντέλα από την κοινότητα Hugging Face, επιτρέποντας την αποτελεσματική λειτουργία AI-agents σε φορητές συσκευές. Αυτή η καινοτομία αυξάνει την αυτονομία και την αντίληψη των AI-agents και ενισχύει τις δυνατότητές τους σε περιβάλλοντα με περιορισμένους πόρους. Οι πρακτικές εφαρμογές αυτών των μεθοδολογιών επεκτείνονται σε διάφορους τομείς, όπως η τεχνολογία, η υγειονομική περίθαλψη και η εξυπηρέτηση πελατών.

## 5.2 Συζήτηση & Μελλοντική Έρευνα

Αυτή η έρευνα συμβάλλει τόσο στη θεωρητική επέκταση όσο και στην πρακτική εφαρμογή της Τεχνητής Νοημοσύνης, ενισχύοντας τις δυνατότητες αλληλεπίδρασης των LLMs με APIs. Επιτρέποντας σε αυτά τα μοντέλα να έχουν πρόσβαση σε δεδομένα σε πραγματικό χρόνο, να αλληλεπιδρούν δυναμικά με άλλα ψηφιακά συστήματα και να εκτελούν σύνθετες εργασίες αυτόνομα, η παρούσα εργασία διευρύνει τις πιθανές χρήσεις των LLMs σε διάφορους τομείς.

Ενώ αυτή η έρευνα παρέχει μια σταθερή βάση για την αλληλεπίδραση των LLMs με APIs, αντιμετωπίζει επίσης περιορισμούς, συμπεριλαμβανομένων των πιθανών προκαταλήψεων στην επιλογή μοντέλων και των προκλήσεων κλιμάκωσης που σχετίζονται με τον όγκο των κλήσεων API. Η εμβέλεια της ενσωμάτωσης API επικεντρώνεται επί του παρόντος σε προκαθορισμένες εργασίες, οι οποίες μπορεί να μην επεκτείνονται σε όλες τις πιθανές εφαρμογές.

Περαιτέρω έρευνα θα πρέπει να διερευνήσει την ενσωμάτωση ενός ακόμη ευρύτερου φάσματος APIs, επεκτείνοντας τη λειτουργική εμβέλεια των LLMs. Επιπλέον, η διερεύνηση της εφαρμογής αυτών των μεθοδολογιών σε διάφορους τομείς και αρχιτεκτονικές LLM θα ήταν επωφελής. Υπάρχει επίσης ανάγκη για μελέτη των μακροπρόθεσμων δυνατοτήτων μάθησης και προσαρμογής των AI-agents σε δυναμικά περιβάλλοντα για την ενίσχυση των εξελικτικών δυνατοτήτων τους με βάση τη συνεχή ανατροφοδότηση.

Λαμβάνοντας υπόψιν το "Νόμο του Moore" και τις ταχείες προόδους στις υπολογιστικές δυνατότητες, αναμένεται ότι η ανάπτυξη LLMs απευθείας σε συσκευές θα γίνει εφικτή στο άμεσο μέλλον. Αυτή η εξέλιξη πιθανότατα θα επιταχύνει τις πρακτικές εφαρμογές των AI-agents, καθιστώντας τους πιο προσιτούς και αποτελεσματικούς.

Οι μεθοδολογίες που αναπτύχθηκαν σε αυτή την εργασία θα μπορούσαν να έχουν σημαντικό αντίκτυπο σε πολλούς κλάδους παρέχοντας λύσεις Τεχνητής Νοημοσύνης που είναι πιο δυναμικές, ευφυείς και προσαρμοστικές. Η μελλοντική έρευνα προς την κατεύθυνση της ενσωμάτωσης αυτών των βελτιωμένων LLMs σε τομείς όπως η υγειονομική περίθαλψη, για εξατομικευμένες αλληλεπιδράσεις, ή τα χρηματοοικονομικά, για λήψη αποφάσεων σε πραγματικό χρόνο, θα μπορούσε να αποδειχθεί μεταμορφωτική.

Αυτή η εργασία όχι μόνο προάγει τον τομέα της Τεχνητής Νοημοσύνης, αλλά επίσης θέτει ένα θεμελιώδες πλαίσιο για μελλοντική έρευνα, με δυνατότητα να οδηγήσει στην κατασκευή πραγματικά αυτόνομων και ευφυών AI-agents ικανών να λειτουργούν σε μεγάλο εύρος περιβαλλόντων και εργασιών.

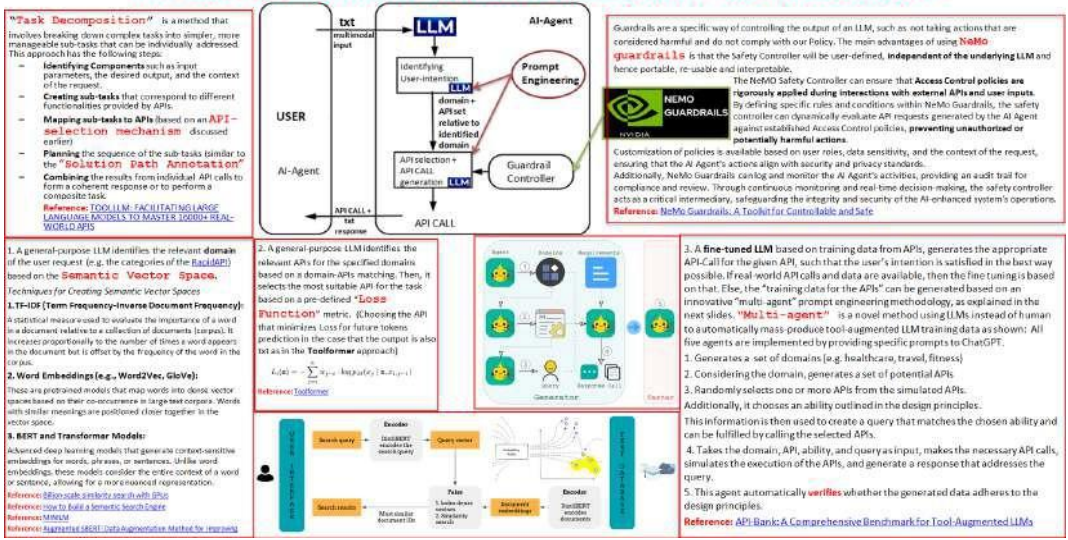

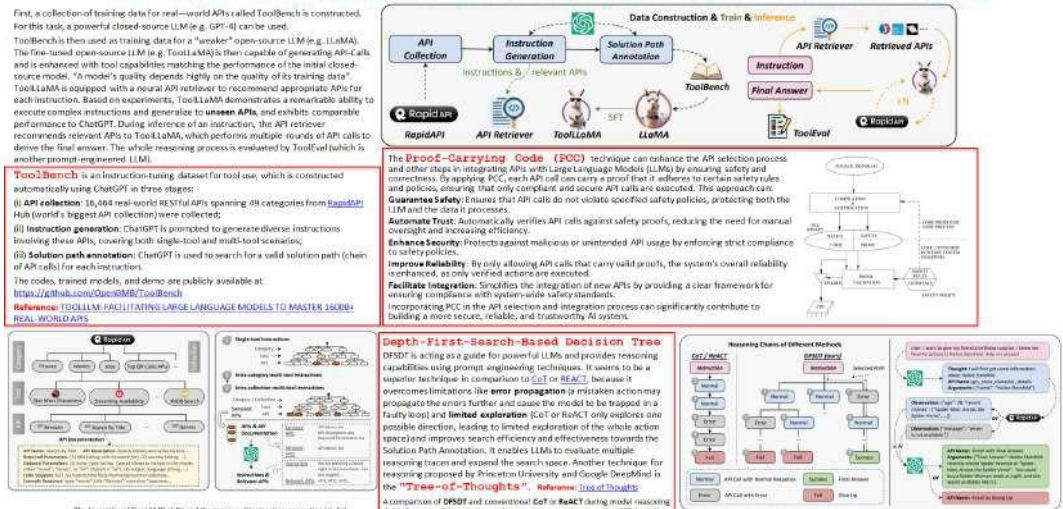

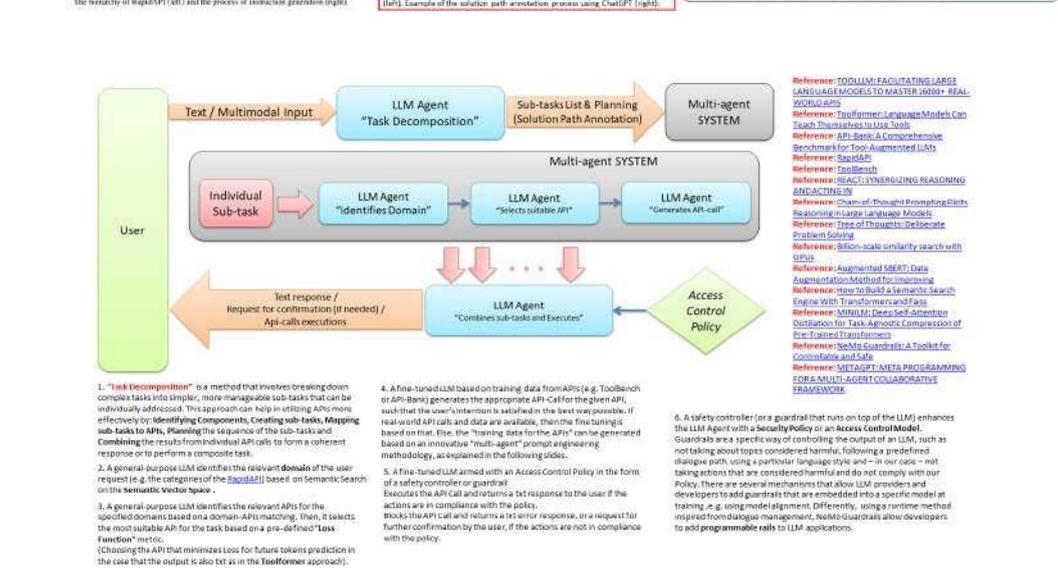

**Κείμενο στα αγγλικά**

# Chapter 1

# Introduction

## 1.1 Why LLM-based AI-agents?

Large Language Models (LLMs) have revolutionized various aspects of engineering and science and are considered a milestone in the area of Artificial Intelligence (AI) and Natural Language Processing (NLP). Beginning with simpler models capable of understanding and generating text based on statistical methods, this field has witnessed the emergence of increasingly sophisticated architectures. These advancements have led to the development of models like GPT-3, GPT-4 [7], BERT [8] and LLaMA [9], which not only grasp the intricacies of human language but also generate coherent, context-aware and domain-specific text. The strengths of these models lie in their deep understanding of language instances, their ability to generate human-like text and their application across diverse tasks, from translation to content creation.

However, despite their impressive capabilities, LLMs face notable limitations. Their knowledge is "frozen" at the time of training, rendering them unable to access or process real-time information or interact dynamically with the environment. Furthermore, they often generate responses based on patterns in their training data, which can lead to hallucinations, inaccuracies, or factual contradictions. Initially, researchers have focused on instruction tuning towards advanced language processing skills, often ignoring the crucial domain of tool use, leading to models that do not replicate the human way of learning when tested in unconstrained, open-domain contexts [10, 11, 12, 13]. Recent studies tend to achieve more human-like results [14, 7, 15, 16, 17, 18, 19, 9], since researchers are enhancing LLMs with memory, planning and reasoning capabilities. Mechanisms that can bridge the gap between static knowledge and dynamic, real-time information retrieval and processing are of predominant interest. The most popular approach towards this direction involves equipping LLMs with the privilege of interacting with external tools and applications. The enhancement of LLMs with Application Programming Interfaces (APIs) gives birth to the research area of LLM-based AI-agents, a field of increasing specialization and exponential growth. It is worth noting that (optimistic) researchers anticipate that Artificial General Intelligence (AGI) will be accomplished via self-directed planning and behaviours and autonomous AI-agents seem to be the most viable method to get there [19, 14].

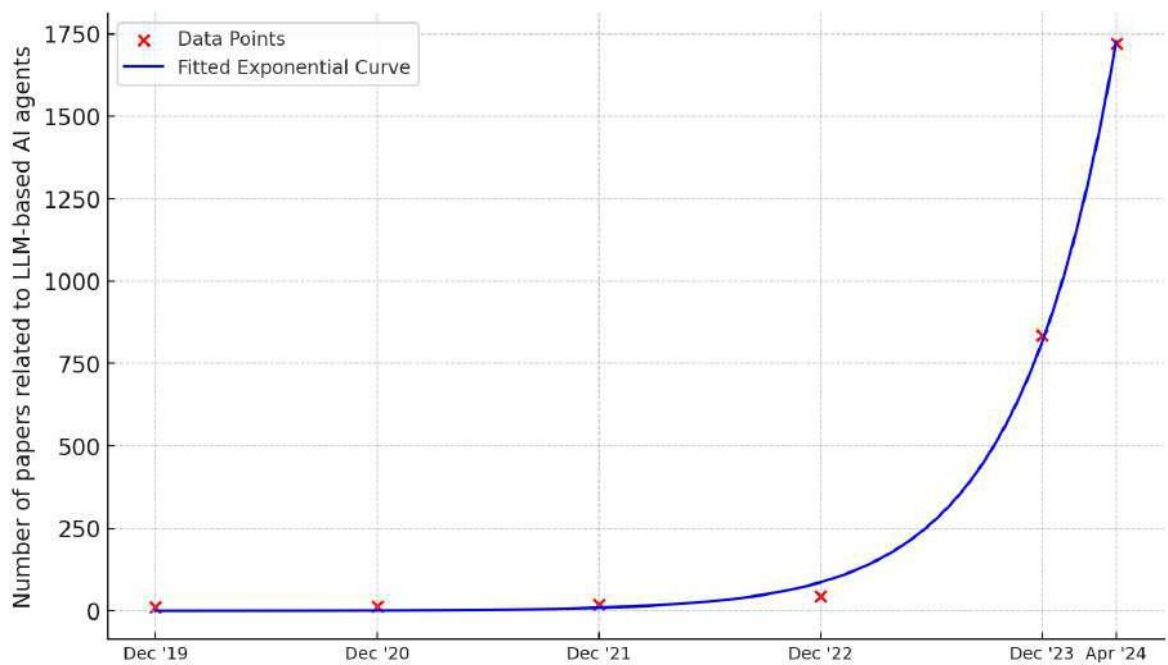

**Figure 1.1:** The exponential growth of the field of LLM-based AI-agents. We present the cumulative number of papers published from December 2019 to April 2024.

## 1.2 LLMs & APIs

APIs serve as a bridge that extends LLM capabilities beyond their inherent limitations by allowing interaction with external systems, databases and services, enabling LLMs to fetch real-time data, perform computations and even execute transactions. This integration opens up new possibilities for LLMs, transforming them from static models to dynamic agents capable of more complex and contextually aware operations [20, 17, 5]. Benchmarks like API-bank have been created to evaluate the increasing amount of new models. The synergy of LLMs and APIs marks a pivotal shift towards creating more intelligent, adaptable and versatile AI systems. While the journey of LLM development has been marked by significant breakthroughs, the path forward involves overcoming the challenges of their limitations through innovative solutions like API integration, setting the stage for the next leap in AI capabilities.

## 1.3 Related Work & Research

Cutting-edge advancements of LLMs and AI include Autonomous AI Software Engineers that can understand high-level human instructions, break them down into steps, research relevant information and write code to achieve a given objective, like OpenDevin [21] and Devika [22]. Also, LLM fora like the Hugging Face community provide a vast variety of specialized pre-trained models oriented towards specific functionalities [23]. With the rapid growth of this research field, multiple extensive surveys have arisen, giving deep insights into a wide range of topics. [24] and [25] provide a comprehensive introduction to the background, significant discoveries and mainstream technologies. On the other hand, [26] focuses particularly on the applications of LLMs in various downstream processes, as well as the obstacles involved with their adoption. Aligning LLMs with human intelligence is a growing topic of study that aims to overcome issues like biases and hallucinations [27]. Reasoning is another important part of intelligence that impacts decision-making, problem-solving and cognitive ability [28]. According to [29], Augmented Language Models (ALMs) can improve language models by including reasoning and tool capabilities. Evaluating the performance of large-scale models becomes more important as their utility grows and there are various metrics that are used for this purpose [30]. This work focuses on the advancement of LLMs through API embracement.

## 1.4 Structure of the Thesis

The thesis is structured as follows: Chapter 1 sets the stage by discussing the significance of LLM-based AI-agents and their integration with APIs. Chapter 2 explores the 7-step methodology employed to enhance LLMs with APIs. It begins with "Model Selection," which details the criteria for choosing LLMs suitable for API integration based on performance, interoperability and adaptability. This is followed by "Enhancing with External Tool Knowledge," where training LLMs with API documentation to execute API-related commands effectively is discussed. "Designing a Multi-Stage

Pipeline" outlines a structured approach for integrating APIs through stages like task identification and API discovery. "API Selection Mechanism" describes the criteria and mechanisms for selecting suitable APIs for specific tasks. "Generating API calls" discusses techniques to ensure API calls are syntactically correct and semantically aligned with user intents. "Task Decomposition" covers strategies for breaking down complex tasks into simpler sub-tasks and "Iterative Improvement and User Feedback" is about refining API interactions through continuous user feedback. Chapter 3 delves into the practical application of the methodologies discussed earlier, introducing an innovative on-device architecture designed for LLMs. This chapter discusses the technical and operational aspects of implementing LLMs on portable devices, emphasizing the storage, RAM and CPU requirements necessary for efficient functionality. It also explores the enhancement of AI-agents with local custom databases, providing a detailed examination of the components and operational flow of the proposed architecture. By addressing the challenges and mitigation strategies, this chapter showcases the advantages of on-device implementation, such as improved response times and reduced dependency on constant connectivity. In Chapter 4, practical applications and use cases are examined to illustrate how these architectures facilitate real-world deployment of AI-agents, paving the way for more robust and context-aware applications. Finally, Chapter 5, "Conclusion, Discussion & Future Research," summarizes the findings, discusses the implications and suggests future research directions. Together, these chapters comprehensively address the novel topic of enhancing LLMs with API capabilities.

# Chapter 2

# Methodology Towards Enhancing LLMs with APIs

This Chapter elaborates on a methodological framework designed to augment Large Language Models (LLMs) with Application Programming Interfaces (APIs), fostering more dynamic and interactive AI systems. This section begins with the strategic selection of LLMs optimized for API integration, emphasizing not only linguistic proficiency but also compatibility with external APIs. The approach progresses through 7 steps, starting with enhancing LLMs with external tool knowledge through targeted training and fine-tuning using API documentation. Further on, the methodology suggests a structured multi-stage pipeline that systematizes API integration, ensuring that interactions are both efficient and effective. Key to this process is the robust mechanism for API selection, that is taking into account both task relevance and operational capabilities to ensure optimal alignment with the functionalities of the LLM. Subsequent steps detail the generation of API calls that ensure precise task execution. At this point, it is important to notice that for addressing complex tasks, the methodology proposes techniques for decomposition into manageable components, thereby streamlining the problem-solving process. The concluding stages focus on iterative improvements and leveraging user feedback, which are vital for continually refining the Human Computer Interactions (HCI) and adaptability in practical applications. This framework not only outlines a pathway for enhancing LLMs with APIs but also underscores the evolving capabilities of AI systems to perform autonomously and robustly in diverse real-world scenarios.

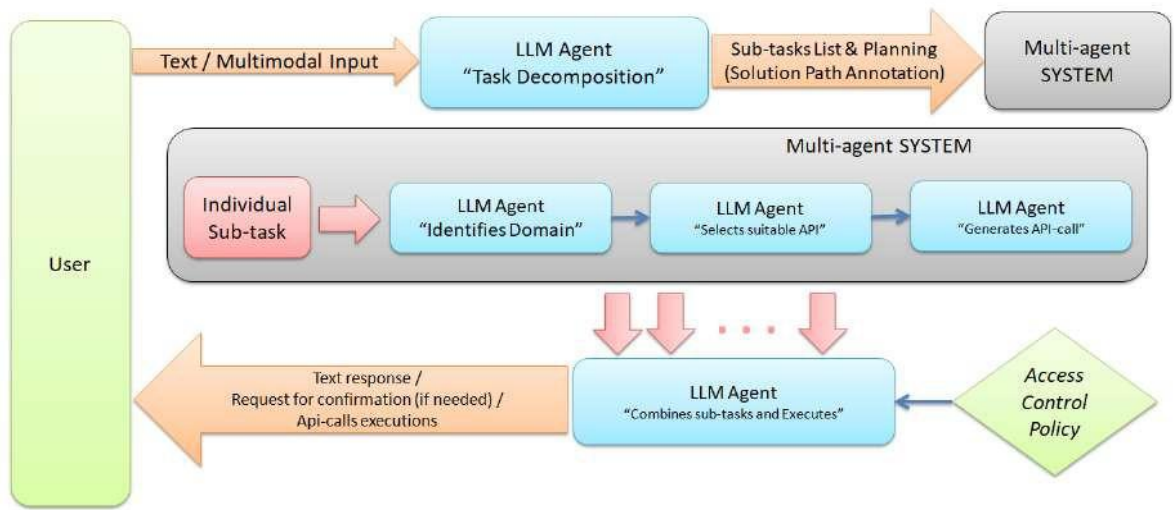

**Figure 2.1:** Overview of an LLM-based AI-agent Architecture

## 2.1 STEP 1: Model Selection

Selecting Large Language Models (LLMs) for API integration involves evaluating several critical criteria to ensure that the models are not only proficient in natural language processing but are also adaptable and efficient in interacting with external APIs. Based on current research [18, 4] and on AI-standards, the following aspects should be taken into consideration:

**Performance on Language Tasks:**
The ability of an LLM to understand and generate human-like text is crucial. The model must exhibit high performance across a variety of language tasks, including (but not limited to) comprehension, translation, summarization and question-answering.

**Interoperability with APIs:**
The chosen LLM should be capable of understanding instructions that involve external tools and translating these instructions into API calls. This requires the model to be able to identify the relevant API and format the request appropriately.

**Adaptability and Learning Capability:**
An ideal LLM for API integration should be adaptable, capable of learning from interactions and improving over time. This involves the model's ability to incorporate feedback from API interactions into future responses, enhancing its performance and accuracy.

**Generalization Across Domains:**
The model should demonstrate strong generalization capabilities, meaning it can effectively apply its knowledge and API integration skills across a wide range of domains and tasks, not limited to the scenarios it was explicitly trained on.

**Scalability and Efficiency:**
Given the potential high volume of API calls and the complexity of processing responses, the selected LLM needs to be scalable and efficient in its operations. This ensures that the integration remains practical and sustainable even as the scope of tasks expands.

**Ethical and Bias Considerations:**
The model must handle data ethically, especially when interfacing with APIs that may access sensitive or personal information. It should also be evaluated for biases to mitigate any negative impacts that could arise from its responses.

**Up-to-dateness:**
The LLM should be trained on a recent dataset, ensuring that its knowledge base is as current as possible. This is crucial for tasks that require up-to-date information, which the model might need to retrieve through APIs.

**Memory and Storage Requirements:**
The LLM should have manageable memory and storage demands to ensure compatibility with existing infrastructure without requiring excessive upgrades. For deployment, the model should efficiently utilize available system memory (RAM) and storage, which is crucial when handling large datasets or multiple simultaneous API calls. This also impacts the speed at which the model can access and process data.

**Response Time and Latency:**
It is important that the LLM exhibits low latency in understanding and processing API calls. Quick response times are essential for maintaining a fluid user experience, especially in applications that require real-time interaction or high-speed data processing. This involves optimizing the model's architecture and deployment setup to minimize delays.

**Prompt Flexibility and Limitations:**
The LLM should accommodate a wide range of prompt lengths and complexities, allowing users to make requests using both concise and detailed descriptions. However, it's important to establish the limitations regarding the length of prompts the model can effectively process without a drop in performance or accuracy, as this impacts its utility across different applications.

**Integration and Compatibility with Existing Systems:**
The model should easily integrate with existing software architectures and technology stacks. This includes compatibility with various programming languages and frameworks commonly used for API integration. It should support standard interfaces and protocols to streamline deployment and maintenance.

**Cost Efficiency:**

Considering the operational costs, including computational resources and potential subscription fees (if using a hosted service), the chosen LLM should offer a balance between performance and cost. This criterion is critical for ensuring that the model remains a viable solution as the scale of API usage grows.

**Support and Community Ecosystem:**

An active support community and the availability of comprehensive documentation are invaluable for troubleshooting and enhancing the model. Access to a robust ecosystem, such as that provided by Hugging Face, can significantly ease the integration process and provide resources for continuous improvement and learning.

The above criteria provide a holistic evaluation of potential LLMs, taking into account not just their intellectual capabilities but also practical considerations that affect their deployment and long-term usability in API integration scenarios. [18] and [5, 9] exemplify these criteria by showcasing models specifically designed for tool usage, including API interactions. They demonstrate superior performance in understanding complex instructions and executing tasks involving external tools, thereby serving as a benchmark for what an LLM should be capable of in the context of API integration.

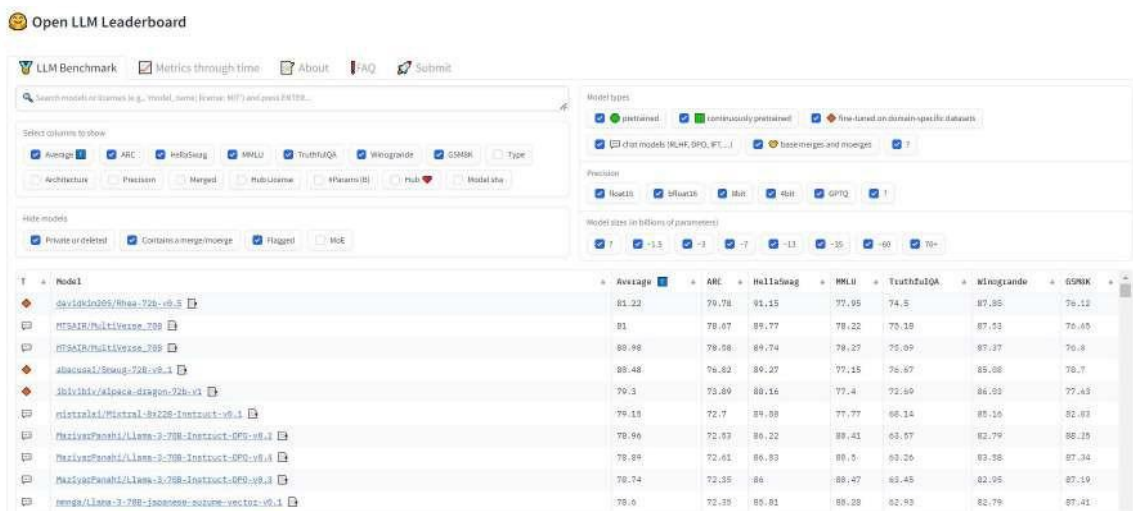

**Figure 2.2:** The Hugging Face Community Leaderboard (as on 9th May 2024)

A very common approach towards selecting an appropriate LLM is using the information available on the leaderboard of the Hugging Face community [23, 17]. As shown in Figure 2.2, there is a wide variety of benchmarks for specialized tasks including summarization, classification, embeddings and question answering.

The page includes a search bar for finding specific models or licenses. Below this, there is a section allowing users to select which data columns to display, such as average performance or results from specific benchmarks like ARC, Hellaswag, MMLU and others. It also features filters to hide certain types of models, including private, deleted, merged, or flagged entries. For further customization, the platform offers filters by model type, such as pretrained or continuously pretrained models, those fine-tuned on domain-specific datasets and other categories. Precision filters allow selection based on the computational precision of the models (e.g. float16, bfloat16) and there are options to filter models by size, categorized by the number of parameters. This leaderboard serves as a comprehensive resource for comparing and analyzing the capabilities and performances of various large language models, catering to researchers, developers and AI enthusiasts.

A useful task before choosing the desired model would be creating an empirical compact comparison table. This table should consist of some candidate models assessed on some useful functionalities. In that way, a rough overview of available options is created. For example, one such table is presented in Figure 2.3.
To fill in this table with accurate data, the following should be taken into consideration:
1. Identifying the tasks that are most relevant to the specific project.
2. Gathering or referencing empirical data from the literature on how each model performs on these tasks. For capabilities, performance metrics and detailed comparisons, consulting specific and recent publications or benchmarks is essential.
3. Updating the table accordingly.

| Model | Language Understanding | Text Generation | Efficiency | Versatility | Ease of Use |
|---|---|---|---|---|---|
| BERT | Very High | Moderate | Low (Resource-intensive) | High | High |
| GPT-3 | High | Very High | Low (Resource-intensive) | High | Moderate |
| RoBERTa | Very High | Moderate | Low (Resource-intensive) | High | High |
| DistilBERT | High | Moderate | High (Resource-efficient) | Moderate | High |
| T5 | Very High | High | Moderate | Very High | Moderate |
| XLNet | Very High | High | Low (Complex training) | High | Moderate |

**Figure 2.3:** Example of a Compact Comparison Table

## 2.2 STEP 2: Enhancing with External Tool Knowledge

Enhancing LLMs with external tool knowledge is the most critical step towards creating more versatile and practical AI systems. This process involves training or fine-tuning LLMs to not only understand natural language but also to interpret and execute commands that involve interacting with external APIs. If real-world data for the API of interest exist, then they are used in order to fine-tune an open-source LLM. This is not always the case. APIs that are not widely used, or that have very recently been released, do not have a large enough amount of real-world data associated with them. This problem is solved using the methodology proposed in [5]. To be more specific, a powerful LLM (like ChatGPT4 [7]) is used in order to create good quality data for the specific APIs, by following a careful multi-agent process. These good quality data are then used for fine-tuning an open-source model. The final model is very well trained in the context of the API of interest. The above method (as also shown in Figure 2.4) reminds of the common data science adage:

*"The quality of the model depends on the quality of the training data."*

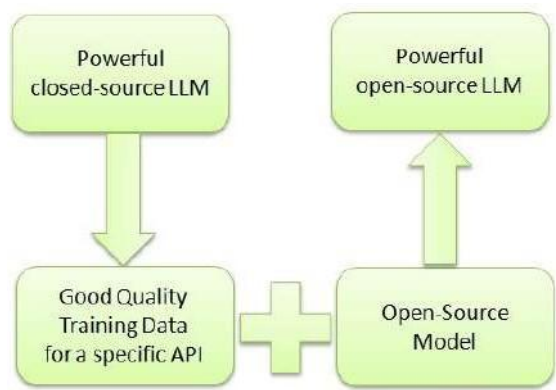

**Figure 2.4:** Training open-source LLMs

More traditional methods include:
**Training with API Documentation and Examples:**
One of the primary methods for incorporating API knowledge into LLMs involves using API documentation and example API calls as part of the training data. This allows the model to learn the structure, syntax and usage patterns of various APIs. By understanding how API requests are formulated and what responses they generate, LLMs can better predict the necessary actions to take when tasked with an operation that requires API interaction.
**Fine-tuning on Task-Specific Scenarios:**
Beyond basic training, fine-tuning LLMs for task-specific scenarios involving API access can drastically improve their ability to communicate with external tools. This requires building datasets that imitate real-world actions involving API calls and then utilizing them to fine-tune the model. This tailored approach teaches the model the intricacies of when and how to leverage APIs to achieve certain goals.
**Simulated API Interaction Environments:**
Another strategy is to use simulated environments where LLMs can interact with mock APIs. These simulated APIs can provide structured responses to the model's requests, allowing it to practice API interactions in a controlled setting. This method helps in reinforcing the model's understanding of API dynamics without the need for actual API calls, which might have rate limits or increased costs.
**Incorporating API Semantic Understanding:**
One of the most important aspects of the LLM-API interaction is the semantic relationship between natural language instructions and API functionality. This involves mapping user intents to specific API actions, a process that can be facilitated through supervised learning techniques where the model is trained on pairs of natural language instructions and the corresponding API calls.

**Leveraging Transfer Learning:**

Transfer learning techniques can be employed to transfer knowledge from pre-trained models that are already proficient in API interactions to new models. This approach can reduce the amount of data required for training and speed up the fine-tuning process, leveraging the pre-existing knowledge embedded in the models.

**Continuous Learning and Adaptation:**

Finally, enabling LLMs to continuously learn from their interactions can further enhance their API knowledge. By analyzing the success and failure of API calls and integrating this feedback into subsequent training cycles, LLMs can adapt and improve their API interaction capabilities over time.

## 2.3 STEP 3: Designing a Multi-Stage Pipeline

Designing a multi-stage pipeline for API integration into Large Language Models (LLMs) is a structured approach that systematically enhances the model's ability to interact with external APIs. Inspired by the processes discussed in [3] and [18], this methodology ensures that LLMs can efficiently understand, select and utilize APIs to perform real-world tasks, thereby significantly expanding their capabilities beyond mere text generation or understanding. Here, we outline the critical stages in this pipeline and the importance of each step.

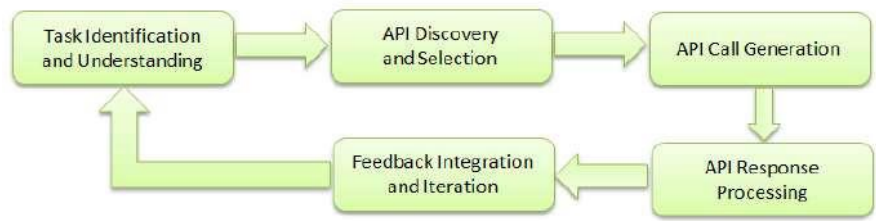

**Figure 2.5:** Multi-Stage Pipeline Overview

**1. Task Identification and Understanding**
The initial stage involves accurately identifying the user's intent and understanding the specific task at hand. This step is crucial as it sets the foundation for selecting the appropriate API(s) to execute the task. It involves natural language understanding techniques to parse the user's request and extract relevant information, such as the desired action and any specific parameters required for the task.

**2. API Discovery and Selection**
Finding and choosing the best API or APIs to complete the task is the next step after it has been properly stated. This entails matching the features provided by the accessible APIs with the requirements of the task. The ability of the API to complete the work, its dependability, response speed and maybe cost are among the decision criteria. This stage might make use of an API database or registry where APIs are arranged according to their functionality and domains (e.g. the "Rapid's API Hub" [31] as used in [5] and [3]).

**3. API call Generation**
With the appropriate API selected, the pipeline then focuses on generating the API call. This step requires transforming the task parameters and requirements identified in the first step into a format that suits to the API's specification. It involves generating the endpoint URL, the method (GET, POST, etc.) and any required headers or body parameters. This step is critical as it directly affects the success of the API interaction.

**4. API Response Processing**
After the API call is executed, the LLM must process the received response. This stage involves parsing the API's response, extracting the relevant data and transforming it into a format that can be easily understood and utilized by the user. Depending on the complexity of the API response, this may involve filtering, sorting, or aggregating data.

**5. Feedback Integration and Iteration**
The final stage involves integrating feedback from the API interaction into future operations. This could involve adjusting the API selection process based on performance or updating the method for generating API calls based on success rates. Continuous learning mechanisms can be implemented to refine each stage of the pipeline based on real-world interaction data.

Inspired by [3], this structured approach towards API integration is paramount for creating LLMs that can effectively utilize external APIs. It ensures a systematic process from understanding user requests to executing tasks via APIs, thereby enhancing the LLMs' functionality and applicability in real-world scenarios. One such scenario that is of practical interest is "booking holidays". Below we explain how

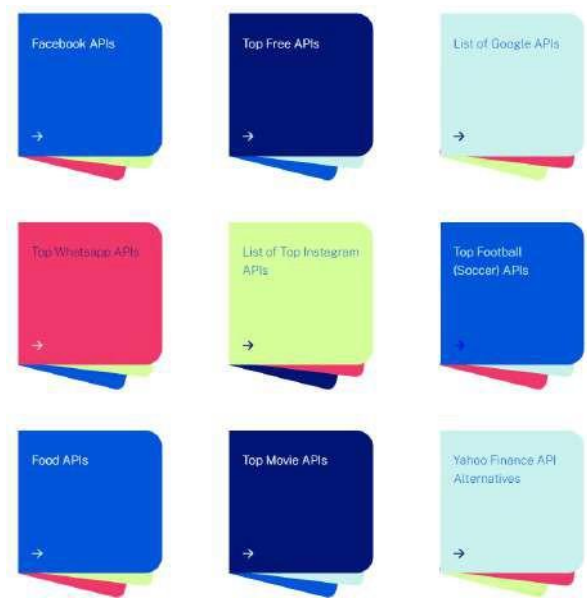

**Figure 2.6:** Some of the categories/domains provided by Rapid's API Hub

every step applies to the task of booking holidays through an API by utilizing an LLM.

If a user inputs "I want to book a beach holiday in Greece for the first week of July," the LLM discerns the task (booking a holiday), the type (beach), location (Greece) and time (first week of July). For booking a holiday, the LLM selects APIs from a travel domain for flights and hotels, ensuring they cover Greek destinations and offer options for beachfront accommodations. To find beach hotels in Greece, the LLM crafts a GET request to the hotel booking API, including parameters for the location ("Greece"), dates ("first week of July") and filters for "beachfront" properties. Then, the LLM processes the list of beachfront hotels, selecting a few options based on ratings, availability and price. Finally, it presents these choices to the user in a concise and informative manner. It is important to notice that based on the user's selection, the system learns preferences, such as prioritizing lower-cost options or specific amenities, enhancing future task executions. This short analysis of each step, demonstrated in the "booking holidays" example, underscores the complexity and potential of integrating LLMs with external APIs to perform real-world tasks, showcasing a significant advancement in the capabilities of AI-agents.

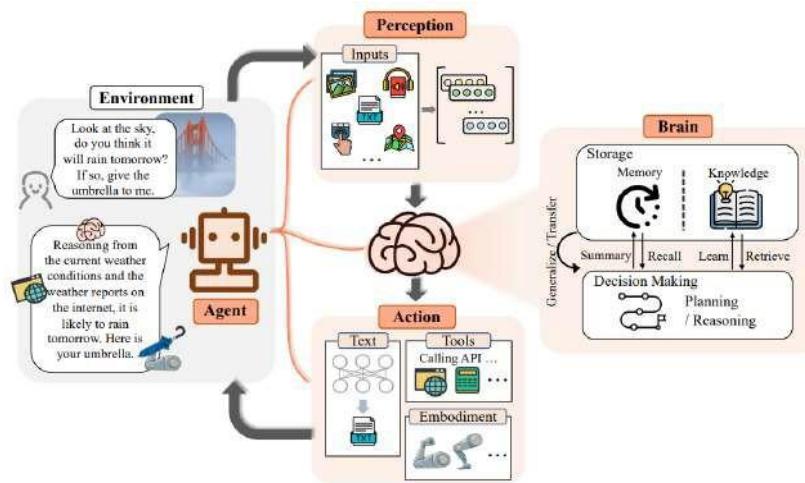

**Figure 2.7:** Overview of the construction of an AI-agent (Image source: [1])

## 2.4 STEP 4: API Selection Mechanism

A crucial step in combining Large Language Models (LLMs) with external APIs to complete certain tasks is the API selection mechanism. This mechanism optimizes for efficiency, reliability and accuracy by going through a set of criteria and steps to make sure the selected API is the most appropriate for the task at hand. The following are the main criteria and mechanisms to be considered:

### 2.4.1 Criteria for API Selection

**1. Relevance to Task:** The goals of the task must be directly supported by the API. For example, an API needed for travel booking needs to be able to handle bookings for flights and hotels, not for unrelated services.
**2. Coverage and Scope:** In order for the API to deliver pertinent data or activities, it must cover all necessary services and geographic areas.
**3. Performance and Reliability:** APIs with high reliability, low latency and rapid response times are preferred to offer a smooth user experience.
**4. Data Quality and Accuracy:** The API should give reliable, up-to-date and precise information, which is essential for jobs that require real-time data or specialized specifics.
**5. Cost and Rate Limits:** The API's fees, as well as any rate constraints, should be in line with the project's budget and estimated usage volume.
**6. Documentation and Support:** Well-documented APIs with clear usage guidelines and active support fora are essential for efficient integration and troubleshooting.
**7. Security and Privacy:** The API must follow strict security and privacy requirements, particularly when dealing with sensitive user data.

Let's examine the example of "Booking Holidays" based on the above considerations:
When a user requests assistance in booking a holiday, the LLM would first identify the task's requirements, such as destination, dates and preferences. It then queries the API catalog, filtering for travel-related APIs that cover the specified destination and dates. The selection mechanism might prioritize APIs based on user preferences for low-cost options, high reliability scores and positive feedback from previous integrations. Once a shortlist of suitable APIs is generated, the LLM selects the API (or APIs) that best match the criteria. Through these criteria and mechanisms, the API selection process becomes a sophisticated, multi-dimensional decision-making process, ensuring that LLMs can effectively leverage external APIs to extend their capabilities and perform a wide range of tasks with higher accuracy, efficiency and user satisfaction.

The **"Relevance to Task" criterion** is foundational in the API selection mechanism for integrating Large Language Models (LLMs) with external tools, particularly APIs. This criterion ensures that the chosen API is intrinsically aligned with the specific objectives of the task at hand, thereby guaranteeing that the LLM can effectively utilize the API to fulfill the user's request. Here, we examine the evaluation of an API's relevance to a given task, using the example of a travel booking task to illustrate the concept in detail.
**Task-Specific Functionalities:**
The primary consideration in determining an API's relevance is its set of functionalities. For a task such as travel booking, the API must offer specific capabilities related to searching for flights, booking accommodations and perhaps even additional services like car rentals or activity reservations. The API's core functions must align with the task requirements to ensure it can serve the intended purpose.
**Geographical Coverage:**
For tasks with a geographical component, such as travel booking, the API's coverage is crucial. It should support queries and transactions for the specific destinations involved in the task. An API that offers extensive hotel listings in Europe but limited options in Asia would not be relevant for planning a trip to Tokyo, for instance.

**Integration Complexity and Compatibility:**
An API's usefulness is also affected by how simple it is to incorporate into the current framework, particularly how well it works with the LLM's data formats and architecture. Though an API may technically support the task's objectives, it may not be as suitable if it needs significant adaption or conversion work.

**Evaluating Relevance**
**Documentation Review:**
A thorough examination of the API's documentation can reveal its capabilities, data models and supported operations, providing insights into whether it meets the task's needs.
**Use Case and Example Analysis:**
Many APIs provide examples or use cases that illustrate typical applications. Reviewing these can help assess how well the API's functionalities map to the task's requirements.
**Trial and Testing:**
In some cases, conducting a trial integration or a test query can offer a direct evaluation of the API's relevance, allowing for an empirical assessment of its suitability for the task.

In our "Booking Holidays" example, consider a user requesting assistance with booking holidays in Greece, including flights, accommodations and activities. The LLM, tasked with fulfilling this request, must select APIs that directly support these objectives:
• **Flight Booking API:**
Must offer a comprehensive database of flights to and from Greece, allowing for searches based on dates, preferences (e.g. non-stop) and pricing.
• **Hotel Booking API:**
Needs to provide access to a wide range of accommodations in Greece, with features allowing users to filter by location, rating, price and amenities.
• **Activities API:**
Should cover recreational activities available in Greece, such as tours, museums and other experiences, with booking capabilities.
Each selected API must not only offer the right functionalities but also ensure up-to-date and comprehensive coverage of options in Greece to be deemed relevant for the task. By rigorously assessing an API's direct support for a task's objectives, one can significantly enhance the user satisfaction of the LLM-powered solutions, ensuring that the external APIs are leveraged to their fullest potential in achieving the desired outcomes.

**Low-Level Details - Cosine Similarity Metric:**
At this point, we give a brief overview of how the concept of cosine similarity is applied in the process of API selection by relevance, based on general principles in machine learning and natural language processing (NLP). Cosine similarity is a metric used to measure how similar two vectors are in a multi-dimensional space, often used in the context of text analysis within NLP. It calculates the cosine of the angle between two vectors, which can represent the textual description of a user's task and the description of an API's functionality. The cosine similarity score ranges from -1 to 1, where 1 indicates identical directionality (or perfect similarity), 0 indicates orthogonality (no similarity) and -1 indicates opposite directionality.
**Application in API Selection by Relevance:**
When selecting APIs based on relevance to a given task, the descriptions of both the task and the APIs can be converted into vectors using techniques like TF-IDF (Term Frequency-Inverse Document Frequency) or embeddings generated by language models. Once in vector form, cosine similarity can be calculated between the task vector and each API vector to determine how closely each API's functionalities align with the task's requirements.

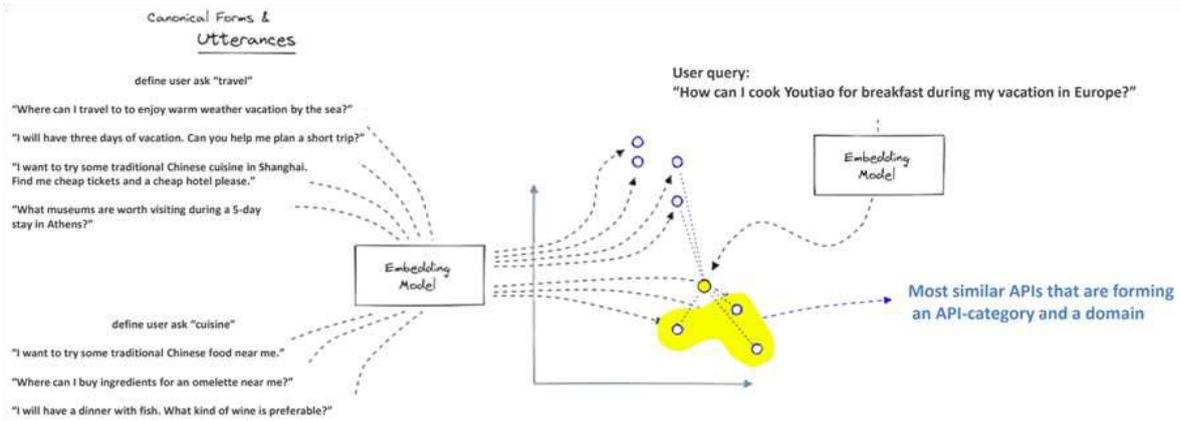

**Figure 2.8:** Illustration of the Semantic Vector Space created based on the Cosine Similarity Metric

**1. Vectorization:**
Convert the textual descriptions of the task and APIs into vectors. This could involve:
• Extracting features using TF-IDF, which emphasizes words that are frequent in a document but not across all documents.
• Utilizing pre-trained word embeddings from models like Word2Vec or BERT [8] to represent words or phrases in vector space.

**2. Cosine Similarity Calculation:**
For each API, calculate the cosine similarity between its vector and the task vector using the formula:

$$cosine\ similarity = \frac{\mathbf{A} \cdot \mathbf{B}}{\|\mathbf{A}\| \|\mathbf{B}\|}$$

where **A** and **B** are the vector representations of the API and the task, respectively and $\|\mathbf{A}\|$ and $\|\mathbf{B}\|$ are their magnitudes.

**3. Ranking and Selection:**
Rank the APIs based on their cosine similarity scores to the task vector. The API(s) with the highest score(s) are considered the most relevant to the task and selected for integration.

**Specialized Details and Considerations**

• **Dimensionality Reduction:**
High-dimensional vectors might necessitate dimensionality reduction techniques (e.g. PCA) to improve computational efficiency without significantly compromising the ability to capture semantic similarity.

• **Contextual Embeddings:**
For complex tasks, contextual embeddings from models like BERT [8] or GPT [7] might be preferred over simpler vectorization methods, as they better capture linguistic entities.

• **Thresholding:**
Implementing a threshold for cosine similarity scores can help in filtering out APIs that fall below a certain level of relevance, ensuring that only the most pertinent APIs are considered for selection.

### 2.4.2 Mechanisms for API Selection

**1. API Cataloging and Metadata:** Maintaining a collection of APIs and metadata about their functionality, coverage, performance metrics and usage terms, speeds-up the selection process.
• Categorization: APIs are organized into categories based on their application domain (e.g. travel, finance, healthcare).
• Metadata Annotation: Each API is annotated with metadata, including its capabilities, performance metrics, usage costs and limitations.

**2. Automated Matching Algorithms:** The selection process can be automated by using algorithms to match task requirements with API metadata. These algorithms are able to rank APIs according to how effectively they fulfill the given requirements.
• Natural Language Processing (NLP): Parsing the task description to extract key requirements.
• Scoring and Ranking: Algorithms evaluate APIs based on match criteria and rank them accordingly. Techniques such as cosine similarity, as discussed earlier, can be employed here.

**3. User Preferences and Historical Data:** The selection process can be improved by including user preferences and past selection data, giving priority to APIs that meet user-specified criteria or have produced effective results in the past.
• Preference Profiles: Users can specify preferences, which are used to filter or prioritize API selections.
• Contextual Analysis: The system considers the context of the request (e.g. geographic location, time of day) to select the most suitable APIs.

**4. Dynamic Evaluation:** Dynamic assessment techniques that test APIs in real-time or almost real-time to evaluate their current performance and data quality can be helpful for activities where numerous APIs could be appropriate.
• API Probing: Sending sample requests to APIs to gauge their performance and the applicability of their responses.
• Adaptive Selection: The system dynamically adjusts its API choices based on the results of these probes.

**5. Feedback Loops:** Continuous improvement is ensured by putting feedback loops in place where the effectiveness and success of API integration attempts are tracked and utilized to update the API catalog and selection algorithms.
• User Feedback Integration: User ratings and comments on API performance are used to adjust future selections.
• Performance Analytics: Analyzing the success rate and efficiency of API integrations to inform adjustments to the selection process.

**6. Security and Compliance Filters**
Applying security standards and compliance requirements as filters in the API selection process to ensure that chosen APIs meet organizational or regulatory standards.
• Compliance Checks: Verifying that APIs adhere to relevant data protection and privacy regulations.
• Security Assessment: Evaluating APIs for security measures and vulnerabilities.

An interesting 2-level selection method is being used for API-selection in recent approaches like [18] and [5]. In the first step of the method, a domain that contains many APIs (e.g. from the 49 categories of the collection of [31] as shown in Figure 2.9) is selected. Then, the most suitable API for the task is selected based on a pre-defined "Loss Function" metric (as explained in detail in the [18]). In a few words, API calls are dynamically filtered based on their contribution to prediction accuracy, quantified by a loss function $L_i(z)$. This function is defined as $L_i(z) = -\sum_{j=i}^{n} w_{j-i} \log p_M(x_j|z, x_{1:j-1})$, where $w_{j-i}$ are weights that decrease with the distance from the position $i$ to emphasize more immediate contributions and $\log p_M(x_j|z, x_{1:j-1})$ denotes the logarithm of the probability of observing token $x_j$ given the context $z$ and all preceding tokens. API calls are generated across various sequence positions and their utility is assessed by the improvement they provide in the model's predictive performance, i.e., a reduction in the computed loss. Only API calls that lead to a significant decrease in loss are retained, allowing the model to utilize external information sources efficiently and enhance its overall decision-making process. This technique reminds of "search space pruning" algorithms. These mechanisms, when combined, form a comprehensive framework for API selection, ensuring that the chosen APIs are not only technically capable of fulfilling the task but also align with user preferences, performance expectations and security requirements.

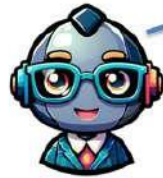

**Figure 2.9:** The AI-agent selects the domain

**Figure 2.10:** The AI-agent selects the specific API of the chosen domain

## 2.5 STEP 5: Generating API calls - Taking Action

This step is the most interesting one in our analysis. In step 5 we take advantage of the API-specific fine-tuned LLMs intoduced in step 2 and we use them in a way such that alignment with user-intention is guaranteed. It is important to understand the nature of the task before going on with the execution. There are "simple tasks" that require only textual output (e.g. "Write a short fairy-tale poem that includes a dragon and a princess..."), there are tasks that require communication with external tools (e.g. "Will it rain tomorrow?") and finally there are tasks that require advanced reasoning and - in some sense - a "juvenile" version of Artificial General Intelligence (AGI) (e.g."Given the image of my fridge, could you suggest a quick recipe that does not require using my oven?" or "Given a piano sheet, could you provide a reasonable fingering so that a pianist can play the song?"). Below we analyze these 3 categories of tasks in detail.

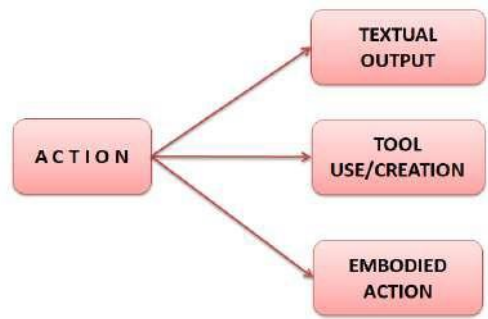

**Figure 2.11:** Types of Action

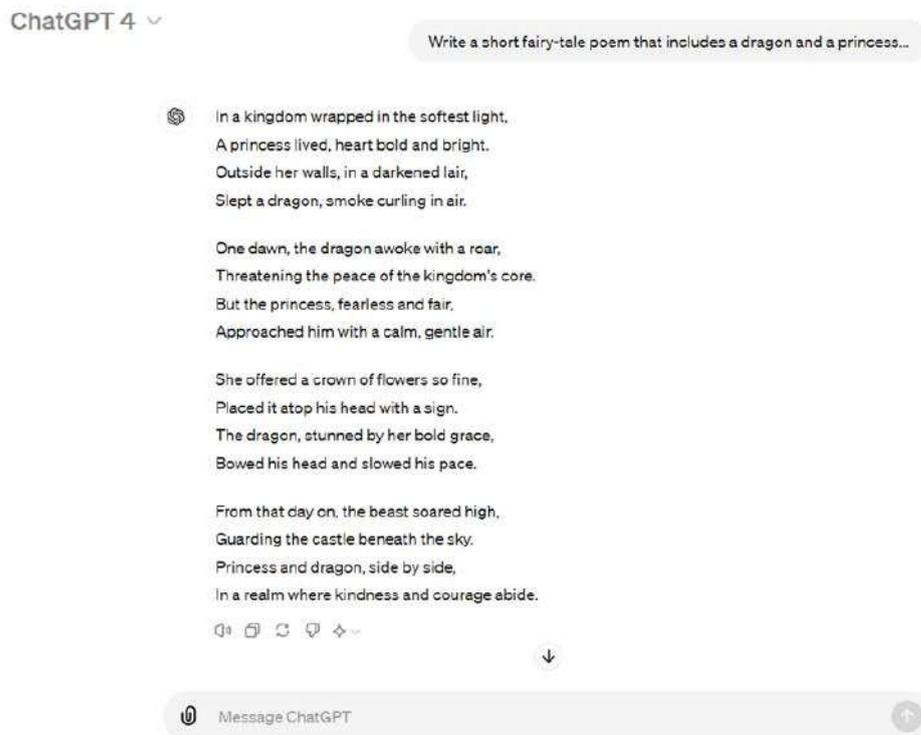

**Figure 2.12:** Example of a "Textual Output" Task:
User query: "Write a short fairy-tale poem that includes a dragon and a princess..."

## 2.5.1 Textual output

Recent generative LLMs have managed to produce human-like text responses. Models like ChatGPT and PanGu are excellent at producing text that is varied, contextually relevant and fluent. This revolution has opened up a wide range of applications where sophisticated, responsive language processing is essential, such as customer service and content creation. One notable characteristic of LLMs is their capacity to produce output that can be guided. This means that users can customize responses by using simple prompt engineering techniques to meet particular needs and situations, which is crucial for highly specialized jobs like technical writing or personalized communication. The versatility and usefulness of the models are demonstrated by the way they are integrated into many industries, which represents a major advancement in the understanding and production of human language by robots. As these technologies advance, they have the potential to completely transform AI interactions by improving the effectiveness and intuitiveness of digital platforms and more smoothly bridging the gap between human and machine communication. To demonstrate this technological breakthrough, we provide a small example of text generation in Figure 2.12.

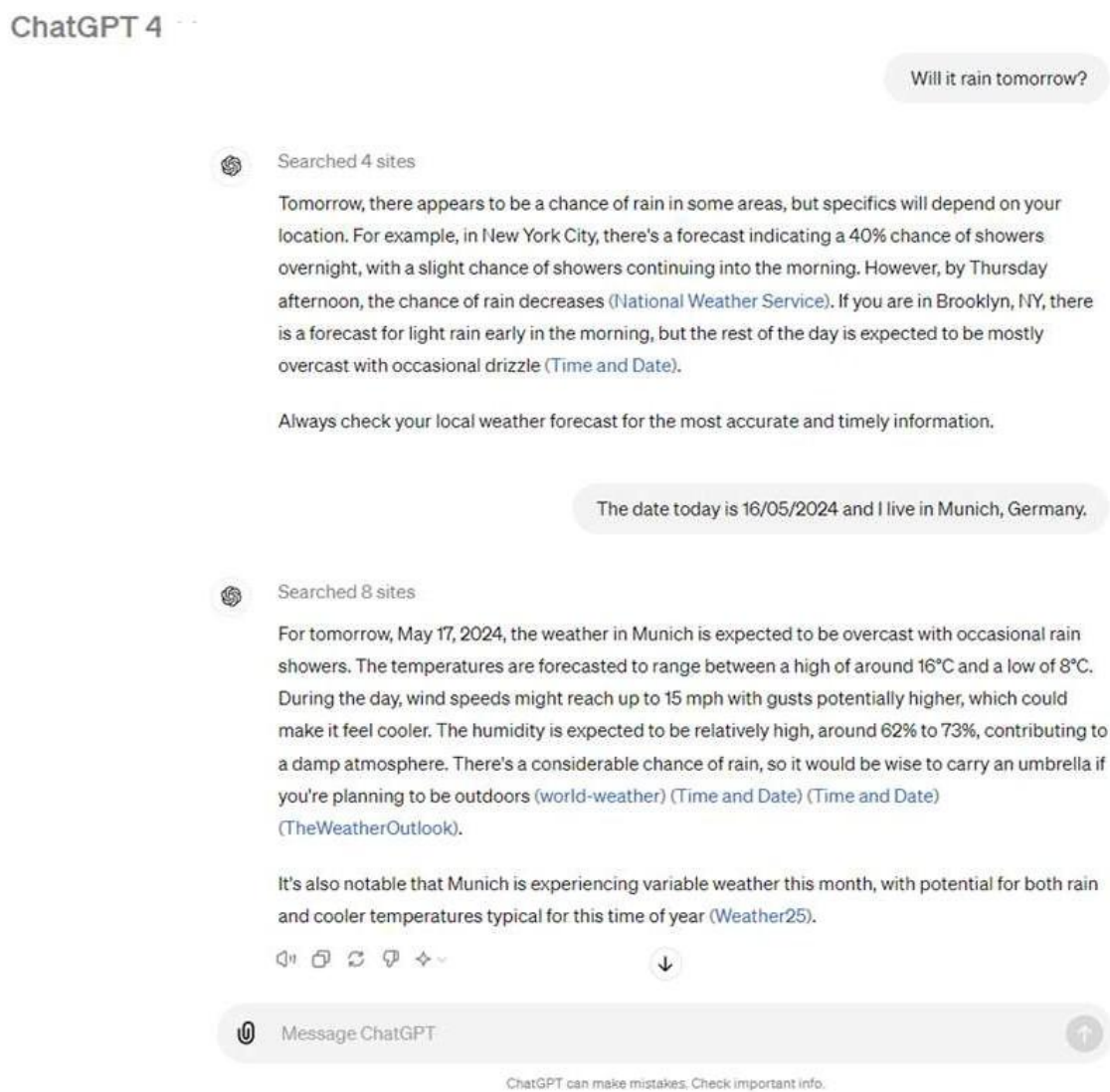

**Figure 2.13:** Example of a "Tool use/creation" Task:
User query: "Will it rain tomorrow?"
ChatGPT 4 is capable of accessing external sites, thus it can provide "tool-augmented" responses in specific scenaria like "Weather Forecasting".

### 2.5.2 Tool use/creation

Tools provide a valuable extention of the LLM capabilities, enabling more efficient and effective and up-to-date task completion. For example, when a user asks for the weather, the LLM should be able to connect to a "Weather Forecasting Tool" in order to provide a reasonable response as demonstrated in Figure 2.13. LLMs have a vast amount of knowledge, but they lack the ability to access real-world on-line data. This limitation can often lead to misunderstanding of user queries or hallucinated knowledge. To mitigate such limitations, we construct LLM-based AI-agents that are able to communicate with the external environment via tools/APIs. Another advantage of the tool-use capability of AI-agents is visible in high-stakes domains (e.g. healthcare, finance). In such applications transparency is critical, but the decision-making process of LLMs is often like a black-box. Equipping LLMs with tools lead to increased interpretability and robustness and provides a more reliable framework for decision-making, reflecting a structured approach to complex problems. These tools are designed to be resilient against adversarial attacks and jailbreaking. Additionally, in order for LLMs to use tools effectively, they must have a thorough awareness of their application contexts and invocation techniques. LLMs can mimic human learning processes (e.g. reviewing manuals or observing other users) and improve their tool-using skills through techniques like zero-shot learning, few-shot learning and prompt-engineering. The transition from LLMs to tool-using agents indicates a change in the direction of increasingly intelligent and autonomous AI systems.

### 2.5.3 Embodied Action

In the era of LLMs and human-like AI technologies, it is reasonable to fantasize the future of AGI. The "Embodied Hypothesis", which contends that intelligence develops via ongoing environmental contact, is a major source of inspiration for Embodied Action in AGI, which emphasizes the critical transition from digital to physical realms. This exact hypothesis reminds the famous philosophical school of Plato, the Ancient Greek Philosopher. Plato developed the theory that through "Anamnesis" (which is an exploration of one's own memory) an individual can gain insight into important truths (or else "Knowledge"). Traditional Machine Learning (ML) algorithms are limited in their operational scope to digital outputs based on structured information. Recent research takes advantage of the digital environment of AI-agents, allowing them to develop human-like learning behaviour through observation, manipulation, navigation and other "Embodied Actions".

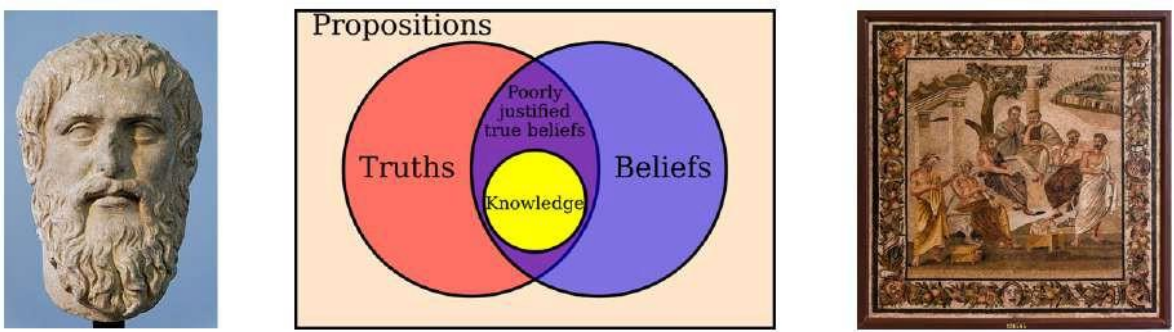

**Figure 2.14:** Left: Plato - Roman copy of a portrait bust c. 370 BC
　　　　　　Middle: A Venn diagram illustrating the classical theory of knowledge
　　　　　　Right: Plato's Academy mosaic in the villa of T. Siminius Stephanus in Pompeii, around 100 BC to 100 CE (Image source: [2])

The long pre-training of LLMs is a prerequisite for increased generalization and reasoning capabilities and enhancement of the AI-agents with real-world dynamics. In order to update agents' memory and internal states, "observation" collects inputs from sensors; "manipulation" entails intricate physical task sequences; and "navigation" provides an execution path for the agent's actions. The combination of "Embodied Actions" with LLM capabilities leads to flexible, independent and effective AI-systems that revolutionize Human-Computer Interaction (HCI).

```
Algorithm 1 API call process
 1: Input: us ← UserStatement
 2: if API Call is needed then
 3:     while API not found do
 4:         keywords ← summarize(us)
 5:         api ← search(keywords)
 6:         if Give Up then
 7:             break
 8:         end if
 9:     end while
10:     if API found then
11:         api_doc ← api.documentation
12:         while Response not satisfied do
13:             api_call ← gen_api_call(api_doc, us)
14:             api_re ← execute_api_call(api_call)
15:             if Give Up then
16:                 break
17:             end if
18:         end while
19:     end if
20: end if
21: if response then
22:     re ← generate_response(api_re)
23: else
24:     re ← generate_response()
25: end if
26: Output: ResponseToUser
```

**Figure 2.15:** API call generation in API-Bank (Image source: [3])

One interesting approach towards evaluating the performance of tool-augmented LLMs is presented in [3]. In this paper, 53 commonly used API tools (diversely selected, including search engines, calculator, calendar queries, smart home control, schedule management, health data management, account authentication workflow and more) are being experimentally studied in the scope of a complete tool-augmented LLM workflow. Datasets involve 264 annotated dialogues that involve 568 API calls. The selection of the appropriate API is done in a similar way as in "step 4" of the current thesis.

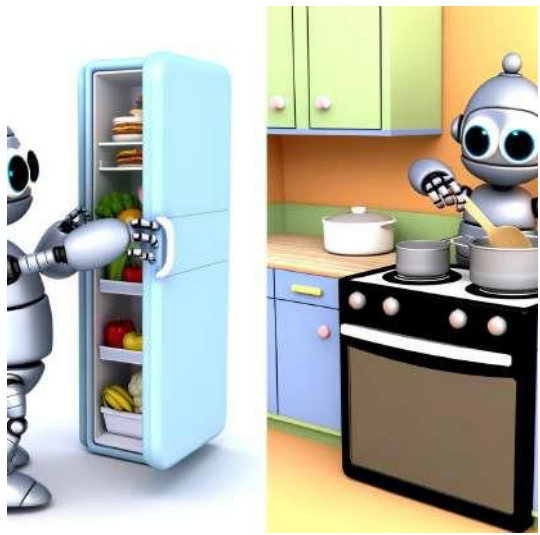

**Figure 2.16:** Example of an "Embodied Action" Task:
User query: "Given the image of my fridge, could you suggest a quick recipe that does not require using my oven?"
Towards Artificial General Intelligence...(?)

The decision-making process is evaluated based on some Metrics and the most important objects of decision are "Whether an API call is needed", "Whether the selected API is good-enough" and "Whether the API call generated is good-enough for the specific task". It is important to note that some tasks like "Schedule a trip to Greece" may involve multiple API calls and multiple APIs to be coordinated. This problem is solved in the next step.

## 2.6 STEP 6: Task Decomposition

As stated in the previous paragraph, some tasks are of high complexity and require multiple API calls and multiple APIs. For such tasks, we create a series of subtasks and we provide a scheduling of them. This is what a human would do as well. Planning is a crucial tactic that people use to overcome difficult situations. Planning aids in cognitive organization, goal setting and determining the course of action to take in order to accomplish greater goals. AI-agents need to be able to plan, reason and "think" in an organized way like humans do. AI-agents use reasoning to break down difficult jobs into smaller, more manageable tasks and then create strategies specifically for each of those activities. In the plan formulation phase, agents typically break down a main task into multiple smaller tasks and different strategies have been suggested for this stage. One stream suggests that the issues should be solved entirely at once by following a plan that is created all at once. Then the plan is executed step by step. Another approach is using adaptive strategies to handle complex tasks more fluidly by planning and addressing subtasks individually. Furthermore, some earlier works highlight the importance of hierarchical planning and structuring the subtasks in a tree-like form. Dependency analysis complements hierarchical decomposition by identifying the dependencies between various sub-tasks. It ensures that tasks are performed in a logical sequence, preventing errors due to unmet prerequisites. This is crucial in tasks where the outcome of one API call influences subsequent calls, like booking flights before hotels in a travel planning application. Sequential and parallel processing strategies determine the most efficient ways to execute the decomposed tasks. Sequential processing is necessary for dependent tasks, ensuring that each step is completed in the right order, while parallel processing speeds up the overall task completion by handling independent tasks simultaneously. Even though LLM-based AI-agents show a wide range of general knowledge, they occasionally run into problems when given tasks requiring specialized knowledge. For this reason, it is often a good idea to integrate specialized planners from certain domains. After a plan is set, it is critical to consider and assess its benefits. LLM-based AI-agents refine and improve their strategies and planning techniques by utilizing internal feedback mechanisms, frequently gaining knowledge from previous models and tasks. Agents actively interact with humans to better fit with human values and preferences. This allows humans to clear up any misunderstandings and incorporate this customized feedback into their planning process. In addition, users might use input from real-world or virtual environments, such indications from completed tasks or after-action observations, to help them edit and improve their plans. Below we mention some key-research approaches towards planning and task decomposition:

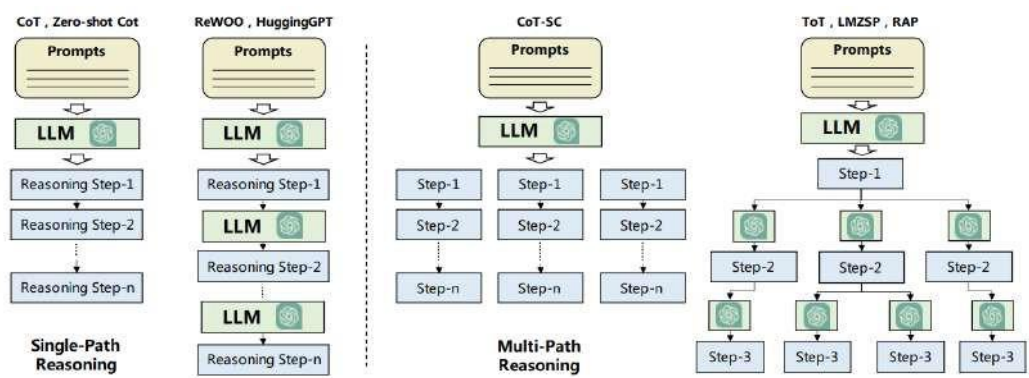

**Figure 2.17:** An overview of the "Task Decomposition" techniques (Image source: [4])

### 2.6.1 Single-path Reasoning

This involves breaking down tasks into sequential steps.
**Chain of Thought (CoT)**: Incorporates reasoning steps into prompts to guide LLMs in a step-by-step manner.

**Zero-shot-CoT**: Uses trigger sentences like "think step by step" to inspire LLMs to generate reasoning processes independently, without preset examples.
**Re-Prompting**: Checks if each step meets prerequisites before continuing, regenerating plans if necessary.
**ReWOO**: Provides the plan first and later integrates observations.
**HuggingGPT**: Sub-goals are solved recursively through multiple queries to LLMs.

### 2.6.2 Multi-path Reasoning

Constructs reasoning steps in a tree-like structure, allowing for multiple subsequent paths at each decision point.
**Self-consistent CoT (CoT-SC)**: Uses CoT to create multiple reasoning paths, selecting the most frequent answer as the final output.
**Tree of Thoughts (ToT)**: Plans are created using a tree structure with nodes representing intermediate steps, finalized through BFS or DFS strategies.
**RecMind**: Implements a self-inspiring mechanism that reuses discarded planning information to generate new steps.
**Graph of Thoughts (GoT) and Algorithm of Thoughts (AoT)**: Expand on ToT by introducing more complex structures and algorithmic examples into prompts.
**RAP**: Employs Monte Carlo Tree Search (MCTS) to simulate and select the best plan from multiple iterations.

### 2.6.3 External Planner

Combines the planning capabilities of LLMs with external tools for specialized tasks.
**LLM+P and LLM-DP**: Transform task descriptions into Planning Domain Definition Language (PDDL) and use external planners to execute plans.
**CO-LLM**: Addresses the limitations of LLMs in executing low-level control by integrating a heuristic-based external planner for action execution.

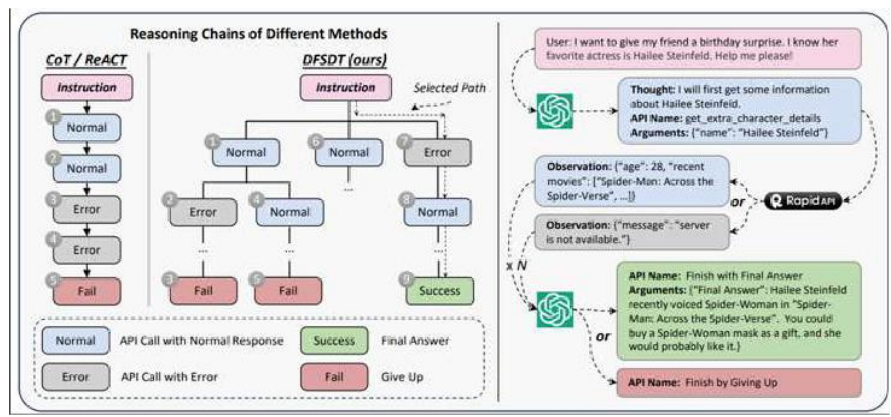

**Figure 2.18:** Left: DFSDT vs CoT & ReACT
Right: The solution path annotation process (using ChatGPT)
(Image source: [5])

At this point, we give some dedicated notice to the method used in the [5]. **Depth-First-Search-Based Decision Tree** DFSDT is acting as a guide for powerful LLMs and provides reasoning capabilities using prompt engineering techniques. It seems to be a superior technique in comparison to CoT ([32]) or REACT ([33]), because it overcomes limitations like error propagation (a mistaken action may propagate the errors further and cause the model to be trapped in a faulty loop) and limited exploration (CoT or ReACT only explores one possible direction, leading to limited exploration of the whole action space) and improves search efficiency and effectiveness towards the Solution Path Annotation. It enables LLMs to evaluate multiple reasoning traces and expand the search space. Another technique for reasoning proposed by Princeton University and Google DeepMind is the "Tree-of-Thoughts" ([34]).

Each strategy addresses different aspects and complexities of planning, from simple sequential steps to complex, recursive and multiple-path reasoning, enhancing the versatility and effectiveness of LLMs in problem-solving.

## 2.7 STEP 7: Iterative Improvement and User Feedback

### 2.7.1 Why we need Feedback

Long-term planning towards completing complex tasks is necessary in numerous real-world situations. This is a very challenging task, since it requires taking into account many environmental parameters and applying advanced reasoning. Because of this, building the initial plan and sticking to it can usually result to failure ([4],[1]). Let's get some inspiration from the human way of creating a plan. People sometimes create and modify their plans repeatedly in response to outside input. Recent research proposes creating external planning modules that allow the agent to receive feedback, allowing the development of human-like competence.
Iterative Improvement and User Feedback is crucial for the continuous learning cycle in AI and ML, involving regular updates and refinements based on user input and performance metrics. This ensures that systems meet and adapt to evolving user needs and environments. User Feedback can guide the AI-agent towards user's preferences and needs. It is also essential for getting rid of hallucinations. The simplest example that demonstrates the necessity of User Feedback is the "Will it rain tomorrow?" prompt, in which the user needed to specify their location, in order for the agent to provide useful information (Figure 2.13).

### 2.7.2 Continuous Learning and Adaptation

In AI and machine learning systems, learning and adapting continuously is crucial for a variety of reasons. It is essential to adjust to user needs because direct input keeps systems updated and in alignment with user expectations. By employing performance metrics monitoring to discover trouble spots, system accuracy can be improved, resulting to improved accuracy and reliability. Iterative updates promote personalization by enabling systems to adjust to unique user preferences and behaviours. Last but not least, preserving technological continuity guarantees that systems may easily utilize latest tools and APIs and - as a result - stay updated.

```
1N73LL1G3NC3
   15 7H3
   4B1L17Y
  70 4D4P7 70
   CH4NG3
- 573PH3N H4WK1NG
```

**Figure 2.19:** "Intelligence is the ability to adapt to change." – Stephen Hawking

### 2.7.3 Other Feedback Collection Mechanisms

Acquiring useful user insights requires the usage of technologies like analytics and feedback forms. AI-agents are efficiently guided by using Key Performance Indicators (KPIs) to monitor performance. Also, A/B testing is used to make data-driven decisions and improve user experience by understanding user preferences and behaviours. Adopting a user-centric design philosophy ensures that systems are customized to match user needs, leading to the creation of user interfaces that enhance system performance and user satisfaction.

### 2.7.4 Environmental Feedback

An essential component of agent-based decision-making is environmental feedback, which affects the planning and actions of the agents. It can be acquired from the virtual or objective world through things like task completion signals and post-action evaluations. Current research is enhanced with this technique. For example, ReAct's high-level thinking and planning exploits thought-act-observation triplets, whereas Voyager follows a "trial-and-error" approach that includes intermediate progress, execution error and self-verification feedback.

### 2.7.5 Model Feedback

Apart from using user and environmental feedback, recent research suggests that internal feedback from the AI-agents themselves is being used. The AI-agent can create output, get feedback and then refine the output based on the feedback received. This process reminds of an infinite-loop prompt engineering technique that is terminated when the intended result is obtained.

### 2.7.6 Challenges of Feedback Integration

Some important considerations when enhancing AI-agents with feedback are:
**Balancing Feedback and Goals**:
It is crucial to align user feedback with the goals of the complete system, while ensuring technical feasibility and system functionality.
**Data Privacy**:
Ethical management and legal compliance are imperative when handling user feedback and data.
**Feedback Quality**:
Unbiased and representative feedback is difficult to be acquired, but vital for the continuous improvement of the system.

# Chapter 3

# Introducing a novel On-Device Architecture

The above 7-step methodology can be utilized towards constructing AI-agents that require a relatively "big" model that is accessed through cloud services. What does "big" mean? Below we discuss the "sizes of importance" in our study. To be more specific, we give a brief overview of the Storage, RAM and CPU requirements of local LLMs ([35], [36], [37]).

## 3.1 Storage and RAM requirements of LLMs

### 3.1.1 Storage Requirements

A 2.7-billion parameter model would normally require a storage capacity that can support its parameters and operating overhead in order to be deployed locally, say on a mobile device. Depending on the datatype of the parameters (floating-point values or integers after quantization), a model with 2.7 billion parameters may still require several gigabytes of storage space after being quantized and optimized for mobile use.

### 3.1.2 RAM Requirements

During inference, the RAM need is closely related to the model's operational features. Substantial model reduction and optimization methods like quantization are required for a model to function properly on mobile devices with as little as 4GB of RAM. By decreasing the model's parameter precision through quantization, less RAM is required to load and operate the model.

### 3.1.3 CPU Utilization

When deploying an LLM locally, CPU capabilities must be carefully considered. CPU use for LLM inference can be high, particularly in the absence of hardware accelerators like GPUs:
**CPU Load**:
Each inference step requires costly computing computations such as matrix multiplications and other tensor operations. The amount of parameters, architectural efficiency and complexity of the model will all affect the CPU load.
**Optimization Techniques**:
Model pruning, quantization and usage of distilled models can lower CPU load by making the computations necessary for inference simpler. These methods assist in modifying the model to fit the limitations of less potent devices, such as tablets or smartphones.

### 3.1.4 Memory Management Strategies

**Flash Memory for Storage**:
Larger models may exceed the DRAM capacity of local devices. A solution used by some engineers is storing the model parameters in flash memory and dynamically load them into DRAM during inference. This strategy involves a trade-off between storage capacity and speed, as flash memory is slower than DRAM.

**Dynamic Loading**:
Dynamic loading techniques involve loading only the necessary parameters into faster memory as needed. This approach can significantly reduce the amount of RAM required, enabling the execution of much larger models on devices with limited RAM.
**Efficiency Enhancements**:
Strategies like using sparsity-aware loading, where only the non-zero parameters are loaded, can optimize both CPU and memory usage. Such methods leverage the good properties of neural networks, where many parameters can have negligible values that do not significantly affect outputs.

## 3.2 Enhancing AI-agents with Local Custom Databases

In the evolving landscape of Artificial Intelligence (AI), the reliance on cloud-based models poses significant challenges in terms of latency, privacy and connectivity. LLMs "almost fit" in devices nowadays, but the need for cloud connectivity is one of the bottlenecks of their capabilities. Our approach aims to enhance mobile devices with LLM-like capabilities, using the ad-hoc concept of "locally-stored previously-defined macros/functions". In this chapter we introduce this novel on-device architecture that aims to integrate LLM-based AI-agents on devices. Local computational resources and custom databases are being used towards this integration. This approach not only aims to mitigate the dependency on cloud services but also ensures efficient, secure and real-time processing of tasks directly on users' devices.

### 3.2.1 On-Device Architecture Overview

The goal is to create an LLM-based AI-agent that can understand the user intention and execute actions based on it. The user input is multimodal (but for simplicity, let's suppose it is only text for the time being). We can achieve the above by using macros/functions that correspond to a series of API calls. Each time a user needs to use the AI-agent for a task that was never executed before, a "Training Mode Interface" is provided to the user. Then, the user verbally describes what goal (what intention) the new task is fulfilling. In this stage, the user and the AI-agent "communicate" through continuous feedback in order to provide a detailed course of actions (essentially a series of API calls) that need to be done in order to fulfill the specific task. At this point, a new entry is added to the Custom Made Database. This entry consists of 3 pieces of information: the Title, the Description and the Series of API calls that correspond to the new function (Figure 3.3). Essentially, the AI-agent can now exploit the database to "remember" the exact API calls that need to be executed for a specific task. In that way, the AI-agent only needs to execute a classification task instead of a generative task. As a result, tasks that are very frequently requested by the user, will be executed by the AI-agent very fast! Finally, one of the most important parts of this approach is matching the text input of the user to one, or a series of functions (predefined tasks). This is done through a semantic vector space or word embeddings.
**A simple use-case**:
Suppose that the user wants to order groceries to his home address and the user input is like:
"Please, order 6 of the cheapest yogurt cups and 0.5 kilogram of any cheese from the closest market to my home".
This task could correspond to the function: *"ORDER_FROM_NEAR-BY_MARKET( X, Y )"*,
where for example X corresponds to *"6 of the cheapest yogurts and 0.5 kilogram of any cheese"* and Y corresponds to *"my home"*. By leveraging semantic similarity techniques, the AI-agent would be able to understand that functions like "PLAY_POPULAR_MUSIC" or "SEND_EMAIL" are not relevant to the specific task and after semantic search the appropriate function mentioned above would be activated.

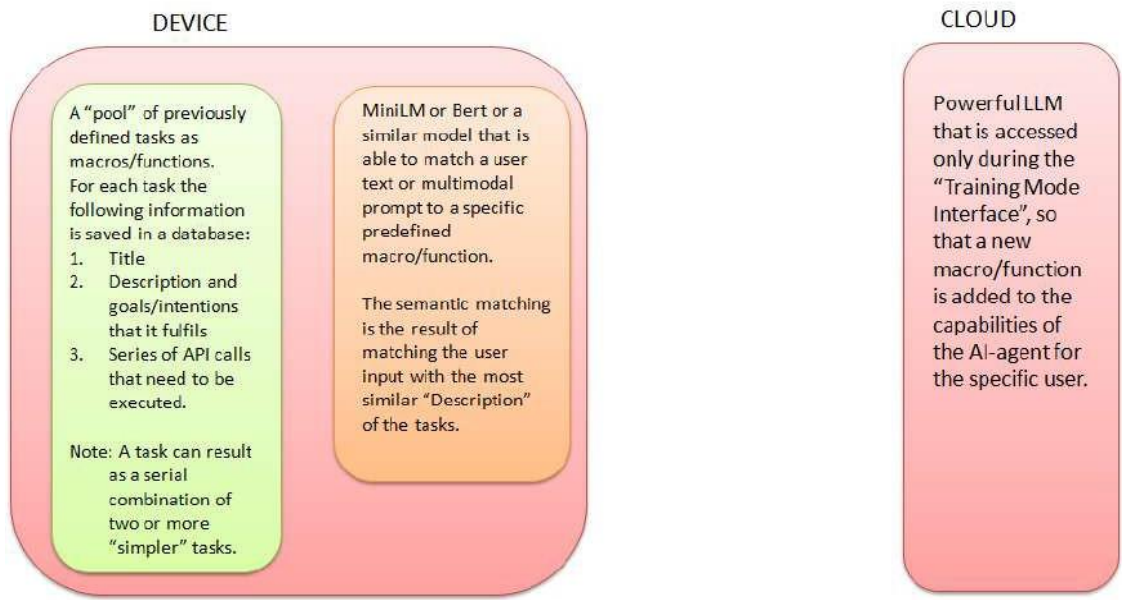

**Figure 3.1:** High-level idea of the proposed On-Device Architecture

**Example Task Definition:**
- **Task:** Order groceries.
- **User Description:** "I want to order certain groceries from the closest market to my home."
- **API Calls:**
  Sequence of calls to a grocery delivery API:
  search for stores near a specified address,
  search for items,
  add items to cart, and
  place order.
- **Macro/Function:**
  ORDER_FROM_NEARBY_MARKET(["6 cheapest yogurts", "0.5kg of any cheese"], "user's home address")

**Example User Input and Action:**
- **Input:** "Order 6 of the cheapest yogurts and 0.5 kilogram of any cheese from the closest market to my home."
- **AI-agent:**
  – Identifies intention,
  – matches it to ORDER_FROM_NEARBY_MARKET, and
  – executes the corresponding API calls.

**Figure 3.2:** "Pseudocode" of the "Order Groceries" example

### 3.2.2 Components

**Local Mini LLMs**:
Smaller, efficient versions of LLMs optimized for running on limited hardware resources without compromising performance.
**Custom Database**:
A locally stored database that contains predefined macros/functions and training data, allowing the AI-agent to perform tasks even when offline.
**Training and Deployment Interface**:
A user-friendly interface that allows users to train the model with new tasks and deploy these tasks as local functions for immediate execution.

| No. | Use Case | Scenario Description | Task Macro/Function | Example API Calls |
|---|---|---|---|---|
| 1 | Personal Finance Management | Track and compare spending on specific categories over time. | TRACK_AND_COMPARE_SPENDING(category, dates) | GET /finance/spending?category={category}&dates={dates} |
| 2 | Smart Home Automation | Adjust home devices based on environmental conditions. | ADJUST_THERMOSTAT_IF_COLD(tempThreshold) | POST /home/thermostat/adjust |
| 3 | Travel Planning | Find travel arrangements within a budget. | PLAN_TRIP(destination, dates, budget) | GET /travel/flights/search |
| 4 | Event Scheduling and Notification | Schedule and remind about meetings based on participant availability. | SCHEDULE_MEETING(participants, time) | POST /calendar/meeting/create |
| 5 | Health and Fitness Tracking | Log workouts and compare fitness progress. | LOG_WORKOUT_AND_COMPARE(details, period) | POST /health/workout/log |
| 6 | Educational Resource Aggregator | Gather educational materials based on topic and level of difficulty. | FIND_STUDY_MATERIALS(topic, level) | GET /education/materials/search |
| 7 | Personalized News Digest | Compile news summaries based on user-selected categories. | GENERATE_NEWS_DIGEST(categories) | GET /news/digest?categories={categories} |
| 8 | Personalized Learning Assistant | Create a learning schedule based on user's skills and availability. | CREATE_LEARNING_SCHEDULE(skill, availability) | POST /learning/schedule/create |
| 9 | Job Hunting Automation | Automate the job search process based on specified criteria. | FIND_JOB_LISTINGS(specifications) | GET /jobs/search?specs={specifications} |
| 10 | Recipe and Meal Planning | Suggest meal plans and recipes based on dietary preferences and nutritional needs. | PLAN_MEALS(dietType, nutrientGoals) | GET /nutrition/meals/plan |
| 11 | Automated Customer Support | Automate responses to common customer queries and escalate unresolved issues. | HANDLE_CUSTOMER_INQUIRIES(topic) | POST /support/query/handle |
| 12 | Real Estate Market Analysis | Analyze real estate market trends and predict investment opportunities. | ANALYZE_MARKET_TRENDS(location, period) | GET /real-estate/market-analysis |
| 13 | Language Learning Practice | Provide daily language practice exercises tailored to user proficiency. | GENERATE_LANGUAGE_EXERCISES(language, level) | GET /language/practice/exercises |
| 14 | Environmental Impact Tracking | Calculate and track an individual's carbon footprint based on daily activities. | CALCULATE_CARBON_FOOTPRINT(activities) | POST /environment/impact/calculate |
| 15 | Virtual Interior Design Consultant | Suggest interior design themes and decorations based on room characteristics and user preferences. | SUGGEST_DESIGN_THEME(roomType, preferences) | GET /design/themes/suggestions |

**Figure 3.3:** Example of a Custom Made Database
(Note: the column "Example API Calls" contains data that does not necessarily correspond to real-world data and is included for illustrative purposes only)

### 3.2.3 Operational Flow

**Input Processing**:
The AI-agent processes user inputs locally, using embedded NLP tools to understand and categorize requests.
**Task Matching and Execution**:
Utilizing the local database, the agent matches user requests to relevant tasks and executes associated actions using the predefined macros/functions.

### 3.2.4 Use Cases and Practical Applications

**Offline Functionality**:
Describes scenarios where the on-device AI-agent can operate entirely offline, such as in remote areas or during network outages, enhancing accessibility and user experience.
**Real-Time Data Processing**:
Explores use cases where immediate processing is crucial, such as in emergency response scenarios or real-time decision-making environments.

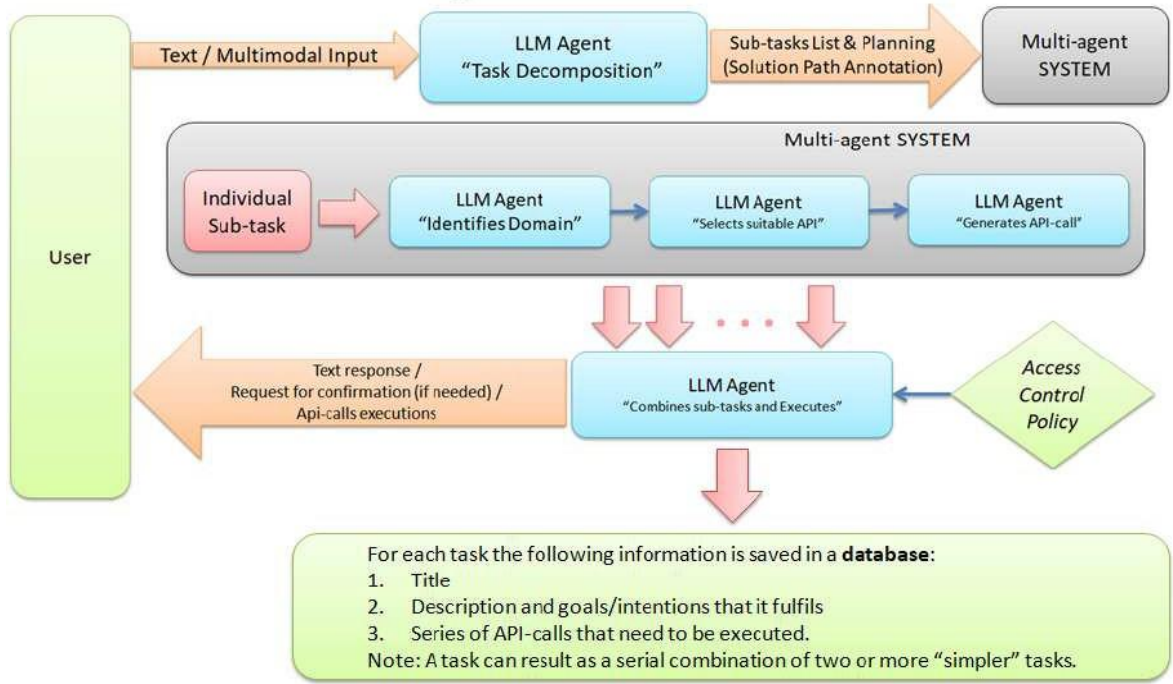

**Figure 3.4:** Training-phase overview of the novel On-Device approach

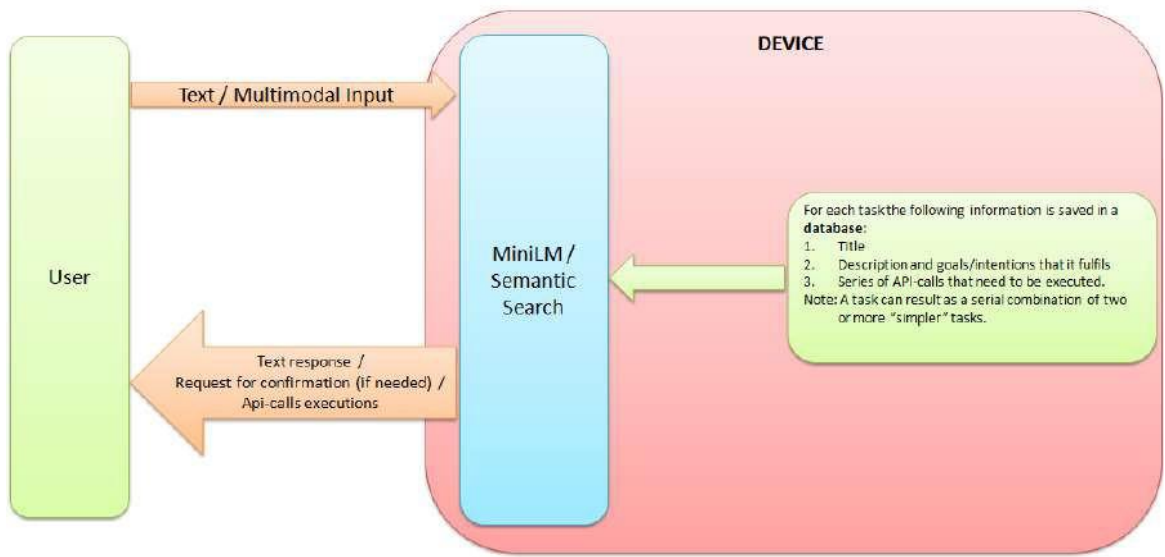

**Figure 3.5:** Run-time Architecture of the novel On-Device approach

### 3.2.5 Challenges and Mitigation Strategies

**Hardware Limitations**:
On-device architectures often face challenges related to limited computational power, storage capacity and energy efficiency. These limitations necessitate the use of optimized AI models through techniques like model pruning, which reduces the complexity and size of the models and knowledge distillation, which transfers knowledge from large models to smaller, more efficient ones. These strategies allow the models to operate effectively within the resource constraints of mobile and embedded devices.

**Security Concerns**:
The danger of data breaches and unauthorized access is central to on-device architectural security and this risk is amplified when sensitive data is processed locally. Secure access mechanisms and strong encryption algorithms are necessary to reduce these dangers. Furthermore, putting in place thorough security measures like secure boot and multi-factor authentication can assist protect the data from future cyber attacks.

**Integration with Existing Frameworks**:
Integrating on-device processing with existing LLM frameworks allows for a hybrid operational mode, combining the advantages of local processing with cloud-based capabilities. This integration enhances the system's reliability and responsiveness, especially in environments with intermittent connectivity. Moreover, it enables the device to perform critical operations offline while still synchronizing with cloud services for updates and advanced processing when online.

**Comparison with Cloud-Based Models**:
This approach provides reduced latency and increased privacy. On the other hand, the need for periodic updates and the initial setup complexity are limiting its usefulness.

### 3.2.6 Advantages of this approach

**Offline Availability**:
The agent's core functionality, including executing predefined macros/functions, does not require a constant online connection to LLMs. This ensures that the agent remains functional even in offline scenarios, provided that the tasks do not require real-time online data.

**Reduced Latency:**
This architecture will provide minimal latency when a user request occurs. This will be because the semantic search for the appropriate predefined task will be much faster than accessing an LLM as a cloud-service and creating the task "on the fly" in the common scenaria.

**Customization**:
Users can tailor the AI-agent to their specific needs by creating or modifying macros/functions. This level of customization means that the agent can evolve to handle a wide array of tasks as dictated by the user's changing requirements.

**Community-Driven Extensions**:
A platform or forum for sharing, discussing and refining macros/functions could significantly enhance the agent's utility. Users without deep coding expertise could leverage the community's collective knowledge to find or adapt macros that fulfill their needs, fostering a collaborative ecosystem.

**Scalability**:
The architecture is designed to accommodate an expanding library of tasks and functions, scaling from personal use cases to - potentially - enterprise-level applications. The training mode interface and intent matching mechanism are key components that facilitate this scalability.

# Chapter 4

# Experimental Part - Proof of Concept

This Chapter consists of the Experimental Part of this thesis. We focus on two aspects. The first one is the demonstration of how step 5 of the 7-step methodology looks like on a music composing scenario. The second part explores the on-device architecture proposed in the previous Chapter and constitutes the proof of concept for this approach.

## 4.1 The Piano Sheet example

Suppose that a user needs the piano sheet of a music piece, or a reasonable fingering for an existing piano sheet. This task requires a specialized AI-agent and is considered to be a very complex one. Below we present the attempts of ChatGPT 4 on these complex tasks and we provide the results of fine-tuning this model on real-world music sheet data and connecting it with a piano sheet API (e.g. MuseScore [6]).

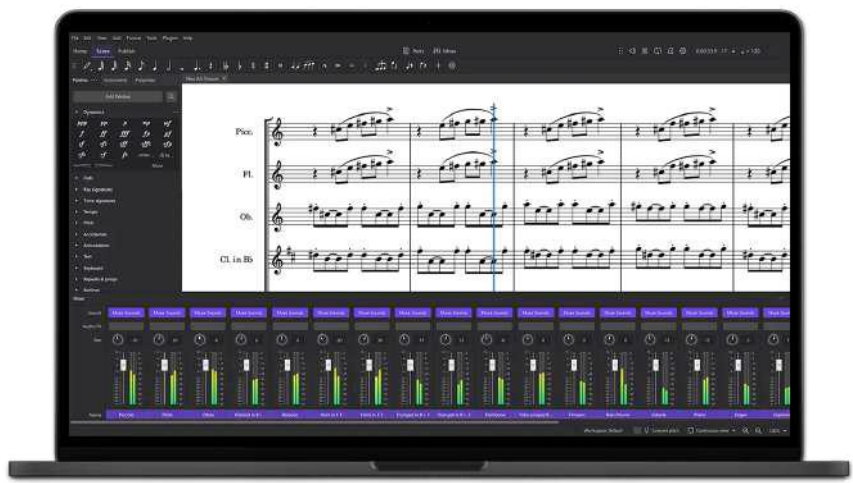

**Figure 4.1:** Providing Piano Sheets using LLMs (Image source: [6])

### 4.1.1 Initial attempt

We asked ChatGPT 4 to provide the transcript of a simple music piece (Figure 4.2).

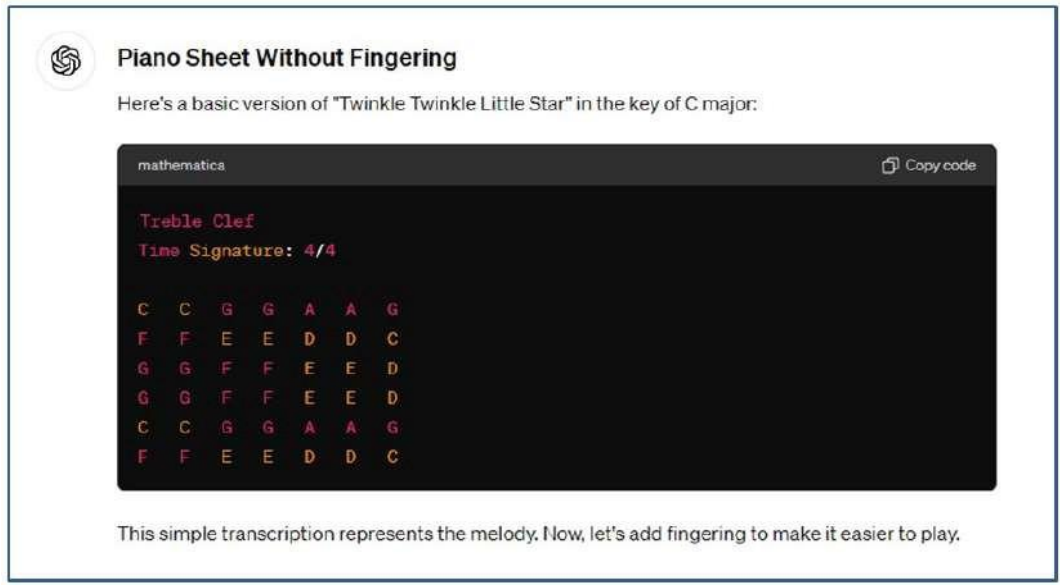

**Figure 4.2:** Piano Transcription of the song "Twinkle Twinkle Little Star" Without Fingering

The melody is completely correct! This is a small proof of the built-in capabilities of modern LLMs. Let's move on with more challenging tasks like providing fingering and trascripting the whole solution/sheet on a music staff/pentagram.

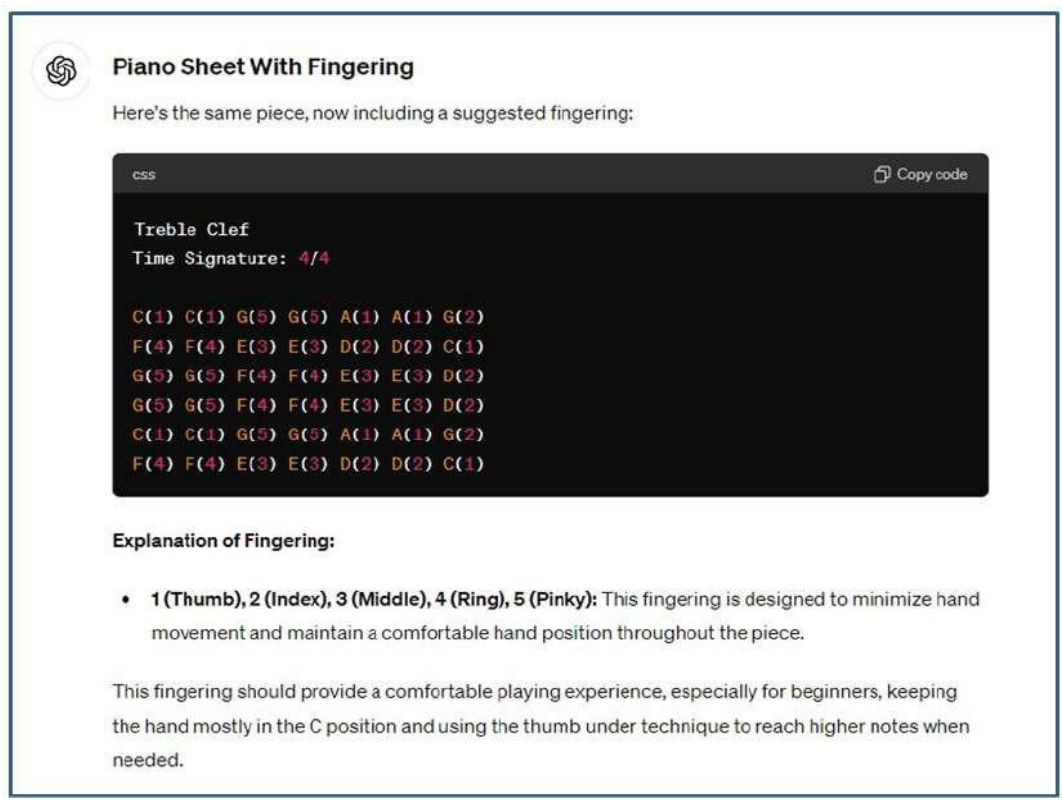

**Figure 4.3:** Piano Transcription of the song "Twinkle Twinkle Little Star" With a Fingering

## 4.1.2 Adding a "fingering"

Providing a reasonable fingering for a piano sheet can be a challenging task, since there are a lot of rules to follow and things to consider (Figure 4.4).

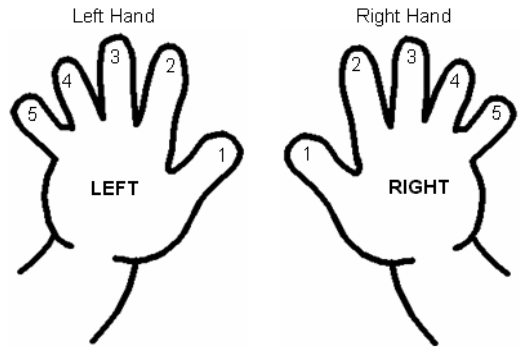

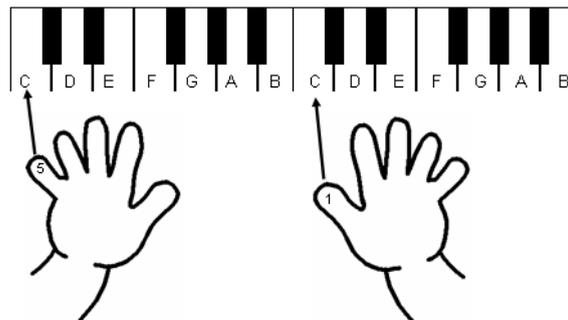

**Figure 4.4:** One of the rules of efficient piano fingering
(Image source: https://www.hearandplay.com/main/major-scale-fingering)

We asked ChatGPT 4 to provide a fingering for this piece (Figure 4.3).
The melody is completely correct and the fingering is reasonable (can be executed by a pianist) but not optimal (there is a more efficient fingering). For our next step, we tried few-shot learning on GPT 4 based on piano sheets from MuseScore and asked for a better fingering. The result was indeed a fingering that seems optimal to a pianist (Figure 4.5).

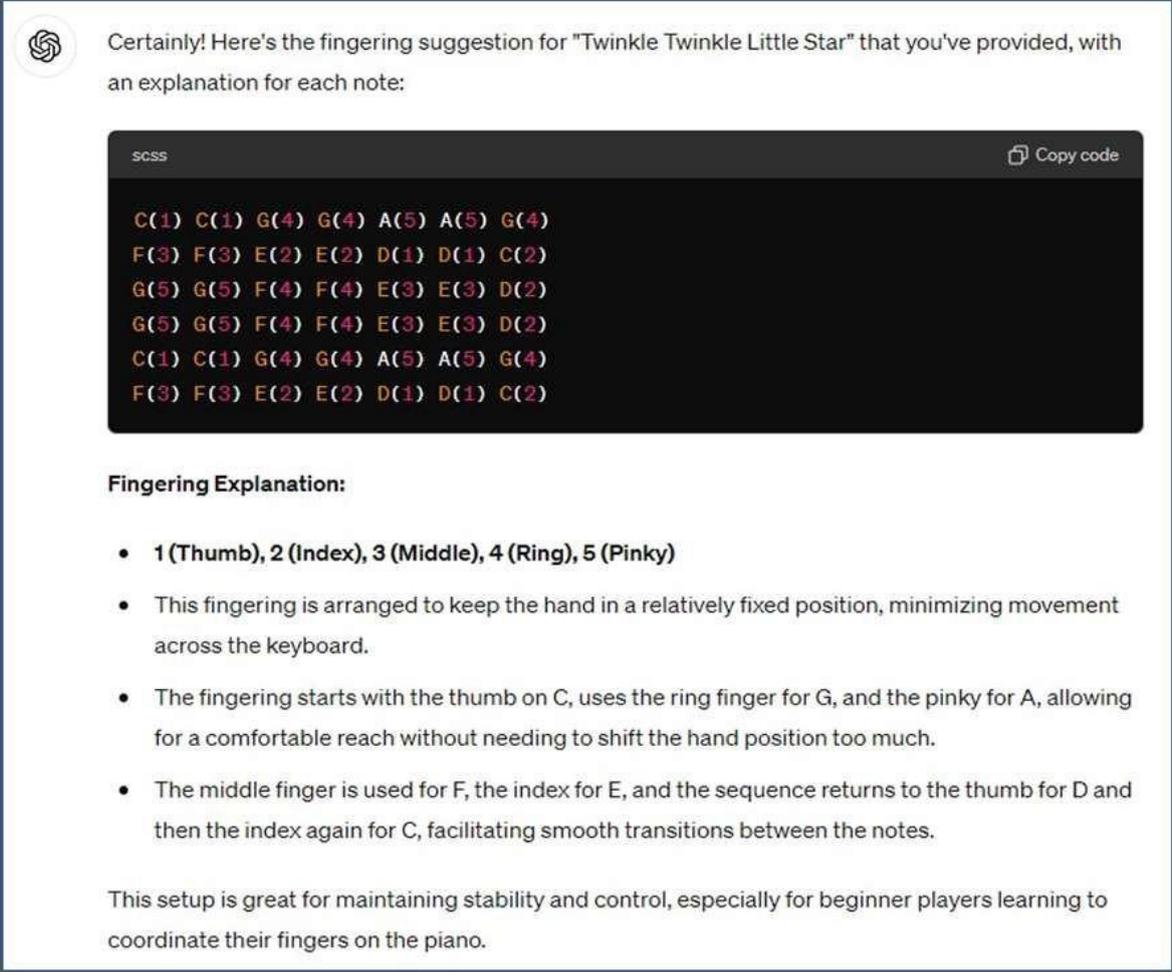

**Figure 4.5:** Optimal fingering using ChatGPT 4 and few-shot learning

### 4.1.3 Providing the notation on a music staff/pentagram

The final step of this task is connecting our LLM to the MuseScore API in order to obtain the desired piano sheet. The results of this integration can be viewed in Figure 4.6.

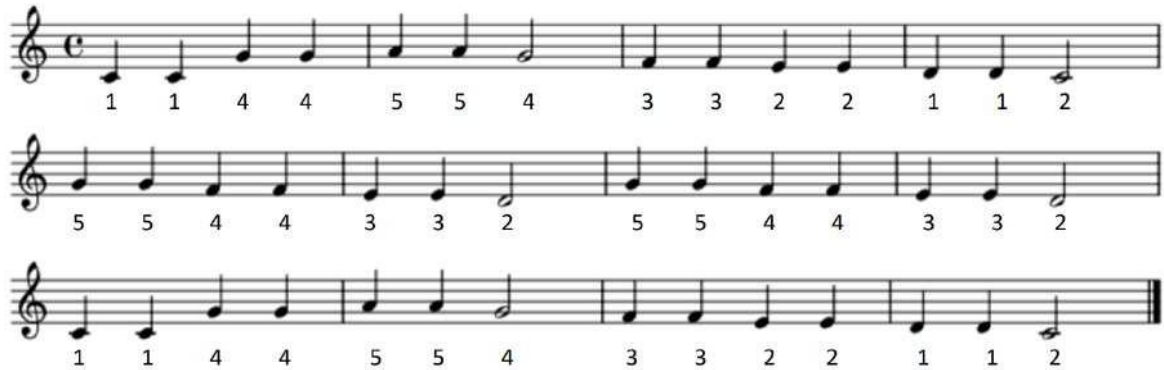

**Figure 4.6:** Piano Sheet with optimal fingering for "Twinkle Twinkle Little Star"

The above experiment is just a "toy-example" and a brief demonstration of the immense capabilities of LLM-based AI-agents through a variety of domains.

### 4.1.4 Challenges & Future Work

It is important to notice that the current example of "Twinkle Twinkle Little Star" is the simplest example of a piano sheet. Advanced piano pieces require fingering for "chords" (playing multiple notes at the same time). For this reason, multiple melodies like the one provided in Figure 4.6 are combined in classical piano pieces. Also, taking into account the timing between consecutive notes is important. There are cases in which it is difficult even for advanced pianists to provide a good fingering, or even choose the appropriate hand with which a note should be played. This area of research is very new and engineers collaborating with musicians are anticipated to overcome the previously mentioned limitations. Fine-tuning LLMs on the appropriate music notation APIs and enhancing them with datasets of music sheets is essential towards achieving the desired automated piano sheet generation.

## 4.2 Utilizing the Local Custom Database architecture towards fulfilling user requests

In this section, a proof-of-concept for the novel on-device methodology introduced in Chapter 3 is provided. First, we use ChatGPT 4 for the generation of human-like user-input. Then, we match the user intention to the pre-defined macros/functions of the custom database of Figure 3.3. For this reason we initially use BERT model for classification. Next, we exploit small generative models. Namely, we use the 8 Billion parameters size version of Llama 3 by Meta and the 2.7 Billion parameter size Phi-2 by Microsoft. The advantage of the generative models is that they can very easily combine two or more pre-defined macros/functions, or generate simple and "obvious" API calls (that do not require a specialized fine-tuned model) towards task execution.

### 4.2.1 Generating human-like user-input

We leverage the abilities of ChatGPT 4 to create 100 examples of user-input that resembles real-world use-cases (Figure 4.9).

### 4.2.2 Leveraging the Local Custom Database

For the aforementioned user-inputs, we mimic the functionality of a local AI-agent that uses the database in Figure 3.3. As the methodology suggests, the attributes of interest are not including the specific list of API calls. We focus on the columns "Use Case", "Scenario Description" and "Task Macro/Function" (Figure 4.7).

| No. | Use Case | Scenario Description | Task Macro/Function |
|---|---|---|---|
| 1 | Personal Finance Management | Track and compare spending on specific categories over time. | TRACK_AND_COMPARE_SPENDING(category, dates) |
| 2 | Smart Home Automation | Adjust home devices based on environmental conditions. | ADJUST_THERMOSTAT_IF_COLD(tempThreshold) |
| 3 | Travel Planning | Find travel arrangements within a budget. | PLAN_TRIP(destination, dates, budget) |
| 4 | Event Scheduling and Notification | Schedule and remind about meetings based on participant availability. | SCHEDULE_MEETING(participants, time) |
| 5 | Health and Fitness Tracking | Log workouts and compare fitness progress. | LOG_WORKOUT_AND_COMPARE(details, period) |
| 6 | Educational Resource Aggregator | Gather educational materials based on topic and level of difficulty. | FIND_STUDY_MATERIALS(topic, level) |
| 7 | Personalized News Digest | Compile news summaries based on user-selected categories. | GENERATE_NEWS_DIGEST(categories) |
| 8 | Personalized Learning Assistant | Create a learning schedule based on user's skills and availability. | CREATE_LEARNING_SCHEDULE(skill, availability) |
| 9 | Job Hunting Automation | Automate the job search process based on specified criteria. | FIND_JOB_LISTINGS(specifications) |
| 10 | Recipe and Meal Planning | Suggest meal plans and recipes based on dietary preferences and nutritional needs. | PLAN_MEALS(dietType, nutrientGoals) |
| 11 | Automated Customer Support | Automate responses to common customer queries and escalate unresolved issues. | HANDLE_CUSTOMER_INQUIRIES(topic) |
| 12 | Real Estate Market Analysis | Analyze real estate market trends and predict investment opportunities. | ANALYZE_MARKET_TRENDS(location, period) |
| 13 | Language Learning Practice | Provide daily language practice exercises tailored to user proficiency. | GENERATE_LANGUAGE_EXERCISES(language, level) |
| 14 | Environmental Impact Tracking | Calculate and track an individual's carbon footprint based on daily activities. | CALCULATE_CARBON_FOOTPRINT(activities) |
| 15 | Virtual Interior Design Consultant | Suggest interior design themes and decorations based on room characteristics and user preferences. | SUGGEST_DESIGN_THEME(roomType, preferences) |

**Figure 4.7:** Example of a Custom Made Database

First we use BERT for classification into the "Use Cases". Then we try the same task using the 8 Billion parameters size version of Llama 3 by Meta and the 2.7 Billion parameter size Phi-2 by Microsoft. The result of the experiments are displayed in Figure 4.8.

**Figure 4.8:** Successful Classification of the 100 Examples of user-input

We observe that in all three cases the classification is successful. Nevertheless, there is an advantage of using a generative model. While BERT was successful at the classification task, Llama 3 was able to handle the user requests with great ease and also provide additional useful explanations while doing so, even in demanding tasks. Phi-2 managed to complete the classification task as well and provided extra information/explanations in simple tasks. The fact that the size of Phi-2 can easily fit on an average smartphone of 2024 is an indicator of the unlimited potential of LLMs and AI-technology. Although this experiment does not constitute a solid proof of the effectiveness of the aforementioned method, it indicates strongly that the integration of LLMs in carry-on devices is a valid, futuristic and useful concept.

1. "Summarize my average workout intensity this month."
2. "Plan a road trip covering major national parks in Utah."
3. "Compare my transportation costs between April and May."
4. "Activate the sprinklers every two days at 7 AM."
5. "How much did I spend on groceries in March?"
6. "List top-rated articles on artificial intelligence basics."
7. "Recommend color schemes for a modern kitchen."
8. "Generate French vocabulary quizz."
9. "Search for eBooks on project management."
10. "Detail my coffee purchases last week."
11. "Show me my entertainment expenses for the last quarter."
12. "Organize a birthday party for my son on July 20th."
13. "Locate a car rental in New York with unlimited mileage."
14. "Find advanced chemistry textbooks for college students."
15. "Provide daily Italian grammar exercises."
16. "Create a meeting agenda for the budget review on Friday."
17. "Calculate the environmental impact of my recent flights."
18. "Remind me to prepare a presentation for Monday's workshop."
19. "Monitor monthly water usage and suggest conservation tips."
20. "Estimate the carbon footprint of my daily commute."
21. "Record a 500-calorie workout session."
22. "Build a study plan for GRE preparation starting next month."
23. "Book a return flight to Berlin for the weekend."
24. "What are the symptoms of dehydration?"
25. "How many calories are in a banana?"
26. "Handle incoming queries about product warranties."
27. "Generate today's top news highlights about the tech industry."
28. "Create a stylish yet functional setup for my home gym."
29. "Design a child-friendly layout for our new nursery."
30. "Direct questions about shipping to the appropriate department."
31. "Assess the impact of recent policy changes on housing markets."
32. "Propose quick breakfast ideas for busy mornings."
33. "Arrange for a pet-friendly hotel in San Francisco next month."
34. "Find a beach resort in Thailand for under $100 per night in July."
35. "Offer meal suggestions for a 1200-calorie diet."
36. "Plan a vegetarian menu for the upcoming week."
37. "Set up an anniversary reminder for June 10th."
38. "List breaking news involving major global economies."
39. "Recommend a three-day itinerary for a Paris trip."
40. "Compile data analyst positions with SQL expertise requirements."
41. "Analyze home price trends in downtown Chicago over the last year."
42. "Compile a weekly digest of political news."
43. "Automate booking confirmations and cancellation responses."
44. "Detail my weight lifting progress over the last three months."
45. "Find entry-level marketing jobs in London."
46. "Summarize today's health and wellness news."
47. "If the outside temperature drops below 18 degrees, turn on the heater."
48. "Generate a shopping list for Italian dinner recipes."
49. "Forecast rental yield for properties near major universities."
50. "Dim the lights at 8 PM every evening."
51. "Close the garage door at 9 PM every night."
52. "Track my heart rate during my morning exercises."
53. "Turn on the garden lights at sunset."
54. "Compare this week's step count with last week."
55. "Gather beginner-level video tutorials on web development."
56. "What's the weather forecast for today?"
57. "Track weekly energy consumption in my household."
58. "Notify all participants about the change in meeting time tomorrow."
59. "Assess the sustainability practices of my lifestyle."
60. "Translate 'Thank you very much' into German."
61. "Schedule daily German language practice sessions."
62. "Log a 30-minute jog in the park today."
63. "Outline a learning path for data science certifications."
64. "Develop a learning timetable for automotive engineering students."
65. "Design a weekly schedule for MBA entrance exam preparation."
66. "Provide a roundup of this week's sports news."
67. "Offer decor ideas for a home office."
68. "Plan a budget-friendly renovation for the guest bathroom."
69. "Adjust the air conditioning to 22 degrees when it's above 30 degrees outside."
70. "Define the term 'blockchain technology'."
71. "Create weekly Japanese speaking tasks."
72. "Provide resources for mastering Python programming."
73. "Identify graphic design vacancies in tech startups."
74. "Schedule a conference call for the sales team next Wednesday at 3 PM."
75. "Activate the security alarm when I leave the house."
76. "Locate customer service jobs that offer evening shifts."
77. "Alert me 15 minutes before any scheduled appointments on Tuesdays."
78. "How do I make a mojito?"
79. "List remote software development positions available in Europe."
80. "What's the current stock price of Apple?"
81. "Search for part-time teaching opportunities in my area."
82. "Compile free online courses for learning Spanish."
83. "Create a monthly digest of advancements in renewable energy."
84. "Suggest furniture arrangements for small living spaces."
85. "Predict future real estate values in the Miami area."
86. "Manage feedback forms submitted through the website."
87. "Sort and prioritize customer complaints received via email."
88. "Can you give me directions to the nearest hospital?"
89. "Compare commercial property prices in different districts."
90. "Evaluate the investment potential of newly developed areas."
91. "Develop a practice schedule for TOEFL preparation."
92. "Arrange daily English pronunciation practice."
93. "Provide quarterly reports on my family's overall environmental footprint."
94. "List my healthcare spending for this year."
95. "Summarize my holiday expenses in December."
96. "Track my gym memberships spending over the past six months."
97. "Respond to FAQs about our refund policy."
98. "Suggest recipes based on ingredients in my fridge."
99. "Offer a monthly Spanish reading comprehension tests."
100. "Can you check if I left the oven on?" (Assuming integration with smart home devices)

**Figure 4.9:** 100 Examples of user-input generated by ChatGPT 4

# Chapter 5

# Conclusion, Discussion & Future Research

## 5.1 Conclusion

This thesis has developed a comprehensive methodology to enhance Large Language Models (LLMs) through the integration of Application Programming Interfaces (APIs), addressing a critical bottleneck in their application: the lack of dynamic interaction with the external digital environment. By introducing a 7-step methodology, this work provides a structured approach for the selection, integration and utilization of APIs with LLMs, thereby expanding their usability and effectiveness in real-world applications. A significant outcome of this research is the proposal of an on-device architecture tailored to utilize small models from the Hugging Face community, enabling efficient AI-agent operations on portable devices. This innovation not only increases the autonomy and context-awareness of AI-agents but also enhances their potential for real-world deployment, particularly in resource-constrained environments. The practical implications of these advancements extend to various sectors, including technology, healthcare and customer service, where AI-agents equipped with enhanced LLMs can perform more complex, timely and relevant tasks, thereby revolutionizing interactions and processes.

## 5.2 Discussion & Future Research

This research contributes to both the theoretical expansion and practical application of AI by enhancing the interaction capabilities of LLMs with APIs. By enabling these models to access real-time data, interact dynamically with other digital systems and execute complex tasks autonomously, the thesis broadens the potential uses of LLMs across various domains.
While this research provides a solid foundation for API integration with LLMs, it also faces limitations, including potential biases in model selection and scalability challenges related to the volume of API calls. The scope of API integration currently focuses on predetermined tasks, which may not extend to all possible applications.
Further research should investigate the integration of an even broader array of APIs, expanding the functional range of LLMs. Additionally, exploring the implementation of these methodologies across various domains and LLM architectures would be beneficial. There is also a vital need to study the long-term learning and adaptation capabilities of AI-agents within dynamic environments to enhance their evolutionary capabilities based on continuous feedback.
Considering Moore's Law and the rapid advancements in computational capabilities, it is anticipated that the deployment of LLMs directly on devices will become feasible in the near future. This progression will likely accelerate the practical applications of AI-agents, making them more accessible and effective.
The methodologies developed in this thesis could significantly impact multiple industries by providing AI solutions that are more dynamic, intelligent and adaptable. Future research focusing on deploying these enhanced LLMs in sectors such as healthcare, for personalized interactions, or finance, for real-time decision-making, could prove transformative.

This thesis not only advances the field of AI but also sets a foundational framework for future research, potentially leading to the construction of truly autonomous and context-aware AI-agents capable of operating across a variety of environments and tasks.